\documentclass[12pt]{article}
\pdfoutput=1
\usepackage[]{hyperref}
\usepackage{xcolor}
\usepackage[]{putex}
\usepackage[utf8]{inputenc}

\usepackage{amsmath,amsthm,amssymb}
\usepackage{tikz}
\usepackage{tikz-cd}
\usetikzlibrary{decorations.markings}
\usetikzlibrary{calc}
\usepackage{subcaption}
\usepackage{enumerate}
\usepackage{enumitem}

\numberwithin{equation}{section}

\makeatletter
\DeclareRobustCommand*{\bfseries}{%
  \not@math@alphabet\bfseries\mathbf
  \fontseries\bfdefault\selectfont
  \boldmath
}
\makeatother
\hypersetup{
  colorlinks=true,
  linkcolor=blue,
  citecolor=green!60!black,
  urlcolor=cyan!70!black,
  linktoc=all
  }

\input{glyphtounicode}
\def\musepic#1{\vcenter{\hbox{\usebox{#1}}}}
\newmuskip\pFqmuskip

\newcommand*\pFq[6][8]{%
  \begingroup 
  \pFqmuskip=#1mu\relax
  \mathchardef\normalcomma=\mathcode`,
  \mathcode`\,=\string"8000
  \begingroup\lccode`\~=`\,
  \lowercase{\endgroup\let~}\pFqcomma
  {}_{#2}F_{#3}{\left[\genfrac..{0pt}{}{#4}{#5};#6\right]}%
  \endgroup
}
\newcommand{\pFqcomma}{{\normalcomma}\mskip\pFqmuskip}

\newsavebox{\figcalWStar}
\savebox{\figcalWStar}{%
}

\begin{document}

\preprint{PUPT-2589}

\title{Propagator identities, holographic conformal blocks, and higher-point AdS diagrams}
\authors{Christian Baadsgaard Jepsen$^1$\footnote{\tt cjepsen@princeton.edu} \& Sarthak Parikh$^2$\footnote{\tt sparikh@caltech.edu}}
\institution{PU}{$^1$Joseph Henry Laboratories, Princeton University, Princeton, NJ 08544, USA}
\institution{Caltech}{$^2$Division of Physics, Mathematics and Astronomy, California Institute of Technology,\cr\hskip0.06in Pasadena, CA 91125, USA}

\abstract{
Conformal blocks are the fundamental, theory-independent building blocks in any CFT, so it is important to understand their holographic representation in the context of AdS/CFT. We describe how to systematically extract the holographic objects which compute higher-point global (scalar) conformal blocks in arbitrary spacetime dimensions, extending the result for the four-point block, known in the literature as a geodesic Witten diagram, to five- and six-point blocks. The main new tools which allow us to obtain such representations are various higher-point propagator identities, which can be interpreted as generalizations of the well-known flat space star-triangle identity, and which compute integrals over products of three bulk-to-bulk and/or bulk-to-boundary propagators in negatively curved spacetime. Using the holographic representation of the higher-point conformal blocks and higher-point propagator identities, we develop geodesic diagram techniques to obtain the explicit direct-channel conformal block decomposition of a broad class of higher-point AdS diagrams in a scalar effective bulk theory, with closed-form expressions for the decomposition coefficients. These methods require only certain elementary manipulations and no bulk integration, and furthermore provide quite trivially a simple algebraic origin of the logarithmic singularities of higher-point tree-level AdS diagrams. We also provide a more compact repackaging in terms of the spectral decomposition of the same diagrams, as well as an independent discussion on the closely related but computationally simpler framework over $p$-adics which admits comparable statements for all previously mentioned results.
}

\maketitle

\setcounter{tocdepth}{2}

{\hypersetup{linkcolor=black}
\tableofcontents
}


\section{Introduction}
\label{INTRO}

The anti-de Sitter/conformal field theory (AdS/CFT) correspondence~\cite{Maldacena:1997re,Gubser:1998bc,Witten:1998qj} provides a powerful repackaging of CFTs in terms of gravitational theories in asymptotically AdS spacetimes and vice versa.
Particularly, conformal correlators in large $N$ CFTs admit a perturbative holographic expansion in $1/N$  in terms of bulk Feynman diagrams (also referred to as Witten diagrams or AdS diagrams).
From the CFT perspective, repeated application of operator product expansion (OPE) in a correlator reduces any higher-point correlator into a combination of two- and three-point functions, which are fixed entirely by conformal invariance, up to an overall constant for the three-point function given by the OPE coefficient.
The resulting decomposition can be interpreted as a conformal block decomposition (CBD) which provides an efficient organization of the kinematic and dynamical information in the correlators, in terms of an expansion in the basis of appropriate conformal blocks (the theory independent, non-perturbative, conformally invariant fundamental building blocks of correlators) and the CFT data (the spectrum of operators in the theory and the associated OPE coefficients).

In the case of four-point correlators, the associativity of taking the OPE provides a powerful constraint, called the crossing equation, which via the conformal bootstrap program~\cite{Ferrara:1973yt,Polyakov:1974gs,Rattazzi:2008pe} has provided one of the strongest numerical and analytical approaches towards solving (higher-dimensional) CFTs (see e.g.\ Refs.~\cite{ElShowk:2012ht,Poland:2018epd}).
Holographically, the AdS diagram expansion of CFT correlators organizes itself such that it solves the crossing equation order by order in $1/N$, as established at leading~\cite{Heemskerk:2009pn} and subleading orders~\cite{Aharony:2016dwx} in $1/N$ in simple cases.
The four-point exchange AdS diagrams in Mellin space~\cite{Mack:2009mi,Penedones:2010ue} (up to certain contact interactions) are also known to be directly related to the four-point conformal block~\cite{Fitzpatrick:2012cg}.\footnote{Subsequently an alternate attractive holographic interpretation for four-point blocks was provided~\cite{Hijano:2015zsa}, which we will comment  on shortly.}
Moreover, these diagrams appear directly as an expansion basis in a variant of the bootstrap approach also in Mellin space~\cite{Gopakumar:2016wkt,Gopakumar:2016cpb,Gopakumar:2018xqi} pioneered in Ref.~\cite{Polyakov:1974gs}.

Given the central role and importance of AdS diagrams in AdS/CFT, they have been the subject of much interest and considerable progress over the past decade.
Arguably the most powerful results so far have been obtained in Mellin space, where Mellin amplitudes in effective scalar field theories on AdS can be written in closed-form series or contour integral representations, for \emph{arbitrary} tree-level AdS diagrams~\cite{Penedones:2010ue,Fitzpatrick:2011ia,Paulos:2011ie,Nandan:2011wc,Kharel:2013mka,Goncalves:2014rfa}, as well as for certain classes of higher-loop diagrams~\cite{Penedones:2010ue,Cardona:2017tsw,Yuan:2017vgp,Yuan:2018qva,Ghosh:2018bgd,Liu:2018jhs}.
 There are also recursive techniques for computing tree-level AdS diagrams in momentum space in four~\cite{Albayrak:2018tam,Albayrak:2019asr} and higher-dimensional~\cite{Raju:2010by,Raju:2011mp,Raju:2012zr,Raju:2012zs} bulk spacetime.
To a limited extent, higher-loop results have also been obtained directly in position space using bulk~\cite{Giombi:2017hpr,Bertan:2018khc,Bertan:2018afl} as well as  CFT techniques~\cite{Hikida:2016wqj,Aharony:2016dwx,Hikida:2017ecj,Alday:2017xua,Aprile:2017bgs,Ponomarev:2019ltz}. However, most position space results have been limited to up to four-point AdS diagrams,\footnote{See however, the recent paper~\cite{Goncalves:2019znr}.} and as such relatively little is known about the position space representation of higher-point diagrams, even at tree-level.

AdS diagrams are by construction conformally invariant, thus like conformal correlators they admit CBDs in any choice of conformal basis.
The CBD is perhaps best understood via harmonic analysis on the (Euclidean) conformal group $SO(d+1,1)$~\cite{Mack:1974jjo,Mack:1974sa,Dobrev:1977qv,Dobrev:1975ru}.
Particularly, the shadow formalism~\cite{Ferrara:1972uq,Ferrara:1972ay,Ferrara:1972xe,Ferrara:1973vz} provides a convenient framework for writing down conformal blocks~\cite{SimmonsDuffin:2012uy} as well as the decomposition of conformal correlators in position space.
The main objects here are the so called conformal partial waves, which are given in terms of linear combinations of conformal blocks and their ``shadow blocks''.
This formalism allows a convenient rewriting of AdS diagrams as spectral integrals, from which the CBD can in principle be obtained by evaluating all (contour) integrals.
However such integrals can get increasingly tedious to evaluate for higher-point diagrams, rendering the path from the spectral decomposition to the explicit CBD somewhat unwieldy.
Thus it remains fruitful to search for methods which can yield the explicit CBD directly.
Furthermore, explicit closed-form expressions for conformal partial waves or global conformal blocks are not known except in a relatively small number of cases, such as for low-point blocks or in low spacetime dimensions.
While closed-form expressions or series representations are known for the global scalar conformal blocks  in general spacetime dimension $d$ at four~\cite{Ferrara:1971vh,Ferrara:1973vz,Ferrara:1974ny,Dolan:2000ut,Dolan:2003hv,Dolan:2011dv} and (only very recently) five points~\cite{Rosenhaus:2018zqn}, at six points and higher, the only  global conformal blocks for which closed-form expressions are available so far are those in the comb channel in $d=1$ and $d=2$~\cite{Rosenhaus:2018zqn}.
Knowledge of higher-point blocks in arbitrary spacetime dimensions thus remains an important missing link in the study of higher-point AdS diagrams.\footnote{See, however, Refs.~\cite{Fortin:2019fvx,Fortin:2019dnq} for recent results on obtaining recursively higher-point conformal blocks and conformal correlators via the embedding space formalism.}

It is useful to study higher-point diagrams  because their decomposition involves multi-twist exchanges.\footnote{A notational remark: The class of double-trace primaries of twist $\Delta_a +\Delta_b+2M-\ell$ and spin $\ell$ is constructed out of scalar operators ${\cal O}_a, {\cal O}_b$ of dimensions $\Delta_a, \Delta_b$ respectively, written schematically as
\eqn{doubletwist}
{
[\mathcal{O}_a\mathcal{O}_b]_{M,\ell}
\approx
\mathcal{O}_a\partial^{2M} \partial_{\mu_1}\ldots \partial_{\mu_\ell} \mathcal{O}_b + {\rm traces}\,.
}
We will interchangeably refer to the operator in~\eno{doubletwist} as ``double-twist'' or double-trace. ``Multi-twist'' operators appearing in this paper will usually arise as double-twists of double-twists and so on. Whenever we refer to ``higher-twist'' operators, we will mean non-zero values of $M$ in~\eno{doubletwist}, and since we will only be dealing with scalar external and exchanged operators, the terminology ``lowest-twist'' operators will refer to the case $M=0$.} Multi-twist exchanges also appear in the conformal perturbation theory of lower-point diagrams such as in the context of the lightcone bootstrap approach~\cite{Komargodski:2012ek,Fitzpatrick:2012yx,Fitzpatrick:2014vua,Kaviraj:2015cxa,Kaviraj:2015xsa,Alday:2015ewa,Alday:2016njk,Simmons-Duffin:2016wlq,Caron-Huot:2017vep,Albayrak:2019gnz}.
Thus understanding various analytic limits of higher-point AdS diagrams can be  useful in gaining further understanding of four-point  crossing symmetry constraints in various regimes.
 Such decompositions  can further be quite useful in setting up an $n$-point analog of the four-point crossing equations and conformal bootstrap with external scalar operators, which collectively  may possibly be sufficient and present  analytical or numerical advantages over the usual four-point program where one must also include all spinning operators in the spectrum~\cite{Rosenhaus:2018zqn}.

The present paper aims to partially fill the gap in the study of higher-point AdS diagrams, particularly in an effective scalar field theory on AdS${}_{d+1}$ by developing a systematic study of higher-point global conformal blocks in arbitrary spacetime dimensions.
Specifically, we will develop tools to obtain the \emph{holographic} representation of higher-point blocks, expressed in terms of geodesic diagrams.
These tools include various integral AdS propagator identities, one of which was used recently to obtain the holographic dual of the five-point block~\cite{Parikh:2019ygo}.
In this paper, we will apply these tools to obtain the six-point block in the so called OPE channel.
Further, we will generalize the geodesic diagram techniques of Ref.~\cite{Hijano:2015zsa} to obtain the explicit direct channel CBD of all   tree-level scalar five-point diagrams with scalar exchanges and a significant subset of six-point diagrams (more precisely, those which admit a direct channel decomposition in the so-called OPE channel).
Like in the case of four-point diagrams~\cite{Hijano:2015zsa}, such calculations will not involve any bulk or contour integrations, but only algebraic steps.
The analysis presented provides the road-map for extensions to conformal blocks beyond six-points.
Moreover, we will also present the parallel story in the closely related framework of $p$-adic AdS/CFT~\cite{Gubser:2016guj,Gubser:2017tsi}, which affords a useful toy model for studying conformal blocks and CBDs.
Here, we will present the corresponding propagator identities on the Bruhat--Tits tree, the holographic duals of the five-point block as well as the six-point block in the OPE channel, and apply geodesic diagram methods to obtain the CBD of five- and six-point diagrams in $p$-adic AdS/CFT.

In the remainder of this section, we expand lightly on the setup of this paper, before ending with an outline.

\paragraph{Holographic conformal blocks and propagator identities.}

At four-points, the scalar contact and exchange diagrams in an effective scalar field theory in AdS admit a direct channel decomposition as a sum over infinitely many four-point conformal blocks, each representing the exchange of an appropriate representation of the conformal group, corresponding to higher-twist double-trace exchanges (more generally ``double-twist'' exchanges) and additionally a single-trace exchange in the case of the exchange diagram, each weighted essentially by factors of OPE coefficients squared~\cite{Liu:1998th}.
    Recent work has focused on alternate efficient means of obtaining such decompositions, including the split representation of bulk-to-bulk propagators~\cite{Costa:2014kfa,Sleight:2017fpc}, the conformal Casimir equation~\cite{Zhou:2018sfz}, and the use of the so-called geodesic Witten diagrams~\cite{Hijano:2015zsa}.\footnote{ Recently, progress has also been made in obtaining relations obeyed by the decomposition coefficients of four-point exchange diagrams in the \emph{crossed} channel~\cite{Hoffmann:2000mx,Sleight:2018epi,Sleight:2018ryu,Cardona:2018dov,Liu:2018jhs,Gopakumar:2018xqi,Zhou:2018sfz,Li:2019dix,Chen:2019gka} but in this paper we will restrict our discussion to only direct channel decomposition.}

The four-point geodesic diagram is a four-point exchange AdS diagram, except with both AdS integrations replaced by geodesic integrals over boundary anchored geodesics joining pairs of  boundary insertion points.
Such a holographic object computes precisely the four-point global conformal block~\cite{Hijano:2015zsa,Nishida:2016vds,Castro:2017hpx,Dyer:2017zef,Chen:2017yia,Gubser:2017tsi,Kraus:2017ezw,Tamaoka:2017jce,Nishida:2018opl,Das:2018ajg} (see also Refs.~\cite{Bhatta:2016hpz,Bhatta:2018gjb} for an alternate point of view).\footnote{In AdS${}_{3}$/CFT${}_2$, various limits of \emph{Virasoro} blocks, obtained by taking particular heavy/light limits of dimensions of external operators, are also interpreted in terms of lengths of bulk geodesics and as geodesic diagrams in defect geometries~\cite{Fitzpatrick:2014vua,Hijano:2015rla, Fitzpatrick:2015zha,Hijano:2015qja,Besken:2016ooo}). In some cases, higher-point results ($n \geq 5$) are also available~\cite{Alkalaev:2015wia,Alkalaev:2015lca,Alkalaev:2015fbw,Banerjee:2016qca,Alkalaev:2016rjl,Alkalaev:2018nik,Alkalaev:2019zhs}.}
The holographic conformal block representation, together with certain crucial two-propagator identities reduce the task of obtaining the decomposition of four-point AdS diagrams in the direct channel to a number of elementary algebraic operations, with no further need to evaluate bulk integrals~\cite{Hijano:2015zsa}.
With some work this approach can be extended to a higher-point setting as is done in this paper; consequently one needs the holographic duals of higher-point conformal blocks, as well as higher-point  generalizations of the two-propagator identities.
In addition to being useful for obtaining the decomposition of AdS diagrams, each of these generalizations is of interest in its own right, as we now briefly describe.

Global conformal blocks are projections of conformal correlators onto the contribution from individual conformal families, associated to representations of the $d$-dimensional global conformal group.
The representations are labeled by the conformal dimensions and spin.
In this paper we will focus only on scalar conformal blocks with scalar intermediate exchanges, so from here on we will suppress the spin label. We leave extensions to external and exchanged spin operators along the lines of Refs.~\cite{Hijano:2015zsa,Nishida:2016vds,Castro:2017hpx,Dyer:2017zef,Chen:2017yia,Kraus:2017ezw,Tamaoka:2017jce,Nishida:2018opl,Das:2018ajg} for the future.

\begin{figure}[!t]
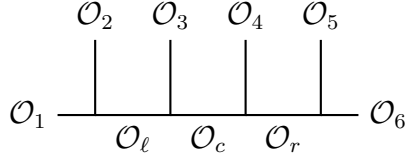

    \centering
    \[ \musepic{\figSixCombChannel}  \]
    \caption{The global scalar six-point block  $W_{\Delta_\ell;\Delta_c;\Delta_r}^{\Delta_1,\ldots,\Delta_6}(x_i)$, in the comb channel. We will not discuss its holographic representation in this paper.}
    \label{fig:6combchannel}
\end{figure}

The five-point conformal block corresponding to the projection onto the conformal multiplets labeled by weights $\Delta_a$ and $\Delta_{b}$ (and zero spin) can be written as\footnote{Following the nomenclature in recent literature, we reserve the term conformal block to refer to objects such as the one in~\eno{ProjectConf5}, which include the entire position space dependence as opposed to dependence merely on the conformal cross-ratios. This is in contrast with the notation used in Ref.~\cite{Parikh:2019ygo} where this object was referred to as a ``conformal partial wave'', a term that in this paper is instead reserved for the object which is given by a linear combination of a conformal block and its shadow blocks, and which has useful orthogonality and single-valuedness properties.}
\eqn{ProjectConf5}{
W_{\Delta_a;\Delta_b}^{\Delta_1,\ldots,\Delta_5}(x_i) \equiv {\langle  {\cal O}_1(x_1) {\cal O}_2(x_2) P_{\Delta_a} {\cal O}_3(x_3) P_{\Delta_{b}} {\cal O}_4(x_4) {\cal O}_5(x_5) \rangle \over C_{\Delta_1\Delta_2\Delta_a} C_{\Delta_a \Delta_3 \Delta_{b}} C_{\Delta_b\Delta_4\Delta_{5}}} \,,
}
where $P_\Delta = \sum_k | P^k {\cal O}_\Delta \rangle \langle P^k {\cal O}_\Delta|$ is the projection operator projecting onto the conformal family of the primary ${\cal O}_\Delta$.
The OPE coefficients, given by $C_{\Delta_i \Delta_j \Delta_k}$, have been quotiented out in the expression to obtain a purely kinematical quantity.
\begin{figure}[!t]
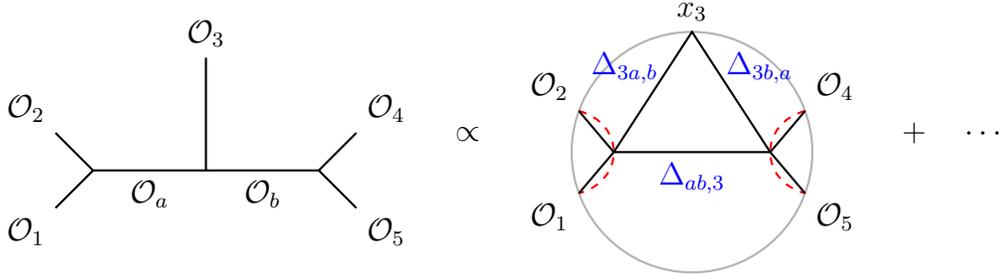

    \centering
    \[ \musepic{\figFiveBlock} \quad \propto \quad \musepic{\figcalWfiveLeading} \quad + \quad \cdots \]
    \caption{Graphical representation of the scalar five-point conformal block, $W_{\Delta_a;\Delta_b}^{\Delta_1,\ldots,\Delta_5}(x_i)$ (LHS), and the leading term in its holographic representation (RHS). All solid lines in the bulk diagram on the RHS are bulk-to-bulk or bulk-to-boundary propagators, with the two bulk vertices integrated over boundary anchored geodesics (red dashed curves) and the conformal dimensions associated with some of the propagators explicitly displayed in blue (which makes use of the shorthand~\eno{DeltaijkDef}). The ellipsis represents contribution from the exchange of descendants in the conformal multiplet of primaries ${\cal O}_a$ and ${\cal O}_b$. The precise relation, obtained in Ref.~\cite{Parikh:2019ygo}, can be found in~\eno{calW5confwave}.}
    \label{fig:5block}
\end{figure}
Graphically, we will often represent the five-point conformal block as shown in the LHS of figure~\ref{fig:5block}. At six points and higher, conformal blocks admit topologically distinct channels not simply related to each other by permutations of operators and boundary insertions or conformal transformations.
The two conformally distinct channels for the six-point block are the so-called comb channel, given by
\eqn{ProjectConf6comb}{
W_{\Delta_\ell;\Delta_c; \Delta_r}^{\Delta_1,\ldots,\Delta_6}(x_i) \equiv {\langle  {\cal O}_1(x_1) {\cal O}_2(x_2) P_{\Delta_\ell} {\cal O}_3(x_3) P_{\Delta_{c}} {\cal O}_4(x_4) P_{\Delta_r} {\cal O}_5(x_5) {\cal O}_6(x_6) \rangle \over C_{\Delta_1\Delta_2\Delta_\ell}  C_{\Delta_\ell \Delta_3 \Delta_c} C_{\Delta_c\Delta_4\Delta_{r}} C_{\Delta_r \Delta_5 \Delta_{6}}} \,,
}
 and the OPE channel, written as
 \eqn{ProjectConf6OPE}{
W_{\Delta_\ell;\Delta_c; \Delta_r}^{\Delta_1,\ldots,\Delta_6}(x_i) &\equiv {1 \over C_{\Delta_1\Delta_2\Delta_\ell}  C_{\Delta_3 \Delta_4 \Delta_c} C_{\Delta_r\Delta_5\Delta_{6}} C_{\Delta_\ell \Delta_c \Delta_{r}}} \sum_{k_\ell, k_c,k_r} \langle  {\cal O}_1(x_1) {\cal O}_2(x_2)| P^{k_\ell} {\cal O}_{\Delta_\ell}\rangle  \cr
& \times \langle P^{k_c} {\cal O}_{\Delta_c}|{\cal O}_3(x_3) {\cal O}_4(x_4) \rangle  \langle  P^{k_\ell} {\cal O}_{\Delta_\ell}|P^{k_c} {\cal O}_{\Delta_c}  P^{k_r} {\cal O}_{\Delta_r}\rangle \langle  P^{k_r} {\cal O}_{\Delta_r}| {\cal O}_5(x_5) {\cal O}_6(x_6) \rangle  \,.
}
The graphical representations of these blocks, shown in figures~\ref{fig:6combchannel}-\ref{fig:6opechannel} are perhaps more illuminating and in fact suggestive of the names for the channels. Later in this paper we will provide an alternative definition of these blocks based on the Casimir equations and appropriate boundary conditions they satisfy.\footnote{The four- and five-point blocks may be interpreted as examples of comb channel blocks, but the four-point block also  qualifies as an OPE channel block.}
\begin{figure}[!t]
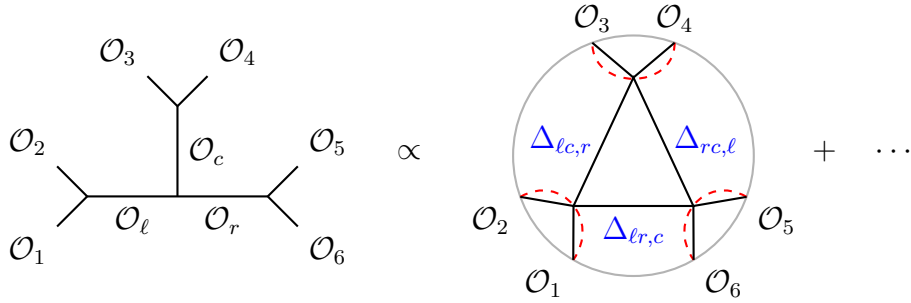

    \centering
    \[ \musepic{\figSixOPEChannel} \quad \propto \quad \musepic{\figcalWsixLeading} + \quad \cdots \]
    \caption{The graphical representation of the global scalar six-point block in the OPE channel, $W_{\Delta_\ell;\Delta_c;\Delta_r}^{\Delta_1,\ldots,\Delta_6}(x_i)$ (LHS), and the leading term in its holographic representation (RHS). To interpret the RHS, see the caption of figure~\ref{fig:5block}. The precise relation can be found in~\eno{calW6starConfwave}.}
    \label{fig:6opechannel}
\end{figure}

For the five-point block, a \emph{series} representation was obtained using the shadow formalism~\cite{Rosenhaus:2018zqn}, while the recently obtained holographic dual of the five-point block provides an alternate mixed \emph{integral} and series representation~\cite{Parikh:2019ygo}.
The first term in the holographic representation is displayed in the RHS of figure~\ref{fig:5block}.
The alternate representation makes the holographic origin of the five-point block more transparent, and in this paper this transparency is leveraged to furnish the CBD of all tree-level scalar five-point AdS diagrams.
Moreover, in this paper we will obtain the general $d$-dimensional holographic representation for the six-point block in the OPE channel (see the RHS of figure~\ref{fig:6opechannel}), for which no other representation, either from the boundary or the bulk perspective, is known at the moment, and apply it in decomposing a class of six-point tree-level AdS diagrams.
These new representations may also be useful in investigating analytically  relatively less understood physical properties of higher-point blocks, such as various non-OPE limits.

This paper also establishes higher-point propagator identities that equate products of bulk-to-bulk and bulk-to-boundary propagators, incident at a common bulk point that is integrated over all of bulk space, with unintegrated expressions involving linear combinations of these propagators.
These identities provide examples of higher-point ``bulk scattering amplitudes''. For instance, we present a three-particle bulk scattering amplitude in AdS (i.e.\ a product of three bulk-to-bulk propagators incident on a cubic contact vertex to be integrated over all of AdS)  as the AdS generalization of the well-known flat space star-triangle identity~\cite{DEramo:1971hnd}.
See figure~\ref{fig:GGGschematic} for a schematic depiction of this identity.
\begin{figure}[!t]
    \centering
    \[ \musepic{\figIntGGG} = C_{\Delta_a\Delta_b\Delta_c} \musepic{\figIntGGGaLeading} + \cdots  \]
    \caption{A schematic representation of a three-propagator identity. The common point of intersection of three bulk-to-bulk propagators, shown as a green disk on  the LHS is to be integrated over all of AdS. The overall factor of $C_{\Delta_a\Delta_b\Delta_c}$ is the OPE coefficient associated with primaries of conformal dimensions $\Delta_a, \Delta_b$ and $\Delta_c$. We have only shown one of a four-fold infinity of terms which appear on the RHS.  The precise identity can be found in~\eno{GGGreal}. See section~\ref{THREEPROPIDS} for variants of this identity involving factors of the bulk-to-boundary propagator.}
    \label{fig:GGGschematic}
\end{figure}
Furthermore, the higher-point AdS propagator identities derived in this paper  enable a physical decomposition of various AdS integrals into terms each of which can be interpreted as corresponding to the contribution to an AdS diagram coming from a particular (multi-twist) operator exchange.
In the future, the identities may also prove useful in evaluating or simplifying various loop-level AdS diagrams.

\paragraph{A $p$-adic toy model.}
Another computational tool we make use of in developing the higher-point holographic functions program is the framework of $p$-adic AdS/CFT~\cite{Gubser:2016guj,Heydeman:2016ldy}.
In this discrete version of holography, boundary operators  are real- or complex-valued maps from the (projective line over) $p$-adic numbers or an algebraic extension thereof.
As a consequence, spinning operators and local derivatives are absent so that not only is the CFT devoid of descendants, but it only contains the lowest-twist operators~\cite{Melzer:1988he,Gubser:2017tsi}.
 For instance, the class of double-trace primaries~\eno{doubletwist} exists only at $M=0, \ell=0$ in such $p$-adic CFTs, and similarly for higher-trace operators.
So the decomposition of AdS diagrams in $p$-adic AdS/CFT is especially simple, with all conformal blocks reduced to scaling blocks given by trivial power laws of conformal cross-ratios (due to the absence of descendants in conformal families), and the presence of only the lowest-twist contributions in the decomposition (due to the absence of local derivative operators); see Ref.~\cite{Gubser:2017tsi} for a demonstration in the case of the  four-point diagrams.
Correspondingly, we will show that the holographic duals of the five- and six-point blocks will be \emph{fully} specified  precisely by the single term shown on the RHS in figures~\ref{fig:5block} and~\ref{fig:6opechannel}.
This is a drastic simplification of the situation in conventional AdS/CFT.
Moreover, the general structure of the CBD and the decomposition coefficients turns out to be strikingly reminiscent of the results from conventional AdS/CFT, as will also be demonstrated for higher-point diagrams in this paper.

Indeed, despite the dramatic simplicity of the $p$-adic setup alluded to above, computations and results closely echo those encountered in the conventional AdS/CFT literature. Some examples include the  (adelically) identical functional forms of OPE coefficients when expressed in terms of local zeta functions~\cite{Gubser:2017tsi}, the similar structure of conformal correlators~\cite{Gubser:2016guj,Gubser:2017vgc,Gubser:2017tsi}, the existence of geodesic bulk diagrams which serve as holographic duals of conformal blocks~\cite{Gubser:2017tsi}, universal real/$p$-adic closed-form expressions for Mellin amplitudes for arbitrary tree-level bulk diagrams~\cite{Jepsen:2018dqp,Jepsen:2018ajn}, and tensor network constructions~\cite{Heydeman:2016ldy,Marcolli:2018ohd,Heydeman:2018qty} to name a few (see Refs.~\cite{Gubser:2016htz,Bhattacharyya:2017aly,Dutta:2017bja,Gubser:2017qed,Qu:2018ned,Gubser:2018cha,Hung:2018mcn,Hung:2019zsk} for other developments).
For this reason, $p$-adic AdS/CFT serves as a convenient toy model, affording significant computational advantages while at the same time informing the more involved setup over reals.
Thus we will find it beneficial to make a brief detour to the $p$-adic setup before turning our attention to conventional AdS${}_{d+1}$/CFT${}_d$ over the reals.

\paragraph{Outline.}
An outline for the rest of the paper is as follows:
\begin{itemize}
    \item In section~\ref{PADIC}, in the context of $p$-adic AdS/CFT, we employ propagator identities to obtain the holographic duals of five- and six-point conformal blocks on the Bruhat--Tits tree. Further, CBD of five-point diagrams is presented. This section is restricted to the $p$-adic setting, but the computations and results find very close analogs with the conventional (real) AdS$_{d+1}$/CFT$_d$ setting discussed in the subsequent sections.
    The discussion is presented such that the rest of the paper can be read independently of this section.

    \item From section~\ref{PROPIDS} onward the paper essentially pans out in the conventional  AdS$_{d+1}$/CFT$_d$ setting over the reals. In section~\ref{PROPIDS} we present new higher-point bulk-to-bulk and bulk-to-boundary propagator identities which relate integrals over all of bulk AdS  of products of propagators to infinite sums over unintegrated combinations of  propagators.

    \item In section~\ref{SIXOPEGEODESIC} we  present new results on the holographic dual of the six-point global scalar conformal block in the OPE channel, and show that it satisfies the correct conformal Casimir equations with the right boundary conditions. Like in the five-point case, the holographic representation of the six-point block is given in terms of an infinite linear combination of six-point geodesic diagrams.

    \item In section~\ref{CBD} we provide a derivation of the CBD for all tree-level five-point diagrams -- the scalar contact diagram, various five-point  exchange diagrams admitting scalar exchanges, as well as a class of six-point diagrams which admits a direct channel decomposition in the OPE channel (which includes the six-point contact diagram, and several six-point exchange diagrams with one, two, or three exchanges). The computation involves a higher-point generalization of the geodesic diagram techniques; the intermediate steps are essentially purely algebraic and no further bulk integration or contour integrals are necessary, although some hypergeometric summation identities will be needed.
    We end the section with comments on the algebraic origins of logarithmic singularities (section~\ref{LOG}) and the relation to the spectral decomposition of AdS diagrams (section~\ref{SPECTRAL}).

    \item Finally, in section~\ref{DISCUSSION}, we end the paper with a discussion of the results  and future directions.

    \item In the appendices we provide the explicit derivation of the spectral decomposition of a few simple diagrams (appendix~\ref{4SPECTRAL}), and proofs of all new propagator and hypergeometric summation identities  (appendix~\ref{PROOFIDS}) utilized in the main text.
\end{itemize}

\vspace{-1em}
\paragraph{Notation.} We introduce a convenient shorthand for conformal dimensions which will be used frequently throughout the paper:
\eqn{DeltaijkDef}{
\Delta_{i_1\ldots i_\ell,i_{\ell+1} \ldots i_k} \equiv {1\over 2} \left( \Delta_{i_1} + \cdots + \Delta_{i_\ell} - \Delta_{i_{\ell+1}} - \cdots - \Delta_{i_k} \right).
}

\section{A $p$-adic \'{e}tude}
\label{PADIC}

In this section we will be focusing on the computationally simpler setup of $p$-adic AdS/CFT.
This section can be read independently from the rest of the paper but serves as a warm-up to the later sections over the reals, and the patient reader may benefit from the general lessons and the less cluttered discussion afforded by the $p$-adic setup.

One of the practical benefits of the $p$-adic AdS/CFT setup of Refs.~\cite{Gubser:2016guj,Heydeman:2016ldy} is that simple bulk theories of massive scalar fields in a fixed negatively curved spacetime are modelled as scalar lattice theories on a regular tree (called the Bruhat--Tits tree) with polynomial contact interactions,\footnote{More generally, higher-order derivative couplings are incorporated as $({\rm next})^k$-to-nearest neighbor interactions in the discrete setting, with $k\geq 0$~\cite{Gubser:2017tsi}.} which dramatically simplifies bulk computations of such objects as the amplitudes associated with bulk Feynman diagrams.
The putative dual conformal field theory lives on the boundary of the Bruhat--Tits tree described by the projective line over the $p$-adic numbers (or some appropriate extension of $p$-adic numbers).
Here we will restrict ourselves to the field $\mathbb{Q}_{p^d}$, which is the unique unramified extension of $p$-adic numbers of degree $d$, which forms a $d$-dimensional vector space over the $p$-adic numbers $\mathbb{Q}_p$ while maintaining a field structure.
(For a review on $p$-adic numbers and their extensions, see, e.g.\ Refs.~\cite[Section 2]{Gubser:2016guj} and~\cite{gouvea1997p}.)
The $p$-adic conformal field theory, with global conformal group ${\rm PGL}(2,\mathbb{Q}_{p^d})$, does not admit local derivative operators; consequently there are no descendants in the conformal family and all operators are the lowest-twist zero-spin single- and multi-trace primary operators~\cite{Melzer:1988he,Gubser:2017tsi}.
Thus the global conformal blocks are trivial, and the conformal block decomposition of CFT correlators is significantly uncomplicated, as will become apparent below.

This section is organized as follows. We will begin in section~\ref{PROPBTTREE} by presenting various propagator identities,  involving bulk integration on the Bruhat--Tits tree of a product of three bulk-to-bulk and/or bulk-to-boundary propagators over a common bulk vertex, adapted from Ref.~\cite{Gubser:2017tsi}. These will then be used  in section~\ref{DUALBTTREE} to obtain the holographic duals of five- and six-point conformal blocks in terms of geodesic diagrams. In section~\ref{CBDBTTREE} we will present a few representative examples demonstrating the geodesic diagram approach to obtaining the CBD of five-point bulk diagrams and interpret the simplicity of the procedure and the final result. More examples are reserved for later in section~\ref{CBD} where we  comment on the close connection between CBD in the $p$-adic and conventional AdS/CFT formalisms.

\subsection{Propagator identities on the Bruhat--Tits tree}
\label{PROPBTTREE}

We collect here various propagator identities which will prove useful in extracting the holographic objects that compute various higher-point global conformal blocks as well as in obtaining the CBD of bulk diagrams in $p$-adic AdS/CFT. These identities were already written down in Ref.~\cite{Gubser:2017tsi}, but here we restate them in a slightly different but mathematically equivalent form, which accommodates a direct analogy with the corresponding propagator identities in real AdS, the subject of section~\ref{PROPIDS}.
The identities described below are set up on the Bruhat--Tits tree ${\cal T}_{p^d}$, a $(p^d+1)$-regular graph with the associated boundary $\partial {\cal T}_{p^d} = \mathbb{P}^1(\mathbb{Q}_{p^d})$.

We first briefly review the propagators of $p$-adic AdS/CFT; for more details refer to Ref.~\cite{Gubser:2016guj}. The normalizable and non-normalizable solutions of the Laplace equation on the Bruhat--Tits tree  will be denoted $\hat{G}_\Delta$ and $\hat{K}_\Delta$, and they represent the bulk-to-bulk and bulk-to-boundary propagators on the Bruhat--Tits tree, respectively. The bulk-to-bulk propagator is given by
\eqn{pG}{
\hat{G}_\Delta(w,z) = p^{-\Delta\: {\mathsf d}(w,z)}
}
where ${\mathsf d}(w,z)$ is the graph-distance on the Bruhat--Tits tree between bulk nodes $w$ and $z$,
and we have chosen the normalization such that $\hat{G}$ is the Green's function of the Bruhat--Tits Laplace equation
\eqn{pGNorm}{
(\square_z + m_\Delta^2)\hat{G}_\Delta(w,z) = {-1\over N_\Delta} \delta(w,z) \qquad N_\Delta \equiv {-\zeta_p(2\Delta) \over 2\nu_\Delta \zeta_p(2\Delta-d)} \qquad 2\nu_\Delta \equiv p^\Delta -p^{d-\Delta}\,,
}
where $\square_z$ is the graph Laplacian acting on the $z$ bulk node.
The $p$-adic mass-dimension relation relates the bulk scalar field mass $m^2_\Delta$ to the conformal dimension of the dual operator $\Delta$ via
\eqn{pMassDim}{
m_\Delta^2 = {-1 \over \zeta_p(-\Delta)\zeta_p(\Delta-d)}\,,
}
where we have defined the ``local zeta function'' for every prime  $p$,
\eqn{zetapDef}{
\zeta_p(s) = {1 \over 1-p^{-s}}\,.
}
The bulk-to-boundary propagator is obtained as a regularized limit of the bulk-to-bulk propagator upon sending one of the bulk nodes to the boundary,
\eqn{pK}{
\hat{K}_\Delta(x,z) = {|z_0|_p^\Delta \over |z_0, z_x- x|_s^{2\Delta}}
}
where $z=(z_0,z_x) \in p^\mathbb{Z} \times \mathbb{Q}_{p^d}$ is the bulk node parametrized by the radial coordinate $z_0$ and the boundary direction $z_x$, and $|\cdot|_p$ is the $p$-adic norm while $|x,y|_s \equiv \sup \{|x|_p, |y|_p\}$ is the supremum norm.

The simplest of the propagator identities is the one involving a product of three bulk-to-boundary propagators, which computes the leading contribution to CFT scalar three-point correlator,
\eqn{KKKpadic}{
\sum_{z\in {\cal T}_{p^d}} \hat{K}_{\Delta_1}(x_1,z) \hat{K}_{\Delta_2}(x_2,z) \hat{K}_{\Delta_3}(x_3,z)   &= {C_{\Delta_1\Delta_2\Delta_3} \over |x_{12}^2|_p^{\Delta_{12,3}} |x_{23}^2|_p^{\Delta_{23,1}} |x_{31}^2|_p^{\Delta_{31,2}}}  \,,
}
 where the OPE coefficient of the putative dual CFT is
\eqn{OPEpadic}{
C_{\Delta_i\Delta_j\Delta_k} = \zeta_p(2\Delta_{ijk,}-d)\,
\frac{\zeta_p(2\Delta_{ij,k})
\zeta_p(2\Delta_{jk,i})
\zeta_p(2\Delta_{ki,j})}{\zeta_p(2\Delta_i)\zeta_p(2\Delta_j)\zeta_p(2\Delta_k)}\,.
}

The following three identities involve replacing more and more factors of bulk-to-boundary propagators $\hat{K}$ with factors of bulk-to-bulk propagators $\hat{G}$, culminating in a purely AdS (more precisely, Bruhat--Tits tree) three-point contact scattering process involving three factors of bulk-to-bulk propagators:
\eqn{KKGpadic}{
\sum_{z\in {\cal T}_{p^d}} \hat{G}_{\Delta_a}(w_a,z) \hat{K}_{\Delta_2}(x_2,z) \hat{K}_{\Delta_3}(x_3,z)   &= C_{\Delta_a\Delta_2\Delta_3}\: {\hat{K}_{\Delta_{a2,3}}(x_2,w_a) \hat{K}_{\Delta_{a3,2}}(x_3,w_a) \over |x_{23}^2|_p^{\Delta_{23,a}}} \cr
 &+ {1 \over m^2_{2\Delta_{23,}}-m^2_{\Delta_a}}{1 \over N_{\Delta_a}} \hat{K}_{\Delta_2}(x_2,w_a) \hat{K}_{\Delta_3}(x_3,w_a)  \,,
}
\eqn{KGGpadic}{
\sum_{z\in {\cal T}_{p^d}} \hat{K}_{\Delta_3}(x_3,z) \hat{G}_{\Delta_a}(w_a,z) \hat{G}_{\Delta_b}(w_b,z)  &= C_{\Delta_3\Delta_a\Delta_b}\: \hat{K}_{\Delta_{3a,b}}(x_3,w_a)\hat{K}_{\Delta_{3b,a}}(x_3,w_b)\hat{G}_{\Delta_{ab,3}}(w_a,w_b) \cr
 &+ {1 \over m^2_{2\Delta_{a3,}}-m^2_{\Delta_b}}{1 \over N_{\Delta_b}} \hat{G}_{\Delta_a}(w_a,w_b) \hat{K}_{\Delta_3}(x_3,w_b) \cr
 &+ {1 \over m^2_{2\Delta_{b3,}}-m^2_{\Delta_a}}{1\over N_{\Delta_a}} \hat{G}_{\Delta_b}(w_a,w_b) \hat{K}_{\Delta_3}(x_3,w_a) \,,
}
and
\eqn{GGGpadic}{
\sum_{z\in {\cal T}_{p^d}} \hat{G}_{\Delta_a}(w_a,z) \hat{G}_{\Delta_b}(w_b,z) \hat{G}_{\Delta_c}(w_c,z)  &= C_{\Delta_a\Delta_b\Delta_c}\: \hat{G}_{\Delta_{ac,b}}(w_a,w_c) \hat{G}_{\Delta_{ab,c}}(w_a,w_b) \hat{G}_{\Delta_{bc,a}}(w_b,w_c) \cr
 &+ {1 \over m^2_{2\Delta_{bc,}}-m^2_{\Delta_a}}{1\over N_{\Delta_a}} \hat{G}_{\Delta_b}(w_b,w_a) \hat{G}_{\Delta_c}(w_c,w_a) \cr
 &+ {1 \over m^2_{2\Delta_{ac,}}-m^2_{\Delta_b}}{1 \over N_{\Delta_b}} \hat{G}_{\Delta_a}(w_a,w_b) \hat{G}_{\Delta_c}(w_c,w_b) \cr
 &+ {1 \over m^2_{2\Delta_{ab,}}-m^2_{\Delta_c}}{1 \over N_{\Delta_c}} \hat{G}_{\Delta_a}(w_a,w_c) \hat{G}_{\Delta_b}(w_b,w_c) \,.
}
The first term in each of these identities has been recast in a form slightly different from that originally written in Ref.~\cite{Gubser:2017tsi}, which made reference to the unique bulk point of intersection of the bulk geodesics joining the three fixed (bulk and/or boundary) points. Such a bulk point always exists on the Bruhat--Tits tree, but the corresponding construction does not carry over to continuum AdS space. On the other hand, the ``star-triangle'' rewriting of the first term in the identities above has a direct analog over the reals, as will become apparent later in section~\ref{THREEPROPIDS}.
We will comment on the relevance and interpretation of both forms of identities at the end of the next subsection, as they relate to the holographic duals of conformal blocks.

\subsection{Holographic duals of five- and six-point conformal blocks}
\label{DUALBTTREE}

\paragraph{Five-point block.} The strategy to extract the holographic duals of conformal blocks will be the same as the one detailed in Ref.~\cite{Parikh:2019ygo} where it was used to obtain the holographic dual of the global five-point block in AdS. Namely, to obtain the holographic dual of the five-point block on the Bruhat--Tits tree, we start with the following bulk Feynman diagram,
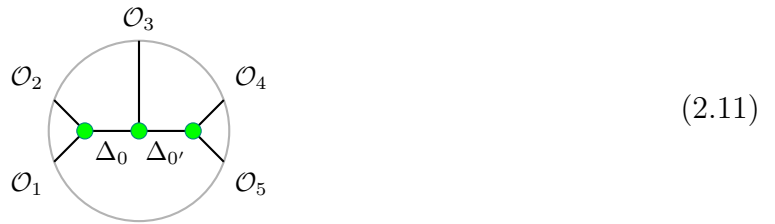
\eqn{p5blockStart}{
\musepic{\figFiveTwoExch}
}
where the green-colored disk-shaped vertices correspond to bulk nodes to be integrated over all nodes of the Bruhat--Tits tree, while the solid lines in the Poincar\'{e} disk represent bulk-to-bulk or bulk-to-boundary propagators~\eno{pG} or~\eno{pK}.
We have suppressed boundary coordinate labels in the diagram; unless otherwise noted, the operator ${\cal O}_i$ of conformal dimension $\Delta_i$ will be inserted on the boundary at position $x_i$.
Further, we will assume that the conformal cross ratios\footnote{On the Bruhat--Tits tree, whose boundary is the projective line over an extension of the $p$-adic numbers, the number of independent cross-ratios that can be formed from five boundary points is precisely two, analogous to the situation in a (real) one-dimensional CFT.}
\eqn{uvDef}{
u \equiv \left|{ x_{12}^2 x_{34}^2 \over x_{13}^2x_{24}^2}\right|_p \qquad v \equiv \left| {x_{23}^2 x_{45}^2 \over x_{24}^2 x_{35}^2 }\right|_p,
}
satisfy $u <1, v<1$.

We can trade the two outside bulk integrations for a geodesic integration, using the identity~\cite{Gubser:2017tsi}
\eqn{pHolOPE}{
\hat{K}_{\Delta_1}(x_1,z) \hat{K}_{\Delta_2}(x_2,z) = {1\over \beta_p(2\Delta_1,2\Delta_2)} \sum_{w \in \gamma_{12}} \hat{K}_{\Delta_1}(x_1,w)\hat{K}_{\Delta_2}(x_2,w) \hat{G}_{\Delta_1+\Delta_2}(w,z)\,,
}
where
\eqn{betapDef}{
\beta_p(s,t) \equiv {\zeta_p(s)\zeta_p(t) \over \zeta_p(s+t) }\,,
}
and $\gamma_{12}$ is the boundary anchored bulk geodesic joining boundary points $x_1$ and $x_2$, over which the bulk point $w$ is to be integrated over.
Pictorially, this is depicted as
\eqn{p5blockKK}{
\musepic{\figFiveTwoExch} &= {1\over \beta_p(2\Delta_1,2\Delta_2) \beta_p(2\Delta_4,2\Delta_5)} \musepic{\figFivePtB} \cr
&= {C_{\Delta_1\Delta_2\Delta_0}\:  C_{\Delta_4\Delta_5\Delta_{0^\prime}} \over \beta_p(2\Delta_{01,2},2\Delta_{02,1})\: \beta_p(2\Delta_{0^\prime4,5},2\Delta_{0^\prime5,4})}\musepic{\figFivePtC} + \cdots
}
where the bulk vertices on the dotted red boundary anchored geodesics are to be integrated over the respective geodesics rather than the entire Bruhat--Tits tree.
The bulk-to-boundary propagators incident, say on $\gamma_{12}$, the boundary anchored geodesic joining boundary points $x_1$ and $x_2$ have associated conformal dimensions $\Delta_1$ and $\Delta_2$ corresponding to the operator insertions ${\cal O}_1$ and ${\cal O}_2$.
In the second equality of~\eno{p5blockKK} we used the two-propagator identity~\cite{Gubser:2017tsi}
\eqn{pGG}{
\sum_{z \in {\cal T}_{p^d}} \hat{G}_{\Delta_a}(a,z) \hat{G}_{\Delta_b}(b,z) =  {1 \over m_{\Delta_a}^2 - m_{\Delta_b}^2} \left({1\over N_{\Delta_b}} \hat{G}_{\Delta_a}(a,b) - {1\over N_{\Delta_a}} \hat{G}_{\Delta_b}(a,b) \right)
}
to write one of in total four terms; the other three are not relevant for the present purposes and are absorbed into the ellipsis. At this point we employ the propagator identity~\eno{KGGpadic} on the first term to obtain
\eqn{p5blockKGG}{
\musepic{\figFiveTwoExch}\!\! &=\!\! \left(\!\! {C_{\Delta_1\Delta_2\Delta_0}\:   C_{\Delta_0\Delta_3\Delta_{0^\prime}}\:C_{\Delta_4\Delta_5\Delta_{0^\prime}} \over \beta_p(2\Delta_{01,2},2\Delta_{02,1})\: \beta_p(2\Delta_{0^\prime4,5},2\Delta_{0^\prime5,4})} \!\! \musepic{\figcalWfivepadicOO} \!\!+ \cdots \!\right) \cr
&\quad + \cdots
}
where the ellipsis inside the parentheses indicates the terms originating from the second and third lines of~\eno{KGGpadic}, which again are unimportant for the purposes of extracting the holographic dual of the five-point block.
As indicated before, we have suppressed boundary coordinate labels, except at $x_3$ where we have explicitly indicated the dimensions of the bulk-to-boundary propagators incident at the boundary point in blue.

We are  interested in the term in the expansion of the exchange diagram proportional to the product of the three OPE coefficients as shown in~\eno{p5blockKGG}, since in the CBD the five-point conformal block representing the exchange of single-trace primaries of weights $\Delta_0$ and $\Delta_{0^\prime}$ in the intermediate channels also appears with an overall factor of the same OPE coefficients. The last observation follows from the holographic version of the CBD for individual diagrams.

We note that this strategy to extract the putative holographic dual to the conformal block is not guaranteed to work, since the inverse problem of solving for the conformal blocks given the five-point function and the OPE coefficients is not well-posed.
Nevertheless, it serves as a heuristic  guide in guessing a natural candidate for the holographic dual, which must then be confirmed by independent means.

Going back to~\eno{p5blockKGG}, we expect from the heuristic argument above that the five-point conformal block is given by
\eqn{calW5CBpadic}{
W_{\Delta_0;\Delta_{0^\prime}}^{\Delta_1,\ldots,\Delta_5}(x_i)  &=   {1 \over \beta_p(2\Delta_{01,2},2\Delta_{02,1})\: \beta_p(2\Delta_{0^\prime4,5},2\Delta_{0^\prime5,4}) } {\cal W}_{\Delta_0;\Delta_{0^\prime}}^{\Delta_1,\ldots,\Delta_5}(x_i)\,,
}
where we have defined the \emph{five-point geodesic bulk diagram} on the Bruhat--Tits tree, the holographic dual to the conformal block as
\eqn{calW5padic}{
{\cal W}_{\Delta_0;\Delta_{0^\prime}}^{\Delta_1,\ldots,\Delta_5}(x_i) &\equiv \musepic{\figcalWfivepadicOO} \cr
 &=  \sum_{\substack{w \in \gamma_{12}\\ w^\prime \in \gamma_{45}}}
\hat{K}_{\Delta_1}(x_1,w) \hat{K}_{\Delta_2}(x_2,w)
\hat{K}_{\Delta_4}(x_4,w^\prime) \hat{K}_{\Delta_5}(x_5,w^\prime) \cr
&\qquad\quad \times \hat{K}_{\Delta_{30,0^\prime}}(x_3,w) \hat{G}_{\Delta_{00^\prime,3}}(w,w^\prime)  \hat{K}_{\Delta_{30^\prime,0}}(x_3,w^\prime) \,.
}
Indeed, explicit evaluation of the geodesic bulk diagram on the Bruhat--Tits tree confirms this expectation. We find
\eqn{W5calc}{
{\cal W}_{\Delta_0;\Delta_{0^\prime}}^{\Delta_1,\ldots,\Delta_5}(x_i) =  \beta_p(2\Delta_{01,2},2\Delta_{02,1})\: \beta_p(2\Delta_{0^\prime4,5},2\Delta_{0^\prime5,4})  W_0^{(5)}(x_i)\: u^{\Delta_0/2} v^{\Delta_{0^\prime}/2} \,,
}
where  the conformal cross-ratios were defined in~\eno{uvDef}
and the ``leg factor'' accounts for the remaining coordinate and external dimensions dependence,
\eqn{W0five}{
W_0^{(5)} \equiv {1 \over |x_{12}^2|_p^{\Delta_{12,}} |x_{34}^2|_p^{\Delta_3/2} |x_{45}^2|_p^{\Delta_{45,}}}
\left|{x_{23}^2 \over x_{13}^2} \right|_p^{\Delta_{1,2}}
\left|{x_{24}^2 \over x_{23}^2} \right|_p^{\Delta_{3}/2}
\left|{x_{35}^2 \over x_{34}^2} \right|_p^{\Delta_{4,5}}.
}
Thus~\eno{calW5CBpadic} with the geodesic bulk diagram given by~\eno{calW5padic}-\eno{W5calc}  reproduces precisely the leading scaling (i.e.\ descendant-free) contribution to the global five-point block in a one-dimensional CFT. In a $p$-adic CFT defined on $\mathbb{P}^1(\mathbb{Q}_{p^d})$, which lacks descendants, the scaling block is the \emph{full} five-point conformal block. This provides the independent check of our proposal.

\paragraph{Six-point block in the OPE channel.} A very similar exercise leads to the holographic dual of the six-point block in the OPE channel, shown in figure~\ref{fig:6opechannel}.
This time, to obtain the geometric representation, the three-bulk-to-bulk-propagators identity~\eno{GGGpadic} will be employed. Detailed calculations are provided in section~\ref{SIXOPEGEODESIC} for the real analog of this result. The $p$-adic result is obtained by following closely the same steps --- which incidentally benefit from considerable calculational simplifications --- leading in this case to a significantly simpler, but related answer as compared with the reals.
The upshot is that the holographic dual of the six-point global conformal block in the OPE channel is given by the following geodesic bulk diagram,
\eqn{calW6padic}{
{\cal W}_{\Delta_a;\Delta_c;\Delta_b}^{\Delta_1,\ldots,\Delta_6}(x_i) &=  \musepic{\figcalWsixpadic} \cr
 &= \!\!\! \sum_{\substack{w \in \gamma_{12}\\ w^\prime \in \gamma_{34}\\ w^{\prime\prime} \in \gamma_{56}}} \!\!\!\!
\hat{K}_{\Delta_1}(x_1,w) \hat{K}_{\Delta_2}(x_2,w)  \hat{K}_{\Delta_3}(x_3,w^\prime) \hat{K}_{\Delta_4}(x_4,w^\prime) \hat{K}_{\Delta_5}(x_5,w^{\prime\prime}) \hat{K}_{\Delta_6}(x_6,w^{\prime\prime}) \cr
&\qquad\quad \times \hat{G}_{\Delta_{ac,b}}(w,w^\prime) \hat{G}_{\Delta_{bc,a}}(w^{\prime\prime},w^\prime)  \hat{G}_{\Delta_{ab,c}}(w,w^{\prime\prime}) \,,
}
and an explicit evaluation of the geodesic integrals (i.e.\ geodesic summations on the Bruhat--Tits tree) above reveals that the diagram is related to the six-point conformal block via a simple relation,
\eqn{calW6CBpadic}{
{\cal W}_{\Delta_a;\Delta_c;\Delta_b}^{\Delta_1,\ldots,\Delta_6}(x_i) &=   \beta_p(2\Delta_{a1,2},2\Delta_{a2,1})\: \beta_p(2\Delta_{c3,4},2\Delta_{c4,3})\: \beta_p(2\Delta_{b5,6},2\Delta_{b6,5}) \:
W_{\Delta_a;\Delta_c;\Delta_b}^{\Delta_1,\ldots,\Delta_6}(x_i) \cr
 &= \beta_p(2\Delta_{a1,2},2\Delta_{a2,1})\: \beta_p(2\Delta_{c3,4},2\Delta_{c4,3})\: \beta_p(2\Delta_{b5,6},2\Delta_{b6,5}) \cr
 & \qquad\qquad\qquad\qquad\qquad\qquad\qquad\qquad\qquad \times W_0^{(6)}(x_i)\: u_1^{\Delta_a/2} u_2^{\Delta_b/2} u_3^{\Delta_c/2}\,,
}
where
\eqn{u1u2u3}{
u_1 \equiv \left| {x_{12}^2 x_{35}^2 \over x_{13}^2 x_{25}^2} \right|_p \qquad u_2 \equiv \left| {x_{13}^2 x_{56}^2 \over x_{15}^2 x_{36}^2} \right|_p \qquad u_3 \equiv \left| {x_{15}^2 x_{34}^2 \over x_{13}^2 x_{45}^2} \right|_p ,
}
and
\eqn{W06}{
W_0^{(6)}(x_i) \equiv
\left|\frac{x_{23}^2}{x_{12}^2x_{13}^2}\right|_p^{\frac{\Delta_1}{2}}
\left|\frac{x_{13}^2}{x_{12}^2x_{23}^2}\right|_p^{\frac{\Delta_2}{2}}
\left|\frac{x_{24}^2}{x_{23}^2x_{34}^2}\right|_p^{\frac{\Delta_3}{2}}
\left|\frac{x_{35}^2}{x_{34}^2x_{45}^2}\right|_p^{\frac{\Delta_4}{2}}
\left|\frac{x_{46}^2}{x_{45}^2x_{56}^2}\right|_p^{\frac{\Delta_5}{2}}
\left|\frac{x_{45}^2}{x_{46}^2x_{56}^2}\right|_p^{\frac{\Delta_6}{2}}\,.
}
Above, we assumed $u_1, u_2, u_3 <1$ and used the fact that in a one-dimensional CFT lacking any descendants, the six-point global conformal block in the OPE channel  is given by a trivial scaling block written in terms of the three independent conformally invariant cross-ratios~\eno{u1u2u3} constructed out of the six insertion points.

The trivial scaling form of $p$-adic global conformal blocks permits an alternate holographic representation on the discrete Bruhat--Tits tree. Such an alternate, but equivalent representation  can be recovered by employing the original forms of the propagator identities~\cite{Gubser:2017tsi} in the derivation of the holographic duals presented above, as was alluded at the end of the previous subsection. This leads to the following dual representations:
\eqn{padicCBdualAlt}{
{\cal W}^{\Delta_1,\ldots,\Delta_5}_{\Delta_0;\Delta_{0^\prime}} = \musepic{\figFiveBlockPAlt} \qquad {\cal W}^{\Delta_1,\ldots,\Delta_6}_{\Delta_a;\Delta_{c};\Delta_b} = \musepic{\figSixBlockPAlt}\,,
}
which satisfy~\eno{calW5CBpadic} and~\eno{calW6CBpadic}, respectively. In the alternate holographic representation for the five-point block above, the unfilled disk-shaped bulk vertex represents the unique Bruhat--Tits tree point of intersection of the bulk geodesics joining the two bulk points being integrated over the boundary anchored (red dashed) geodesics, and the boundary insertion point $x_3$. This vertex remains \emph{fixed} even as the bulk points are integrated over the geodesics.\footnote{This vertex is precisely the same vertex on the Bruhat--Tits tree as the one at which geodesics from all five boundary insertion points meet.} Such a representation is equivalent to the one obtained in~\eno{calW5padic} as long as the conformal cross-ratios~\eno{uvDef} satisfy $u, v<1$. The equivalence is easily established by appealing to the tree-like geometry of the Bruhat--Tits tree. Likewise, the holographic dual of the six-point block admits an alternate representation as shown above, for cross-ratios as defined in~\eno{u1u2u3} satisfying $u_1, u_2, u_3 <1$. The unfilled disk-shaped bulk vertex in the holographic representation is now the unique point of intersection on the Bruhat--Tits tree of geodesics joining the three bulk points being integrated over respective boundary anchored geodesics.\footnote{Alternatively, it is unique the vertex at which geodesics from \emph{all} six boundary insertion points meet in the bulk.} As we will show later, the real analogs of the holographic representations resemble closely the forms in~\eno{calW5padic} and~\eno{calW6padic}, but it is not entirely clear what the real analog of~\eno{padicCBdualAlt} should be, or whether such a representation even exists.
It is intriguing to note that the unfilled disk-shaped vertices in~\eno{padicCBdualAlt} can be interpreted as Fermant-Torricelli points on the Bruhat--Tits tree.
Curiously, such points play an important role in the holographic interpretation of perturbative large-$c$ Virasoro conformal blocks~\cite{Alkalaev:2018nik}.
It would be interesting to explore whether Fermat-Torricelli points also appear in the context of holographic representations of (real) global conformal blocks.

\subsection{Conformal block decomposition of bulk diagrams}
\label{CBDBTTREE}
In this section, we  work out the CBDs of scalar five-point bulk diagrams using the previously discussed propagator identities, and the holographic representation of the five-point conformal block. With these tools in hand, no additional bulk integrations need be explicitly performed and all steps are purely algebraic.
This approach is a generalization of the geodesic diagram techniques~\cite{Hijano:2015zsa,Gubser:2017tsi} to evaluate higher-point diagrams.
We obtain explicit closed-form expressions for the decomposition in the basis of five-point conformal blocks involving the exchange of single- or multi-trace primaries in the intermediate channels, and we find that the CBD coefficients themselves are given by very simple analytic expressions in terms of the CFT data. Further, the CBD coefficients of various five-point exchange diagrams share simple relations that repackage the pole structure of the diagrams into simple algebraic relations.

We begin with the five-point contact diagram,\footnote{Throughout this section, without loss of generality, we will assume the cross-ratios defined in~\eno{uvDef} satisfy $u, v \leq 1$. Such a situation can always be arranged up to a relabelling of boundary coordinates.}
\eqn{pD5}{
D_5 \equiv \musepic{\figFiveContact}\,.
}
To obtain a CBD in the channel shown in figure~\ref{fig:5block}, we apply the propagator identity~\eno{pHolOPE} to the boundary insertion points $(x_1,x_2)$ and $(x_4,x_5)$ to get
\eqn{pD5step2}{
D_5 = {1\over \beta_p(2\Delta_1,2\Delta_2) \beta_p(2\Delta_4,2\Delta_5)} \musepic{\figFivePtExchA}\,.
}
The bulk integration over the green vertex is immediately carried out using identity~\eno{KGGpadic}, to give a sum of three terms
\eqn{pD5step3}{
D_5 &= {1\over \beta_p(2\Delta_1,2\Delta_2) \beta_p(2\Delta_4,2\Delta_5)} \left( C_{2\Delta_{12,}\,\Delta_3\,2\Delta_{45,}}
\musepic{\figFiveCBDePadic} \right.
\cr
&  \quad
\left. +\frac{(N_{2\Delta_{45,}})^{-1}}{m^2_{2\Delta_{123,}}-m^2_{2\Delta_{45,}}}
  \musepic{\figFiveCBDgPadic}
+
 \frac{(N_{2\Delta_{12,}})^{-1}}{m_{2\Delta_{345,}}^2-m^2_{2\Delta_{12,}}}
 \musepic{\figFiveCBDiPadic} \right).
}
We remind the reader that our convention is that the operator ${\cal O}_i$ is inserted at position $x_i$, but the position label is usually suppressed. In the first term in~\eno{pD5step3}, we have indicated explicitly the position coordinate $x_3$ and the conformal dimensions of the bulk-to-boundary propagators incident at the boundary point in blue. Consistent with this notation, we have also indicated the dimensions of the bulk-to-boundary propagators incident at $x_3$ in the second and third terms, although here the conformal dimension label $\Delta_3$ is redundant since we have already specified the operator insertion ${\cal O}_3$. This notation will come handy when we eventually discuss the analogous computation in standard AdS/CFT over the reals later in this paper.

Using~\eno{calW5CBpadic}-\eno{calW5padic} we recognize each of the geodesic diagrams in~\eno{pD5step3} as a five-point conformal block associated with the exchange of a particular set of (higher-trace) primary operators in the intermediate channels. This leads to
\eqn{pD5CBD}
{
&\musepic{\figFiveContact}=
P_{\rm cont}^{(12,45)}  \left(\!\!\musepic{\figFiveContCBDaPadic} \!\!\right)
\cr
&
+   P_{\rm cont}^{(12,123)} \left(\!\!\musepic{\figFiveContCBDbPadic}\!\!\right)
+  P_{\rm cont}^{(345,45)} \left(\!\!\musepic{\figFiveContCBDcPadic}\!\!\right),
}
where we are using the graphical notation for the $p$-adic five-point block as defined in figure~\ref{fig:5block}, with the CBD coefficients given by
\eqn{pD5CBDcoeffs}{
P_{\rm cont}^{(12,45)} &= C_{2\Delta_{12,}\,\Delta_3\,2\Delta_{45,}} \cr
P_{\rm cont}^{(12,123)} &= {\beta_p(2\Delta_{1234,5},2\Delta_{1235,4}) \over \beta_p(2\Delta_4,2\Delta_5)} \frac{(N_{2\Delta_{45,}})^{-1}}{m^2_{2\Delta_{123,}}-m^2_{2\Delta_{45,}}} \cr
 &= C_{2\Delta_{123,}\, \Delta_4 \, \Delta_5} \cr
 P_{\rm cont}^{(345,45)} &= {\beta_p(2\Delta_{3451,2},2\Delta_{3452,1}) \over \beta_p(2\Delta_1,2\Delta_2)}  \frac{(N_{2\Delta_{12,}})^{-1}}{m_{2\Delta_{345,}}^2-m^2_{2\Delta_{12,}}} \cr
  &= C_{\Delta_1\,\Delta_2\,2\Delta_{345,}}\;.
}
In simplifying the coefficients above, we made use of the identity
\eqn{BetaToC}
{
\frac{\beta_p(2\Delta_{01,2},2\Delta_{02,1})}{\beta_p(2\Delta_1,2\Delta_2)N_{2\Delta_{12,}}(m^2_{\Delta_0}-m_{2\Delta_{12,}}^2)}=C_{\Delta_0\Delta_1\Delta_2}\,,
}
which is easily verified by substituting for each function using their definitions.

Now let's turn to a five-point diagram with a single scalar field exchange of mass $m^2_{\Delta_0}$,
\eqn{}
{
D_5^{\rm 1-exch}=\musepic{\figFiveOneExch}\,.
}
Applying~\eno{pHolOPE} to the pairs~$(x_1,x_2)$ and $(x_4,x_5)$ like before, we recast the exchange diagram as a diagram involving two geodesic integrals, and two full bulk integrals. One of the full bulk integrals is of the form~\eno{pGG} while the other is of the form~\eno{KGGpadic}. Using~\eno{pGG} to evaluate the former, we end up with
\eqn{}{
D_5^{\rm 1-exch}&= \frac{\beta_p(2\Delta_1,2\Delta_2)^{-1}\beta_p(2\Delta_4,2\Delta_5)^{-1}}{m_{2\Delta_{12,}}^2-m^2_{\Delta_0}} \cr
&\quad \times \left(\frac{1}{N_{\Delta_0}}\!\!
\musepic{\figFivePtExchA}
-\frac{1}{N_{2\Delta_{12,}}}\!\!
\musepic{\figFivePtExchB}
\right).
}
Applying~\eno{KGGpadic} to both the terms, we get
\eqn{}{
& D_5^{\rm 1-exch}= \frac{\beta_p(2\Delta_1,2\Delta_2)^{-1}\beta_p(2\Delta_4,2\Delta_5)^{-1}}{m_{2\Delta_{12,}}^2-m^2_{\Delta_0}}\left(\!\!
\frac{C_{2\Delta_{12,}\,\Delta_3\,2\Delta_{45,}}}{N_{\Delta_0}}
\musepic{\figFiveCBDePadic} \right.
\cr
&
+\frac{(N_{\Delta_0}N_{2\Delta_{45,}})^{-1}}{m^2_{2\Delta_{123,}}-m^2_{2\Delta_{45,}}}
  \musepic{\figFiveCBDgPadic}
 \!\!
+
 \frac{(N_{\Delta_0}N_{2\Delta_{12,}})^{-1}}{m_{2\Delta_{345,}}^2-m^2_{2\Delta_{12,}}}
 \musepic{\figFiveCBDiPadic}
 \cr
 &-
 \frac{C_{\Delta_0\,\Delta_3\,2\Delta_{45,}}}{N_{2\Delta_{12,}}}
 \musepic{\figFiveCBDfPadic}
 -
 \frac{(N_{2\Delta_{12,}}N_{2\Delta_{45,}})^{-1}}{m^2_{2\Delta_{03,}}-m^2_{2\Delta_{45,}}}
\musepic{\figFiveCBDhPadic}
 \cr
 &- \left.   \frac{(N_{2\Delta_{12,}}N_{\Delta_0})^{-1}}{m_{2\Delta_{345,}}^2-m^2_{\Delta_0}}
 \musepic{\figFiveCBDiPadic}
\!\!\right).
}
Using~\eno{calW5CBpadic}-\eno{calW5padic}, we recognize five independent conformal blocks above (with the third and sixth terms above proportional to the same block).
Indeed, with simple substitutions, the exchange diagram CBD can be re-expressed as
\begingroup
\allowdisplaybreaks
\begin{align*}
&\musepic{\figFiveOneExch}=
P_{\rm 1-exch}^{(12,45)}  \left(\!\!\musepic{\figFiveContCBDaPadic} \!\!\right)
\\
&
+  P_{\rm 1-exch}^{(0,45)} \left(\musepic{\figFiveCBDaPadic} \right)
+   P_{\rm 1-exch}^{(12,123)} \left(\!\!\musepic{\figFiveContCBDbPadic}\!\!\right)
\\
&
+  P_{\rm 1-exch}^{(0,03)} \left(\musepic{\figFiveCBDbPadic}\right)
+  P_{\rm 1-exch}^{(345,45)} \left(\!\!\musepic{\figFiveContCBDcPadic}\!\!\right), \stepcounter{equation}\tag{\theequation}\label{pD5exch1}
\end{align*}
\endgroup
where\footnote{In simplifying the CBD coefficients, we used the identity~\eno{BetaToC}, and
\eqn{Elephant}
{
\frac{1}{m^2_{2\Delta_{12,}}-m^2_{\Delta_0}}\left(
\frac{1}{m^2_{2\Delta_{345,}}-m^2_{2\Delta_{12,}}}
-\frac{1}{m^2_{2\Delta_{345,}}-m^2_{\Delta_{0,}}}
\right)=
\frac{1}{m^2_{2\Delta_{345,}}-m^2_{\Delta_0}}\,.
}}
\eqn{pD5exch1CoeffA}
{
P_{\rm 1-exch}^{(12,45)} &=
\frac{C_{2\Delta_{12,}\,\Delta_3\,2\Delta_{45,}}}{N_{\Delta_0}(m_{2\Delta_{12,}}^2-m^2_{\Delta_0})}
\cr
P_{\rm 1-exch}^{(12,123)} &=
\frac{\beta_p(2\Delta_{1234,5},2\Delta_{1235,4})\beta_p(2\Delta_4,2\Delta_5)^{-1}}{N_{\Delta_0}(m_{2\Delta_{12,}}^2-m^2_{\Delta_0})N_{2\Delta_{45,}}(m^2_{2\Delta_{123,}}-m^2_{2\Delta_{45,}})} \cr
&=\frac{C_{2\Delta_{123,}\,\Delta_4\,\Delta_5}}{N_{\Delta_0}(m_{2\Delta_{12,}}^2-m^2_{\Delta_0})}
\cr
P_{\rm 1-exch}^{(345,45)} &=
\frac{\beta_p(2\Delta_{1345,2},2\Delta_{2345,1})
\beta_p(2\Delta_1,2\Delta_2)^{-1}
}{N_{\Delta_0}(m^2_{2\Delta_{345,}}-m_{\Delta_0}^2)N_{2\Delta_{12,}}(m^2_{2\Delta_{345,}}-m^2_{2\Delta_{12,}})} \cr
&=\frac{C_{\Delta_1\,\Delta_2\,2\Delta_{345,}}}{N_{\Delta_0}(m^2_{2\Delta_{345,}}-m^2_{\Delta_0})}\,,
}
and
\eqn{pD5exch1CoeffB}
{
P_{\rm 1-exch}^{(0,45)} &=
\frac{\beta_p(2\Delta_{01,2},2\Delta_{02,1})}{\beta_p(2\Delta_1,2\Delta_2)}
\frac{C_{\Delta_{0}\,\Delta_3\,2\Delta_{45,}}}{N_{2\Delta_{12,}}(m^2_{\Delta_0}-m_{2\Delta_{12,}}^2)} \cr
&= C_{\Delta_0\Delta_1\Delta_2}\:C_{\Delta_{0}\,\Delta_3\,2\Delta_{45,}}
\cr
P_{\rm 1-exch}^{(0,03)} &=
\frac{\beta_p(2\Delta_{01,2},2\Delta_{02,1})\beta_p(2\Delta_{034,5},2\Delta_{035,4})
\beta_p(2\Delta_1,2\Delta_2)^{-1}\beta_p(2\Delta_4,2\Delta_5)^{-1}
}{N_{2\Delta_{12,}}(m^2_{\Delta_0}-m_{2\Delta_{12,}}^2)N_{2\Delta_{45,}}(m^2_{2\Delta_{03,}}-m^2_{2\Delta_{45,}})} \cr
&=C_{\Delta_0\Delta_1\Delta_2}\:C_{2\Delta_{03,}\,\Delta_4\,\Delta_5}\,.
}

Three of the five conformal blocks which appear in the direct channel CBD of the exchange diagram in~\eno{pD5exch1} also appeared in the CBD of the contact diagram, and are related to exchanges of double- and higher-trace primaries in the two intermediate channels. Interestingly, the associated CBD coefficients in the decomposition are also very simply related:
\eqn{pContExchCoeffs}{
{P_{\rm 1-exch}^{(12,45)} \over P_{\rm cont}^{(12,45)}} &= {1 \over N_{\Delta_0}(m_{2\Delta_{12,}}^2-m^2_{\Delta_0})}\cr
{P_{\rm 1-exch}^{(12,123)} \over P_{\rm cont}^{(12,123)}} &= {1 \over N_{\Delta_0}(m_{2\Delta_{12,}}^2-m^2_{\Delta_0})} \cr
{P_{\rm 1-exch}^{(345,45)} \over P_{\rm cont}^{(345,45)}} &= {1 \over N_{\Delta_0}(m_{2\Delta_{345,}}^2-m^2_{\Delta_0})}\,.
 }
Such relations were noted in the case of the four-point contact and exchange diagrams earlier~\cite{Hijano:2015zsa,Gubser:2017tsi}. We will comment on this observation further in the next subsection.

Finally, for the tree-level five-point diagram built from three cubic bulk interaction vertices, the procedure to obtain its direct-channel CBD is identical to the one demonstrated for the contact and exchange diagrams above. We omit the details of the straightforward computation, but present the final result:
\eqn{5TwoExchCBDpadic}{
& \musepic{\figFiveTwoExch} = P_{\rm 2-exch}^{(0,0^\prime)}\!\! \left(\!\!\musepic{\figFiveCBD}\!\!\right)\!\! +  P_{\rm 2-exch}^{(12,45)} \!\! \left(\!\!\musepic{\figFiveContCBDaPadic} \!\!\right) \cr
&+ P_{\rm 2-exch}^{(0,45)} \left(\musepic{\figFiveCBDaPadic} \right) +  P_{\rm 2-exch}^{(0,03)} \left(\musepic{\figFiveCBDbPadic}\right)  \cr
&+   P_{\rm 2-exch}^{(12,0^\prime)} \left(\musepic{\figFiveCBDcPadic} \right) +  P_{\rm 2-exch}^{(0^\prime 3,0^\prime)} \left(\musepic{\figFiveCBDdPadic}\right)  \cr
&+   P_{\rm 2-exch}^{(12,123)} \left(\!\!\musepic{\figFiveContCBDbPadic}\!\!\right) +  P_{\rm 2-exch}^{(345,45)} \left(\!\!\musepic{\figFiveContCBDcPadic}\!\!\right),
}
with the coefficients
\eqn{Fivept2ExchOPEpadic}{
{P}_{\rm 2-exch}^{(0,0^\prime)} =  C_{\Delta_1 \Delta_2 \Delta_0}\:  C_{\Delta_0 \Delta_3 \Delta_{0^\prime}}\:  C_{ \Delta_4 \Delta_5 \Delta_{0^\prime}} \qquad
{P}_{\rm 2-exch}^{(12,45)}  =  { {P}_{\rm 1-exch}^{(12,45)} \over  N_{\Delta_{0^\prime}} (m^2_{2\Delta_{45,}} - m_{\Delta_{0^\prime}}^2)}  \cr
{P}_{\rm 2-exch}^{(0,45)}  =   {P_{\rm 1-exch}^{(0,45)} \over  N_{\Delta_{0^\prime}}(m^2_{2\Delta_{45,}} - m_{\Delta_{0^\prime}}^2)} \qquad
{P}_{\rm 2-exch}^{(0,03)} =   { P_{\rm 1-exch}^{(0,03)} \over  N_{\Delta_{0^\prime}}(m^2_{2\Delta_{03,}} - m_{\Delta_{0^\prime}}^2)}  \cr
{P}_{\rm 2-exch}^{(12,123)} =   { {P}_{\rm 1-exch}^{(12,123)} \over N_{\Delta_{0^\prime}}(m^2_{2\Delta_{123,}} - m_{\Delta_{0^\prime}}^2)} \qquad
{P}_{\rm 2-exch}^{(345,45)} =   {{P}_{\rm 1-exch}^{(345,45)} \over N_{\Delta_{0^\prime}}(m^2_{2\Delta_{45,}} - m_{\Delta_{0^\prime}}^2)}
 \,.
 }
The remaining two coefficients are obtained by symmetry:
\eqn{pFiveSymOPE}{
P_{\rm 2-exch}^{(12,0^\prime)} =  {P_{\rm 1-exch}^{(12,0^\prime)} \over  N_{\Delta_{0}}(m^2_{2\Delta_{12,}} - m_{\Delta_{0}}^2)} \qquad
P_{\rm 2-exch}^{(0^\prime 3,0^\prime)} =  { P_{\rm 1-exch}^{(0^\prime 3,0^\prime)} \over  N_{\Delta_{0^\prime}} (m^2_{2\Delta_{0^\prime 3,}} - m_{\Delta_{0}}^2) }
\,,
}
where, analogous to~\eno{pD5exch1CoeffB} we have defined
\eqn{}{
P_{\rm 1-exch}^{(12,0^\prime)} =  C_{\Delta_3\,\Delta_{0^\prime}\, 2\Delta_{12,}}\: C_{\Delta_4\Delta_5\Delta_{0^\prime}} \qquad
P_{\rm 1-exch}^{(0^\prime 3,0^\prime)} = C_{\Delta_1\,\Delta_2\,2\Delta_{0^\prime 3,}} \: C_{\Delta_4\Delta_5\Delta_{0^\prime}}\,.
}
The coefficients corresponding to exchanges of solely higher-trace primaries further admit an expression in terms of the contact diagram coefficients:
\eqn{Fivept2ExchOPEAgainpadic}{
{P}_{\rm 2-exch}^{(12,45)}  &= {P_{\rm cont}^{(12,45)} \over N_{\Delta_0}(m_{2\Delta_{12,}}^2-m_{\Delta_0}^2) N_{\Delta_{0^\prime}} (m^2_{2\Delta_{45,}} - m_{\Delta_{0^\prime}}^2)} \cr
{P}_{\rm 2-exch}^{(12,123)} &=  { {P}_{\rm cont}^{(12,123)} \over N_{\Delta_0}(m_{2\Delta_{12,}}^2-m_{\Delta_0}^2) N_{\Delta_{0^\prime}}(m^2_{2\Delta_{123,}} - m_{\Delta_{0^\prime}}^2)} \cr
{P}_{\rm 2-exch}^{(345,45)} &=  {P^{(345,45)}_{\rm cont} \over    N_{\Delta_0} (m_{2\Delta_{345,}}^2-m_{\Delta_0}^2)  N_{\Delta_{0^\prime}}(m^2_{2\Delta_{45,}} - m_{\Delta_{0^\prime}}^2)}\,.
}

A similar procedure also leads to the CBD of six-point diagrams, but we refrain from presenting the details here. Instead, the six-point case is discussed in more detail for the usual real AdS/CFT (with a few remarks on their $p$-adic analogs) in section~\ref{SIXCBD}.
In section~\ref{SPECTRAL} we will comment on the form of the decompositions obtained and match against the expectations from the shadow formalism.

The simplicity of the CBD in the $p$-adic setting essentially stems from the lack of spinning and descendant operators, so that there are no infinite series (corresponding to higher-twist contributions) to be summed in the decomposition.
Mathematically, this drastic simplicity arises because the poles of the spectral density function in the spectral representation of the bulk diagrams are governed by poles of the local zeta function~\eno{zetapDef}, which only has a single simple pole on the real axis, as opposed to the case in the conventional (real) AdS/CFT where the pole structure is governed by arguments of the Euler Gamma function, which has a semi-infinite sequence of poles along the real axis.
Such a simplification is manifest in the logarithmic singularity structure of the CBDs, which we now discuss.

\subsubsection{Logarithmic singularities}
\label{PLOGARITHMS}

Relations similar to~\eno{pContExchCoeffs} and~\eno{Fivept2ExchOPEpadic}-\eno{Fivept2ExchOPEAgainpadic} between the CBD coefficients of five-point bulk diagrams also exist for \emph{four-point} contact and exchange diagrams~\cite{Gubser:2017tsi} (see also Ref.~\cite{Hijano:2015zsa} for four-point diagrams in real AdS/CFT). These, together with the form of the contact diagram CBD coefficients provide an algebraic origin~\cite{Hijano:2015zsa} of the integrality conditions for non-generic conformal dimensions, which signal the appearance of logarithmic singularities in bulk Feynman diagrams~\cite{Liu:1998th} (arising when anomalous dimensions of exchanged operators  contribute at tree-level). Due to the lack of higher-twist primaries in the CBD, such ``integrality'' conditions are in fact more restrictive in $p$-adic AdS/CFT~\cite{Gubser:2017tsi}.

In this subsection we show that this continues to hold true for the five-point diagrams discussed above. In the case of the five-point contact diagram  decomposed in the channel depicted in figure~\ref{fig:5block}, using~\eno{pD5CBDcoeffs}-\eno{BetaToC} to isolate poles coming from the factor of $m^2_{\Delta_a} - m^2_{\Delta_b}$ in the denominator for various pairs of conformal dimensions $\Delta_a, \Delta_b$,  we find that the following poles of the CBD coefficients at non-generic values of external dimensions,
\eqn{pD5masses}{
 m^2_{2\Delta_{12,}} - m^2_{2\Delta_{345,}} = 0 \qquad\quad  m^2_{2\Delta_{123,}} - m^2_{2\Delta_{45,}} = 0 \,,
}
correspond to the appearance of logarithmic terms. Equivalently\footnote{The algebraic condition $m^2_{\Delta_a} - m^2_{\Delta_b}$ has in general two solutions, $\Delta_a -\Delta_b=0$ or $\Delta_a + \Delta_b = d$ in $p$-adic AdS/CFT. This follows directly from the mass-dimension relation~\eno{pMassDim} and the fact that the local zeta function $\zeta_p$ has a single simple pole on the real axis. However, the latter solution in conjunction with unitarity (and some convergence conditions) reduces to the former solution~\cite{Gubser:2017tsi}.}
\eqn{pD5integrality}{
\Delta_{1}+\Delta_{2} - \Delta_{3}-\Delta_{4}-\Delta_5 = 0 \qquad\qquad \Delta_{1}+\Delta_2+\Delta_3-\Delta_{4}-\Delta_5=0\,.
}
These are the $p$-adic ``integrality'' conditions for the five-point contact diagram.
For the exchange diagram~\eno{pD5exch1}, in addition to the conditions above,  some more conditions are possible. These are easily obtained from~\eno{pD5exch1CoeffB}-\eno{pContExchCoeffs}, to be
\eqn{pD5exch1masses}{
m_{\Delta_0}^2 - m_{2\Delta_{12,}}^2 = 0 \qquad\qquad m_{\Delta_0}^2 - m_{2\Delta_{345,}}^2 = 0  \qquad\qquad m_{2\Delta_{03,}}^2 - m_{2\Delta_{45,}}^2 = 0\,.
}
These give rise to the integrality conditions,
\eqn{pD5exch1integrality}{
\Delta_0 - \Delta_1 - \Delta_2 = 0 \qquad \Delta_0 -\Delta_3-\Delta_4-\Delta_5=0 \qquad \Delta_0 + \Delta_3 - \Delta_4 - \Delta_5 = 0\,.
}
Finally, for the five-point diagram~\eno{5TwoExchCBDpadic} involving the exchange of two scalar fields, in addition to~\eno{pD5masses},~\eno{pD5exch1masses}, we also have
\eqn{pD5exch2masses}{
m_{\Delta_{0^\prime}}^2 - m_{2\Delta_{45,}}^2 = 0 \qquad\qquad  m_{\Delta_{0^\prime}}^2 - m_{2\Delta_{123,}}^2 = 0 \qquad\qquad m_{2\Delta_{0^\prime 3,}}^2 - m_{2\Delta_{12,}}^2 = 0 \cr
m_{\Delta_{0^\prime}}^2 - m_{2\Delta_{03,}}^2 = 0 \qquad\qquad  m_{\Delta_0}^2 - m_{2\Delta_{0^\prime 3,}}^2 = 0\,. \qquad\qquad
}
which translate straightforwardly to five obvious integrality conditions, which we do not write down.
For any such choice of conformal dimensions, the contributions from the anomalous dimensions of double- or higher-trace operators will be visible at tree-level.

\vspace{0.5em}
This concludes our discussion of the toy model of $p$-adic AdS/CFT. From the next section onward, we begin the analysis in the usual Euclidean signature ${\rm AdS}_{d+1}/{\rm CFT}_d$ over \emph{real numbers}.

\section{Propagator identities in AdS}
\label{PROPIDS}

 This section is devoted to various crucial propagator identities in continuum AdS$_{d+1}$ space which allow the extraction of holographic duals of conformal blocks in real CFT$_d$, as well as the CBD of individual bulk diagrams, which will be the subject of the next section. The new identities, collected in section~\ref{THREEPROPIDS} are proven in appendix~\ref{PROOFPROPAG}.

We first recall the basic building blocks of perturbative bulk dynamics in EAdS$_{d+1}$, i.e.\ the propagators used to construct the AdS Feynman diagram expansion perturbatively in $1/N$. The normalizable solution to the bulk Klein-Gordon equation is the scalar bulk-to-bulk propagator $\hat{G}$, which we normalize according to
\eqn{GNorm}{
(-\nabla_{\rm AdS}^2 + m_\Delta^2)\hat{G}_\Delta(w,z) = {-1\over N_\Delta} \delta^{d+1}(w,z) \qquad N_\Delta \equiv {-\zeta_\infty(2\Delta) \over 2\nu_\Delta \zeta_\infty(2\Delta-d)} \qquad 2\nu_\Delta \equiv 2\Delta-d\,,
}
where we have defined the ``local zeta function'',
\eqn{zetainftyDef}{
\zeta_\infty(s) = \pi^{-s/2} \Gamma(s/2)\,,
}
and the classic mass-dimension relation relating the bulk scalar field mass $m^2_\Delta$ to the conformal dimension of the dual operator $\Delta$ is $m_\Delta^2 = \Delta(\Delta-d)$. The expression for the propagator in Poincar\'{e} coordinates takes the form
\eqn{Gform}{
	\hat{G}_\Delta(w,z)  &= \left({\xi(w,z) \over 2}\right)^\Delta\: {}_2F_1\left[{\Delta\over 2},{\Delta+1 \over 2};\Delta-{d\over 2}+1;\xi(w,z)^2\right] \cr
	\xi(w,z) &= {2 w_0 z_0 \over w_0^2 + z_0^2 + (w^i-z^i)^2 }\,,
}
where $z=(z_0,z^i) \in \mathbb{R}^+ \times \mathbb{R}^d$ is the bulk point in Poincar\'{e} coordinates, parametrized by the radial coordinate $z_0$ and the boundary direction $z^i$ (and similarly for $w$). We will sometimes abuse terminology to refer to the factor of $\xi(w,z)/2$ as the chordal distance.\footnote{Strictly speaking, $\xi(w,z)^{-1} = \cosh \sigma(w,z) = 1 + u(w,z)$ where $\sigma(w,z)$ is the geodesic distance between the bulk points $w$ and $z$ and $u(w,z)$ is the true chordal distance-squared between them.}

The bulk-to-boundary propagator is obtained as a regularized limit of the bulk-to-bulk propagator upon sending one of the bulk points to the boundary:
\eqn{Kdef}{
\hat{K}_\Delta(x^i,z) = {z_0^\Delta \over (z_0^2 + (z^i- x^i)^2)^{\Delta}}\,.
}
 We will drop the spatial index on $x^i$ and it should be clear from the context whether the variable refers to the boundary direction or a bulk coordinate.

The simplest of the propagator identities is the well-known one involving a product of three bulk-to-boundary propagators, which computes the leading contribution to the scalar three-point conformal correlator,
\eqn{KKK}{
\int_{z\in {\rm AdS}} \hat{K}_{\Delta_1}(x_1,z) \hat{K}_{\Delta_2}(x_2,z) \hat{K}_{\Delta_3}(x_3,z)   &= {C_{\Delta_1\Delta_2\Delta_3} \over (x_{12}^2)^{\Delta_{12,3}} (x_{23}^2)^{\Delta_{23,1}} (x_{31}^2)^{\Delta_{31,2}}}  \,,
}
where the AdS integral $\int_{z \in AdS}$ represents the integration of bulk point $z$ over all of AdS, such that in Poincar\'{e} coordinates the measure takes the form $\int d^{d+1} z/z_0^{d+1}$, with
 the OPE coefficient of the putative dual CFT given by
\eqn{OPEreal}{
C_{\Delta_i\Delta_j\Delta_k} &= {1\over 2}\, \zeta_\infty(2\Delta_{ijk,}-d)\,
\frac{\zeta_\infty(2\Delta_{ij,k})
\zeta_\infty(2\Delta_{jk,i})
\zeta_\infty(2\Delta_{ki,j})}{\zeta_\infty(2\Delta_i)\zeta_\infty(2\Delta_j)\zeta_\infty(2\Delta_k)} \cr
 &=\frac{\pi^{\frac{n}{2}}}{2} \, \Gamma\left(\Delta_{ijk,}-\frac{d}{2}\right)
\frac{\Gamma(\Delta_{ij,k})
\Gamma(\Delta_{jk,i})
\Gamma(\Delta_{ki,j})}{\Gamma(\Delta_i)\Gamma(\Delta_j)\Gamma(\Delta_k)}\,.
}
We invite the reader to notice the strong functional similarity between the OPE coefficients in the real and $p$-adic (equation~\eno{OPEpadic}) setups.

\subsection{Propagator identities involving two propagators}
\label{TWOPROPIDS}

In this subsection, we recall two propagator identities which we will employ in the subsequent sections.
First, we have the identity~\cite{Hijano:2015zsa}
\eqn{KKexpansion}{
\hat{K}_{\Delta_1}(x_1,z) \hat{K}_{\Delta_2}(x_2,z)  &= 2 \sum_{M=0}^\infty  a_M^{\Delta_1;\Delta_2}
\int_{w\in \gamma_{12}} \hat{K}_{\Delta_{1}}(x_1,w)\: \hat{K}_{\Delta_{2}}(x_2,w) \: \hat{G}_{2\Delta_{12,}+2M}(w,z) \,,
}
where
\eqn{aDef}{
a_M^{\Delta_1;\Delta_2} = {1 \over B(\Delta_1+M,\Delta_2+M)} {(-1)^M \over M!} {(\Delta_1)_M (\Delta_2)_M \over (\Delta_1+\Delta_2+M-h)_M }\,.
}
Here, it will be convenient to give the following combination a compact name:
\eqn{alphaDef}{
\alpha_M^{\Delta_1;\Delta_2} \equiv B(\Delta_1+M,\Delta_2+M)\:  a_M^{\Delta_1;\Delta_2}\,.
}
Second, we will utilize the identity~\cite{Hijano:2015zsa}
\eqn{GGid}{
\int_{z \in {\rm AdS}} \hat{G}_{\Delta_a}(w_a,z) \hat{G}_{\Delta_b}(w_b,z) = { {1\over N_{\Delta_b}} \hat{G}_{\Delta_a}(w_a,w_b) - {1\over N_{\Delta_a}} \hat{G}_{\Delta_b}(w_a,w_b)  \over m_{\Delta_a}^2-m_{\Delta_b}^2}\,,
}
where $N_\Delta$ was defined in~\eno{GNorm}.

Graphically, we express \eno{KKexpansion} as
\eqn{KKexpFig}{
\musepic{\figKKexpA} = 2 \sum_{M_L=0}^\infty a_{M_L}^{\Delta_1;\Delta_2}  \musepic{\figKKexpB}\,,
}
where the red-dashed arc denotes a boundary anchored geodesic $\gamma_{12}$ joining boundary points $x_1, x_2$, along which the trivalent bulk point situated on it must be integrated over, and
\eqn{DeltaADef}{
\Delta_L \equiv \Delta_{1}+\Delta_2+2M_L\,.
}
In~\eno{KKexpFig} and below, we will often suppress position space labels; unless otherwise indicated, the operator insertion ${\cal O}_i$ will be understood to be at boundary point $x_i$.

Identity~\eno{KKexpFig} is to an extent reminiscent of taking an OPE of the operators ${\cal O}_1$ and ${\cal O}_2$ on the boundary~\cite{Hijano:2015zsa,Czech:2016xec,deBoer:2016pqk,daCunha:2016crm,Guica:2016pid}, thus we will often refer to it as the ``holographic OPE'' identity.
In a similar vein~\eno{GGid} is expressed graphically as
\eqn{GGidFig}{
\musepic{\figGGidA} = {1  \over m_{\Delta_a}^2-m_{\Delta_b}^2} \left({1\over N_{\Delta_b}} \musepic{\figGGidB} - {1\over N_{\Delta_a}} \musepic{\figGGidC} \right),
}
where the green bulk vertex denotes a bulk point to be integrated over the entire AdS space, while the filled black discs are fixed bulk points. The bulk-to-bulk propagators are represented as solid lines starting and ending inside the Poincar\'{e} disk while bulk-to-boundary propagators are represented as solid lines starting on the boundary and ending in the bulk.
For future reference, we also point out that factors of powers of chordal distance $(\xi/2)^\Delta$ (see equation~\eno{Gform}) will be represented by dotted-black lines joining bulk points.

\subsection{Bulk/boundary three-point scattering}
\label{THREEPROPIDS}
In this subsection we collect new propagator identities involving a bulk integration over three AdS propagators. The graphical representation will be based on the conventions explained in the previous subsection, with one additional piece of notation explained below. We encourage the reader to notice the close similarities between the following identities and their $p$-adic analogs written in section~\ref{PROPBTTREE}.

\subsubsection{$\int \hat{K}\hat{K}\hat{G}$}

The simplest of the mixed bulk/boundary three-point scattering process, represented by a bulk integration over a product of two factors of the bulk-to-boundary propagator and one factor of bulk-to-bulk propagator, can be re-expressed, using a version of an AdS star-triangle-like identity, as follows:
	\eqn{KKGreal}{
	 \musepic{\figIntKKG} &= C_{\Delta_a\Delta_2\Delta_3} \sum_{k_a=0}^\infty {c_{k_a}^{\Delta_2;\Delta_a;\Delta_3} \over (x_{23}^2)^{\Delta_{23,a}-k_a}} \musepic{\figIntKKGa}
    \cr
 &\quad + {1 \over m_{2\Delta_{23,}}^2-m_{\Delta_a}^2} {1 \over N_{\Delta_a}} \sum_{k_a=0}^\infty  {d_{k_a}^{\Delta_2;\Delta_a;\Delta_3} \over (x_{23}^2)^{-k_a}} \musepic{\figIntKKGb}\,,
 	}
 where explicit expressions for the coefficients $c_{k_a}^{\Delta_2;\Delta_3;\Delta_a}$ and $d_{k_a}^{\Delta_2;\Delta_3;\Delta_a}$ are provided in appendix~\ref{KKGPROOF} in equations~\eno{c1Def}-\eno{d1Def}. In the same appendix this identity is written explicitly in terms of bulk-to-bulk and bulk-to-boundary propagators, and an equivalent contour integral form is also presented, along with proofs for each of these using the embedding space formalism.

\subsubsection{$\int \hat{K}\hat{G}\hat{G}$}

The tree-level three-point bulk/boundary scattering amplitude with precisely two bulk legs can be evaluated as follows
\eqn{KGGreal}{
\musepic{\figIntKGG} &= C_{\Delta_3\Delta_a\Delta_b} \sum_{k_a,k_b=0}^\infty c_{k_a;k_b}^{\Delta_a;\Delta_3;\Delta_b} \musepic{\figIntKGGa} \cr
 &\quad  + \left(
\sum_{k_a,k_b=0}^\infty
d_{k_a;k_b}^{\Delta_a;\Delta_3;\Delta_b}  \musepic{\figIntKGGb}
   + (a\leftrightarrow b)\right),
   }
where the expansion coefficients are written in terms of Lauricella functions and are given in~\eno{c2Def}-\eno{d2Def}. The dotted-black lines denote factors of chordal distance rather than the full bulk-to-bulk propagator; for instance in the first term on the RHS, the dotted line stands for a factor of $(\xi(a,b)/2)^{\Delta_{ab,3}+k_a+k_b}$.
Here and below, the moniker ``$(a \leftrightarrow b)$'' represents switching all instances of $a$ and $b$, including all labels where $a$ and $b$ appear as subscripts or superscripts. This identity, rewritten explicitly in~\eno{KGGreal2}, is proven in appendix~\ref{KGGPROOF}.

\subsubsection{$\int \hat{G}\hat{G}\hat{G}$}

Finally, the purely AdS three-particle scattering provides the following variant of the flat space star-triangle-identity~\cite{DEramo:1971hnd}:
\eqn{GGGreal}{
 \musepic{\figIntGGG} &= C_{\Delta_a\Delta_b\Delta_c} \sum_{k_a,k_b,k_c=0}^\infty c_{k_a;k_b;k_c}^{\Delta_a;\Delta_b;\Delta_c} \musepic{\figIntGGGa} \cr
 &\quad + \left(
\sum_{k_a,k_b,k_c=0}^\infty
d_{k_a;k_b;k_c}^{\Delta_a;\Delta_b;\Delta_c}  \musepic{\figIntGGGb}
   + (a\leftrightarrow b) +(a\leftrightarrow c) \right),
   }
 where the expansion coefficients are given in~\eno{c3Def}-\eno{d3Def} and the identity is proven in appendix~\ref{GGGPROOF}.  Here we are using the shorthand
 \eqn{kabc}{
k_{i_1 \ldots i_\ell, i_{\ell+1} \ldots i_{j}} \equiv k_{i_1} + \cdots + k_{i_\ell} - k_{i_{\ell+1}} - \cdots - k_{i_j}
}
for the integral parameters $k_a$, where we stress the factor of two difference compared with the corresponding shorthand for conformal dimensions defined in~\eno{DeltaijkDef}.

\vspace{1.5em}

We end this section with some remarks.
\begin{itemize}
    \item The inverse relation~\cite{Parikh:2019ygo}
\eqn{G2xi}{
\left({\xi(w,z) \over 2}\right)^\Delta = \sum_{k=0}^\infty {(-1)^k \over k!} {(\Delta)_{2k} \over (\Delta-d/2+k)_k} \hat{G}_{\Delta+2k}(w,z)\,,
}
can be used to re-express the dotted-black lines denoting chordal distance factors on the right-hand-side of the various propagator identities in terms of the bulk-to-bulk propagator, at the cost of introducing an additional sum for each such factor.
    \item As will become clear later, when we use the identities above to obtain the CBD of bulk diagrams, the terms in the second lines of~\eno{KKGreal},~\eno{KGGreal} and~\eno{GGGreal} will correspond physically to the exchange of various combinations of higher-trace operators in the intermediate channels.
\end{itemize}

\section{Holographic dual of the six-point block in the OPE channel}
\label{SIXOPEGEODESIC}

Before presenting new results for the six-point conformal block, let us briefly review what is known in the literature. The holographic dual for the global four-point conformal block with external scalar insertions and an arbitrary operator in the intermediate channel was worked out in Ref.~\cite{Hijano:2015zsa}.
As mentioned in section~\ref{INTRO}, the geometric representation is given in terms of a four-point exchange diagram, except with both bulk integrals replaced by geodesic integrals over boundary anchored geodesics.
Using the pictorial conventions of sections~\ref{TWOPROPIDS}-\ref{THREEPROPIDS}, the geodesic diagram  given by
\eqn{calW4}{
{\cal W}_{\Delta_a}^{\Delta_1,\Delta_2,\Delta_3,\Delta_4}(x_1,x_2,x_3,x_4) = \musepic{\figcalWfour}
}
computes the scalar four-point block, via
\eqn{calWconfwave4}{
W_{\Delta_a}^{\Delta_1,\Delta_2,\Delta_3,\Delta_4}(x_i) = {4 \over
B(\Delta_{a1,2},\Delta_{a2,1})\: B(\Delta_{a3,4},\Delta_{a4,3})}\:
{\cal W}_{\Delta_a}^{\Delta_1,\Delta_2,\Delta_3,\Delta_4}(x_i) \,,
}
where $B(\cdot,\cdot)$ is the Euler Beta function.\footnote{The four-point block is normalized such that for $(x_{12}^2 x_{34}^2)/(x_{13}^2 x_{24}^2) \ll 1$ and $(x_{14}^2 x_{23}^2)/(x_{13}^2 x_{24}^2) \approx 1$, it has the leading order behavior
\eqn{W4norm}{
W_{\Delta_a}^{\Delta_1,\ldots,\Delta_4}(x_i) \approx {1 \over (x_{12}^2)^{\Delta_{12,}} (x_{34}^2)^{\Delta_{34,}}}
\left({x_{24}^2 \over x_{14}^2} \right)^{\Delta_{1,2}}
\left({x_{14}^2 \over x_{13}^2} \right)^{\Delta_{3,4}} \left({x_{12}^2 x_{34}^2 \over x_{13}^2 x_{24}^2}\right)^{\Delta_a/2}.
}
}
 Further generalizations to spinning external and exchanged operators were considered in Refs.~\cite{Nishida:2016vds,Castro:2017hpx,Dyer:2017zef,Chen:2017yia,Gubser:2017tsi,Kraus:2017ezw,Tamaoka:2017jce,Nishida:2018opl,Das:2018ajg}.
 It is helpful to consider an alternate representation of~\eno{calW4}, which makes the comparison with the higher-point blocks discussed below more transparent. Essentially, we replace the bulk-to-bulk propagator in~\eno{calW4} with its series representation in terms of the chordal distance function (see~\eno{Gform}), to yield
 \eqn{4ptlimInit}{
{\cal W}_{\Delta_a}^{\Delta_1,\ldots,\Delta_4}(x_i) =
\sum_{k_a=0}^\infty c_{k_a}^{\Delta_a} \musepic{\figcalWfourxi}\,,
}
where
\eqn{c0Def}{
c_{k_a}^{\Delta_a} = {1\over k_a!} {(\Delta_a)_{2k_a} \over (\Delta_a-d/2+1)_{k_a}}\,.
}

\paragraph{The five-point case.} More recently, in Ref.~\cite{Parikh:2019ygo} it was shown that the object
\eqn{calW5}{
{\cal W}_{\Delta_a;\Delta_b}^{\Delta_1,\ldots,\Delta_5}(x_i) &= \sum_{k_a,k_b=0}^\infty c_{k_a;k_b}^{\Delta_a;\Delta_3;\Delta_b} \musepic{\figcalW},
}
with $c_{k_a;k_b}^{\Delta_a;\Delta_3;\Delta_b}$ given by~\eno{c2Def}, computes the global five-point conformal block in arbitrary spacetime dimensions, according to
\eqn{calW5confwave}{
{W}_{\Delta_a;\Delta_b}^{\Delta_1,\ldots,\Delta_5}(x_i) = {4 \over
B(\Delta_{a1,2},\Delta_{a2,1})\: B(\Delta_{b4,5},\Delta_{b5,4})} \:
{\cal W}_{\Delta_a;\Delta_b}^{\Delta_1,\ldots,\Delta_5}(x_i)\,,
}
where $W_{\Delta_a;\Delta_b}^{\Delta_1,\ldots,\Delta_5}(x_i)$ is the five-point conformal block~\eno{ProjectConf5} corresponding to external scalar insertions $\Delta_1,\ldots,\Delta_5$, representing the contribution coming from the conformal families of operators with highest-weight spin-$0$ representations labelled by dimensions $\Delta_a$ and $\Delta_b$.
 The only bulk-integrations in~\eno{calW5} are geodesic integrations over boundary anchored geodesics $\gamma_{12}$ and $\gamma_{45}$ joining boundary points $x_1$ to $x_2$ and $x_4$ to $x_5$ respectively.
Graphically, we will represent the five-point conformal block~\eno{calW5confwave} itself by the diagram shown in figure~\ref{fig:5block}.
The five-point block was also recently computed purely within the CFT framework using  the shadow formalism~\cite{Rosenhaus:2018zqn}.

It is instructive to compare the $p$-adic result~\eno{calW5padic}-\eno{W5calc} with its real analog above. The holographic dual over the reals takes essentially the same form as the $p$-adics, the difference being it is written as a sum over a two-parameter semi-infinite families of diagrams parametrized by two integers, which account for the descendant contribution originating from the conformal families of the two single-trace primaries being exchanged in the intermediate channels. The similarities are even more apparent when one notes that the Euler Beta function $B(s,t)$ in~\eno{calW5confwave}, which is to be compared with~\eno{calW5CBpadic}, can be rewritten as
\eqn{betainftyDef}{
B(s,t) = \beta_\infty(2s,2t) \equiv {\zeta_\infty(2s)\zeta_\infty(2t) \over \zeta_\infty(2s+2t)}
}
in terms of the local zeta function $\zeta_\infty$ defined in~\eno{zetainftyDef}.

\paragraph{Six-point block in the OPE channel.}
We now turn our attention to the six-point conformal block in the so-called OPE channel, depicted on the LHS of figure~\ref{fig:6opechannel}, where one first takes pairwise OPEs between operator insertions at $(x_1, x_2)$, $(x_3,x_4)$ and $(x_5,x_6)$ to isolate the contribution from the conformal families associated with highest weight representations labelled by $\Delta_a, \Delta_b$ and $\Delta_c$ in the intermediate channels.
    The procedure leading to its holographic representation, as discussed below, follows closely the strategy discussed in Ref.~\cite{Parikh:2019ygo} in the context of the five-point block.

The starting point is the six-point exchange diagram in the star configuration, with three internal scalar field exchanges, to which we apply the holographic OPE identity~\eno{KKexpFig} at insertion-point pairs $(x_1,x_2)$, $(x_3,x_4)$ and $(x_5,x_6)$:
\eqn{SixOPEstep1}{
\musepic{\figSixExchThreeA} = \sum_{M_LM_R,M_C=0}^\infty 8\, a_{M_L}^{\Delta_1;\Delta_2}\, a_{M_R}^{\Delta_5;\Delta_6}\, a_{M_C}^{\Delta_3;\Delta_4} \musepic{\figSixOPEstepB}\,,
}
where
\eqn{DeltaLCRDef}{
\Delta_{L} \equiv 2\Delta_{12,}+2M_L \qquad \Delta_{C} \equiv 2\Delta_{34,}+2M_C \qquad \Delta_{R} \equiv 2\Delta_{56,}+2M_R\,.
}
From here on,  the upper and lower limits of sums over the variables $M,M_A, M_B,\ldots$ etc will be suppressed and implicitly assumed to be $\infty$ and $0$, respectively, unless otherwise stated.
Now use~\eno{GGidFig} to evaluate the three ``outside'' bulk integrations, but focus solely on one  of the  resulting eight terms:
\eqn{SixOPEstep2}{
& \musepic{\figSixExchThreeA} \cr
&= \sum_{M_LM_R,M_C} {8\,a_{M_L}^{\Delta_1;\Delta_2}\, a_{M_R}^{\Delta_5;\Delta_6}\, a_{M_C}^{\Delta_3;\Delta_4} \over N_{\Delta_L}  (m^2_{\Delta_0} -m^2_{\Delta_L})  N_{\Delta_R}(m^2_{\Delta_{0^\prime}} -m^2_{\Delta_R})N_{\Delta_C}(m^2_{\Delta_c} -m^2_{\Delta_C})} \musepic{\figSixOPEstepC} \cr
&\quad + ({\rm 7\ other\ terms}) \,.
}
In fact, the three sums in the first term of~\eno{SixOPEstep2} can be performed analytically, using
\eqn{OPEsumId}
{
\sum_{M=0}^\infty
\frac{1}{N_{\Delta_{1}+\Delta_{2}+2M}}
\frac{a_M^{\Delta_1,\Delta_2}}{m^2_{\Delta_0}-m^2_{\Delta_{1}+\Delta_{2}+2M}}
=\frac{C_{\Delta_0\Delta_1\Delta_2}}{B(\Delta_{01,2},\Delta_{02,1})}\,;
}
see appendix \ref{OPEsumIdProof} for a derivation.\footnote{A special case of this identity, for $d=2$, was noted in Ref.~\cite{Gubser:2017tsi}. The equation \eno{OPEsumId} itself is a special case of the identity \eno{key1} that we use below to carry out the CBD of 5-point diagrams. And \eno{key1} is a special case of the identity \eno{key6pt} that we use to perform the CBD of 6-point diagrams. We can think of these equations as analogs of the simpler equation \eno{BetaToC} that we used in the parallel $p$-adic computations.} This leads to
\eqn{SixOPEstep3}{
 \musepic{\figSixExchThreeA}
&= {8\,C_{\Delta_0\Delta_1\Delta_2}\,  C_{\Delta_c\Delta_3\Delta_4}\, C_{\Delta_0^\prime \Delta_5\Delta_6} \over B(\Delta_{01,2},\Delta_{02,1})\,  B(\Delta_{c3,4},\Delta_{c4,3})\, B(\Delta_{0^\prime 5,6},\Delta_{0^\prime 6,5})}\! \musepic{\figSixOPEstepC} \cr
&\quad + ({\rm 7\ other\ terms}) \,.
}
At this point the three-particle scattering identity~\eno{GGGreal} is employed on the first term to obtain
\eqn{SixOPEstep4}{
 \musepic{\figSixExchThreeA}
&= \left( {8\,C_{\Delta_0\Delta_1\Delta_2}\, C_{\Delta_c\Delta_3\Delta_4}\,C_{\Delta_0^\prime \Delta_5\Delta_6}\, C_{\Delta_0\Delta_c\Delta_0^\prime} \over B(\Delta_{01,2},\Delta_{02,1})\, B(\Delta_{c3,4},\Delta_{c4,3})\, B(\Delta_{0^\prime 5,6},\Delta_{0^\prime 6,5})} {\cal W}_{\Delta_0;\Delta_c;\Delta_{0^\prime}}^{\Delta_1,\ldots,\Delta_6}(x_i) \right. \cr
&\quad + ({\rm 3\ more\ terms}) \bigg) + ({\rm 7\ other\ terms}) \,,
}
where  only one of the four terms arising from applying~\eno{GGGreal} are explicitly shown, and  we have defined the \emph{six-point geodesic bulk diagram in the OPE channel} (more precisely, a \emph{weighted sum} over six-point geodesic bulk diagrams)\footnote{Note the slight change in the names for internal scaling dimensions moving forward.}
\eqn{calW6star}{
{\cal W}_{\Delta_a;\Delta_c;\Delta_b}^{\Delta_1,\ldots,\Delta_6}(x_i) \equiv \sum_{k_a,k_b,k_c=0}^\infty c^{\Delta_a;\Delta_b;\Delta_c}_{k_a;k_b;k_c} \musepic{\figcalWStar}\,,
}
where, explicitly
\eqn{6ptStarIntegral}{
H_{\Delta_L,\Delta_C,\Delta_R} \! \equiv\!\!\!\! \musepic{\figSixIntegral} \!\!\!\! \equiv \!\!\!\! \begin{array}{l} \displaystyle{\iiint_{\substack{w \in \gamma_{12}\\ w^\prime \in \gamma_{34}\\ w^{\prime\prime} \in \gamma_{56}}}
 \hat{K}_{\Delta_1}(x_1,w) \hat{K}_{\Delta_2}(x_2,w)
\hat{K}_{\Delta_3}(x_3,w^\prime)} \cr
\qquad  \times  \hat{K}_{\Delta_4}(x_4,w^\prime) \hat{K}_{\Delta_5}(x_5,w^{\prime\prime}) \hat{K}_{\Delta_6}(x_6,w^{\prime\prime}) \cr
\qquad  \times \left(\displaystyle{\xi(w,w^\prime) \over 2}\right)^{\color{blue}\Delta_L}
\left(\displaystyle{\xi(w,w^{\prime\prime}) \over 2}\right)^{\color{blue}\Delta_C}
\left(\displaystyle{\xi(w^\prime,w^{\prime\prime}) \over 2}\right)^{\color{blue} \Delta_R}
    \end{array}
}
and $c^{\Delta_a;\Delta_b;\Delta_c}_{k_a;k_b;k_c}$ is given in~\eno{c3Def}.

Following the strategy of Ref.~\cite{Parikh:2019ygo}, we are now in a position to make an educated guess for the holographic representation of the six-point global conformal block.
We conjecture that the coefficient of the product of OPE coefficients is the holographic object which computes the six-point block, that is
\eqn{calW6starConfwave}{
{W}_{\Delta_a;\Delta_c;\Delta_b}^{\Delta_1,\ldots,\Delta_6}(x_i) &= {8 \over
B(\Delta_{a1,2},\Delta_{a2,1})\:
 B(\Delta_{c3,4},\Delta_{c4,3}) \:
B(\Delta_{b5,6},\Delta_{b6,5})}\: {\cal W}_{\Delta_a;\Delta_c;\Delta_b}^{\Delta_1,\ldots,\Delta_6}(x_i)
\,,
}
where ${W}_{\Delta_a;\Delta_c;\Delta_b}^{\Delta_1,\ldots,\Delta_6}(x_i)$ is the six-point block in the OPE channel, given in~\eno{ProjectConf6OPE} and depicted graphically in figure~\ref{fig:6opechannel}.

In the rest of this section, we  prove this conjecture by establishing that  the geodesic bulk diagram satisfies the correct differential equations with the right boundary conditions. More precisely, we will show that~\eno{calW6starConfwave} is an eigenfunction of the appropriate multi-point Casimirs of the global conformal group $SO(d+1,1)$ with the right eigenvalues and has the right limiting behavior in the OPE limit. These checks are sufficient to establish the conjecture~\cite{SimmonsDuffin:2012uy}.

Before proceeding, we make some additional remarks.
\begin{itemize}
 \item Using~\eno{G2xi}, one can easily re-express the holographic representation~\eno{calW6star} (and~\eno{calW5}) in terms of bulk-to-bulk propagators rather than chordal distances. However, for computational convenience we prefer to use the representation given in~\eno{calW6star}.

 \item It is reassuring to observe that the correspondence between the real and $p$-adic holographic representations of four- and five-point blocks continues to hold at six-points; at six-points the comparison is between~\eno{calW6star}-\eno{calW6starConfwave} and~\eno{calW6padic}-\eno{calW6CBpadic}, with~\eno{betainftyDef} in mind.

 \item Equipped with the knowledge of the holographic representation of the six-point block, a full analysis of all the terms in~\eno{SixOPEstep4} leads to the full CBD of the six-point star-shaped bulk diagram in the OPE channel; look forward to section~\ref{SIXCBD} for more details.
\end{itemize}

\subsection{OPE limit}

In the OPE limit $x_2 \to x_1$, the leading contribution to the conformal block defined in~\eno{calW6starConfwave} comes from the term with $k_a=0$, and takes the form
\eqn{calW6starOPE1}{
&\lim_{x_2\to x_1} {8\:{\cal W}_{\Delta_a;\Delta_c;\Delta_b}^{\Delta_1,\ldots,\Delta_6}(x_i) \over B(\Delta_{a1,2},\Delta_{a2,1})
 B(\Delta_{c3,4},\Delta_{c4,3})
B(\Delta_{b5,6},\Delta_{b6,5}) } \cr
 &=   {4 (x_{12}^2)^{\Delta_{a,12}} \over B(\Delta_{c3,4},\Delta_{c4,3}) B(\Delta_{b5,6},\Delta_{b6,5})
 } \sum_{k_b,k_c=0}^\infty c_{0;k_b;k_c}^{\Delta_a;\Delta_b;\Delta_c} \!\!\!\!\!\! \musepic{\figcalWOPElim} \cr
 &= (x_{12}^2)^{\Delta_{a,12}} \:  W^{\Delta_3,\Delta_4,\Delta_a,\Delta_5,\Delta_6}_{\Delta_c;\Delta_b}(x_3,x_4,x_1,x_5,x_6)\,,
}
where in the last equality we used the fact that $c_{0;k_b;k_c}^{\Delta_a;\Delta_b;\Delta_c}= c_{k_b;k_c}^{\Delta_b;\Delta_a;\Delta_c}$ (using the definitions~\eno{c2Def} and~\eno{c3Def}), as well as the relation between the linear combination of five-point geodesic diagrams and the five-point conformal block~\eno{calW5}-\eno{calW5confwave}. The subleading terms above scale with an overall factor of $(x_{12}^2)^{\Delta_{a,12}+k_a}$ for positive integral $k_a$, thus are suppressed in the limit under consideration.

By symmetry, an identical analysis is possible in the other OPE limits, $x_4\to x_3$ and $x_6 \to x_5$. Thus in each case, we establish that the six-point conformal block in the OPE channel, as defined in~\eno{calW6starConfwave}, obeys the correct boundary conditions:
\eqn{W6bdy}{
\lim_{x_2\to x_1} W_{\Delta_a;\Delta_c;\Delta_b}^{\Delta_1,\Delta_2,\Delta_3,\Delta_4,\Delta_5,\Delta_6}(x_1,x_2,x_3,x_4,x_5,x_6) = (x_{12}^2)^{\Delta_{a,12}} \:  W^{\Delta_3,\Delta_4,\Delta_a,\Delta_5,\Delta_6}_{\Delta_c;\Delta_b}(x_3,x_4,x_1,x_5,x_6)\,,
}
with the other limits obtained from permuting the labels appropriately.

This provides a check that the right boundary conditions are obeyed by our conjecture. Next we prove that the conformal block as defined by the right hand side of~\eno{calW6starConfwave} obeys the correct differential equations as well.

\subsection{Eigenfunction of conformal Casimirs}

In this section we will  prove that the six-point block given by~\eno{calW6starConfwave} is an eigenfunction of the appropriate multi-point Casimirs with the right eigenvalues. Particularly, in the OPE channel of figure~\ref{fig:6opechannel}, we will show that the geodesic diagram~\eno{calW6star} satisfies
\eqn{CasimirEqn}{
({\cal L}^{(1)}+{\cal L}^{(2)})^2\: {\cal W}_{\Delta_a;\Delta_c;\Delta_b}^{\Delta_1,\ldots,\Delta_6}(x_i) &= C_2(\Delta_a)\: {\cal W}_{\Delta_a;\Delta_c;\Delta_b}^{\Delta_1,\ldots,\Delta_6}(x_i) \cr
({\cal L}^{(3)}+{\cal L}^{(4)})^2\: {\cal W}_{\Delta_a;\Delta_c;\Delta_b}^{\Delta_1,\ldots,\Delta_6}(x_i) &= C_2(\Delta_c)\: {\cal W}_{\Delta_a;\Delta_c;\Delta_b}^{\Delta_1,\ldots,\Delta_6}(x_i) \cr
({\cal L}^{(5)}+{\cal L}^{(6)})^2\: {\cal W}_{\Delta_a;\Delta_c;\Delta_b}^{\Delta_1,\ldots,\Delta_6}(x_i) &= C_2(\Delta_b)\: {\cal W}_{\Delta_a;\Delta_c;\Delta_b}^{\Delta_1,\ldots,\Delta_6}(x_i) \,,
}
where
\eqn{CasimirEigenvalue}{
C_2(\Delta) = m_\Delta^2 = \Delta(\Delta-n)\,.
}
Here ${\cal L}_{AB}^{(1)}$ are the differential generators of the global conformal algebra constructed out of the spacetime coordinate and conformal dimension associated with the operator insertion at $x_1$. The Casimir is constructed as ${\cal L}^2 \equiv {1\over 2}{\cal L}_{AB} {\cal L}^{AB}$, while the Casimirs considered above are analogously defined multi-point Casimirs.

In fact we simply need to prove the first equation in~\eno{CasimirEqn}; the remaining two follow trivially from symmetry arguments. The proof is most convenient in embedding space. To keep the discussion short, we refer the reader to Refs.~\cite{Hijano:2015zsa,Parikh:2019ygo} for more details on the embedding space formalism as it relates to the proof by conformal Casimir. Indeed, the proof presented here follows closely the procedure used in Ref.~\cite{Parikh:2019ygo} in the context of the five-point block.

In embedding space, define
\eqn{FDef}{
F_{\Delta_L,\Delta_C}(P_1,P_2,W^\prime,W^{\prime\prime}) \equiv  \int_{W \in \gamma_{12}}
 \hat{K}_{\Delta_1}(P_1,W) \hat{K}_{\Delta_2}(P_2,W)
\left({\xi(W,W^\prime) \over 2}\right)^{\Delta_L}
\left({\xi(W,W^{\prime\prime}) \over 2}\right)^{\Delta_C}\!\!\!,
}
where $P^2_i = 0$ denote the null coordinates and $W^2 = W^{\prime 2} = W^{\prime\prime2} = -1$ define the AdS hypersurface. The bulk-to-boundary propagator and chordal distance factors take simple power law forms in embedding space, while the conformal generators act linearly making computations especially easier.
In fact, using the fact that $F_{\Delta_L,\Delta_C}$ is conformally invariant under simultaneous rotations of $P_1, P_2, W^\prime, W^{\prime\prime}$, one can trade the action of the multi-point Casimir constructed out of $P_1, P_2$ for one constructed out of $W^\prime, W^{\prime\prime}$~\cite{Hijano:2015zsa,Parikh:2019ygo}:
\eqn{FCasimir}{
({\cal L}^{(1)}+{\cal L}^{(2)})^2\: F_{\Delta_L,\Delta_C}(P_1,P_2,W^\prime,W^{\prime\prime}) &= ({\cal L}^{(W^\prime)}+{\cal L}^{(W^{\prime\prime})})^2\: F_{\Delta_L,\Delta_C}(P_1,P_2,W^\prime,W^{\prime\prime}) \,.
}
Using the definition of $F_{\Delta_C,\Delta_R}$ and the explicit forms for $\hat{K}$ and $\xi$ in embedding space (see appendix~\ref{PROOFPROPAG}), this can easily be computed to give (suppressing the arguments of $F_{\Delta_C,\Delta_R}$)~\cite{Parikh:2019ygo},
\eqn{FCasimirAgain}{
({\cal L}^{(W^\prime)}+{\cal L}^{(W^{\prime\prime})})^2\: F_{\Delta_L,\Delta_C} &= m_{\Delta_L+\Delta_C}^2\: F_{\Delta_L,\Delta_C} - 4\Delta_L \Delta_C   \left({\xi(W^\prime,W^{\prime\prime}) \over 2}\right)^{-1} F_{\Delta_L+1,\Delta_C+1} \cr
&\quad  - 4\Delta_L(\Delta_L+1)\: F_{\Delta_L+2,\Delta_C} - 4\Delta_C(\Delta_C+1)\: F_{\Delta_L,\Delta_C+2}\,.
}
Using this we immediately obtain
\eqn{HCasimir}{
({\cal L}^{(1)}+{\cal L}^{(2)})^2\: H_{\Delta_L,\Delta_C,\Delta_R} &=  m_{\Delta_L+\Delta_C}^2\: H_{\Delta_L,\Delta_C,\Delta_R} - 4\Delta_L \Delta_C \:   H_{\Delta_L+1,\Delta_C+1,\Delta_R-1} \cr
&\quad  - 4\Delta_L(\Delta_L+1)\: H_{\Delta_L+2,\Delta_C,\Delta_R} - 4\Delta_C(\Delta_C+1)\: H_{\Delta_L,\Delta_C+2,\Delta_R}\,,
}
where $H_{\Delta_L,\Delta_C,\Delta_R}$ was defined in~\eno{6ptStarIntegral}.
Due to~\eno{HCasimir}, we conclude that the geodesic bulk diagram in~\eno{calW6star} satisfies
\eqn{calWStarCasimir}{
({\cal L}^{(1)}+{\cal L}^{(2)})^2\: {\cal W}_{\Delta_a;\Delta_c;\Delta_b}^{\Delta_1,\ldots,\Delta_6}(P_i) &= ({\cal L}^{(1)}+{\cal L}^{(2)})^2 \!\!\sum_{k_a,k_b,k_c=0}^\infty \!\! c^{\Delta_a;\Delta_b;\Delta_c}_{k_a;k_b;k_c} \musepic{\figcalWStar} \cr
&= \sum_{k_a,k_b,k_c=0}^\infty \widetilde{c}_{k_a;k_b;k_c} \musepic{\figcalWStar},
}
where (suppressing in the symbols $c^{\Delta_a;\Delta_b;\Delta_c}_{k_a;k_b;k_c}$ the superscripts, which remain fixed) we have defined
\eqn{ctilde}{
\widetilde{c}_{k_a;k_b;k_c} &\equiv m_{\Delta_a+2k_a}^2\: c_{k_a;k_b;k_c} - 4 (\Delta_{ac,b} + k_{ac,b}-1) (\Delta_{ab,c}+k_{ab,c}-1) \: c_{k_a-1;k_b;k_c} \cr
&\quad -4(\Delta_{ac,b}+k_{ac,b}-2)(\Delta_{ac,b}+k_{ac,b}-1)\: c_{k_a-1;k_b;k_c-1} \cr
&\quad -4(\Delta_{ab,c}+k_{ab,c}-2)(\Delta_{ab,c}+k_{ab,c}-1)\:
c_{k_a-1;k_b-1;k_c}\,.
}
In the last step in~\eno{calWStarCasimir}, we integer-shifted the dummy variables $k_a, k_b, k_c$ to bring the diagram on the RHS to the same form as the one on the LHS, leading to the coefficient in~\eno{ctilde}.

All we need to do now is to show that
\eqn{ToShowStar}{
\widetilde{c}_{k_a;k_b;k_c} - m_{\Delta_a}^2 c_{k_a;k_b;k_c} = 0 \qquad \forall\: k_a;k_b;k_c\,.
}

We first recall from~\eno{c3Def} that the $c_{k_a;k_b;k_c}$ coefficients involve Lauricella functions of three variables.
So, to show~\eno{ToShowStar} we use the series representation of the Lauricella function given in~\eno{LauricellaDef} to rewrite the LHS of~\eno{ToShowStar} as a triple sum over dummy integral variables $n_a,n_b$ and $n_c$, where we have paired $k_a$ with $n_a$ and so on.
Next, we evaluate the sums over $n_a$ and $n_b$, to obtain a summand for the $n_c$ sum,  which can be simplified using a basic hypergeometric ${}_3F_2$ identity~\cite{wolfram2} to give that $\widetilde{c}_{k_a;k_b;k_c} - m_{\Delta_a}^2 c_{k_a;k_b;k_c}$ is equal to
\eqn{}{
& {4(-1)^{k_a+k_b+k_c+1} (\Delta_{ab,c})_{k_{ab,c}} (\Delta_{ac,b})_{k_{ac,b}} (\Delta_{bc,a})_{k_{bc,a}} \over k_b! (k_a-1)!  (\Delta_a-{n\over 2}+1)_{k_a-1} } \sum_{n_c=0}^{k_c}  \Bigg[ { (-1)^{n_c} - \sin \pi k_c \csc \pi(k_c-n_c)  \over  n_c! \: \Gamma(k_c-n_c) }  \cr
& \qquad \qquad  \times { \left(\Delta_{abc,}-{n \over 2}\right)_{n_c} \!\! (\Delta_{a,bc}-n_c)_{k_a}  \over (\Delta_{bc,a}+n_c)\: (\Delta_c-{n\over 2}+1)_{n_c} }\: \pFq{3}{2}{-k_b,\Delta_{bc,\!\!a}+n_c,\Delta_{abc,}\!\!\!-{n\over 2}+n_c}{\Delta_b-{n\over 2}+1, \Delta_{bc,\!\!a}+n_c-k_a}{1}\Bigg].
}
But we notice that the summand vanishes for each $n_c$ since it is proportional to a factor of
\eqn{}{
(-1)^{n_c} - \sin \pi k_c \csc \pi(k_c-n_c)  = (-1)^{n_c} - {\sin \pi k_c \over (-1)^{n_c} \sin \pi k_c} = 0\,.
}
This proves~\eno{ToShowStar}, thus establishing~\eno{CasimirEqn} and in turn the fact that the geometric RHS of~\eno{calW6starConfwave} indeed computes the six-point conformal block.

\section{Conformal block decomposition via geodesic diagrams}
\label{CBD}

In this section we work out the CBD of several tree-level AdS diagrams in the direct channel.
We will use the propagator identities of section~\ref{PROPIDS}, where each individual term in identities such as~\eno{KKGreal},~\eno{KGGreal} and~\eno{GGGreal} will be physically reinterpreted as the contribution to a given bulk diagram from the conformal families of a specific set of intermediate (single-trace and/or multi-twist) primary exchanges.
This computation will also rely on the knowledge of the holographic representations for five- and six-point conformal blocks in the decomposition channel of interest, as discussed in the previous section.
At the end of this section we will provide a repackaging of the CBDs in terms of spectral integrals, which reaffirms the agreement with the expected results from the shadow formalism.

We begin by recalling  the CBD of four-point contact and exchange diagrams, rederived in Ref.~\cite{Hijano:2015zsa} with the help of two-propagator identities~\eno{KKexpFig} and~\eno{GGidFig}. Like in the rest of the paper, we will restrict ourselves to external scalar operators with scalar exchanges. Generalizations to spinning four-point bulk diagrams with spin exchanges can be found in Refs.~\cite{Nishida:2016vds, Castro:2017hpx, Dyer:2017zef, Chen:2017yia, Tamaoka:2017jce, Nishida:2018opl, Das:2018ajg}.

The CBD of the four-point contact diagram is given by
\eqn{FourptContact}{
D_4 &\equiv \int_{z \in {\rm AdS}} \hat{K}_{\Delta_1}(x_1,z) \hat{K}_{\Delta_2}(x_2,z) \hat{K}_{\Delta_4}(x_4,z) \hat{K}_{\Delta_5}(x_5,z) \cr
 &= \sum_{M_A} P_{\rm cont}^{(12)}(M_A)\: W_{\Delta_{A}}^{\Delta_1,\Delta_2,\Delta_4,\Delta_5}(x_i)
 + \sum_{M_B} P_{\rm cont}^{(45)}(M_B)\: W_{\Delta_{B}}^{\Delta_1,\Delta_2,\Delta_4,\Delta_5}(x_i)\,,
}
where we have defined
\eqn{DeltaABDef}{
\Delta_A \equiv \Delta_{1}+\Delta_2+2M_A \qquad \Delta_B \equiv \Delta_{4}+\Delta_5+2M_B\,,
}
and like before $W_{\Delta}^{\Delta_1,\Delta_2,\Delta_4,\Delta_5}$ is the four-point conformal block~\cite{Dolan:2000ut} for external scalar operators $\Delta_1,\Delta_2,\Delta_4$ and $\Delta_5$ with the exchanged scalar primary labelled by the dimension $\Delta$, and the decomposition coefficients are\footnote{\label{fn:notationtypo}We thank E.~Perlmutter for pointing out a normalization convention typo in Eqn.~(4.8) of Ref.~\cite{Hijano:2015zsa}. Fixing the typo leads to an extra normalization factor in~\eno{FourptOPE}, as compared to Eqn.~(4.12) of Ref.~\cite{Hijano:2015zsa}.}
\eqn{FourptOPE}{
P_{\rm cont}^{(12)}(M_A) &=  \alpha_{M_A}^{\Delta_1;\Delta_2}  \left( B(\Delta_{A4,5}, \Delta_{A5,4})\sum_{M_B} {1\over N_{\Delta_{B}}}{a_{M_B}^{\Delta_4;\Delta_5} \over m_{\Delta_{A}}^2-m_{\Delta_{B}}^2} \right) \cr
P_{\rm cont}^{(45)}(M_B) &= \alpha_{M_B}^{\Delta_4;\Delta_5} \left( B(\Delta_{B1,2}, \Delta_{B2,1})\sum_{M_A} {1\over N_{\Delta_{A}}}{a_{M_A}^{\Delta_1;\Delta_2} \over m_{\Delta_{B}}^2-m_{\Delta_{A}}^2} \right),
}
where $\alpha_M^{s;t}$ was defined in~\eno{alphaDef}.
We remind the reader that the upper and lower limits on the $M_i$ sums are understood to be $\infty$ and $0$ respectively.
In fact, we can analytically perform the sums in~\eno{FourptOPE} using~\eno{OPEsumId}
to obtain  the compact expressions
\eqn{FourptOPEAgain}{
P_{\rm cont}^{(12)}(M_A) &= \alpha_{M_A}^{\Delta_1;\Delta_2}  \: C_{\Delta_4\Delta_5\Delta_{A}} \cr
P_{\rm cont}^{(45)}(M_B) &= \alpha_{M_B}^{\Delta_4;\Delta_5}  \: C_{\Delta_1\Delta_2\Delta_{B}}\,,
}
where the OPE coefficient $C_{ijk}$ was written down in~\eno{OPEreal}.
The primary operators being exchanged in the CBD~\eno{FourptContact} are interpreted as double-trace primaries,  written schematically as
\eqn{doubleTrace}{
[{\cal O}_i {\cal O}_j]_M \approx {\cal O}_i \partial^{2M} {\cal O}_j\,,
}
with conformal dimensions (the same as twists since the external operators are scalars and we have set $\ell =0$ in~\eno{doubletwist})
\eqn{doubleTraceDim}{
\Delta^{(ij)}(M) = \Delta_i + \Delta_j + 2M + \gamma^{(ij)}(M)\,,
}
where $\gamma^{(ij)}$ is the anomalous dimension. For generic external conformal dimensions, these anomalous dimensions do not contribute to the CBD at this order in $1/N$, a characteristic of large $N$ CFTs~\cite{Liu:1998th}.

Graphically, we write~\eno{FourptContact} as
\eqn{D4schematic}{
\musepic{\figFourContact} \!\!\!\!= \sum_{M_A} P_{\rm cont}^{(12)}(M_A) \left(\musepic{\figFourContCBDa} \right) + \sum_{M_B} P_{\rm cont}^{(45)}(M_B) \left(\musepic{\figFourContCBDb}\right),
}
where the green disc-shaped bulk point is a quartic contact interaction vertex to be integrated over all of AdS, and the diagrams in parentheses denote the four-point conformal blocks in a chosen channel with double-trace primaries  exchanged in the intermediate channels.

We now turn to the four-point exchange diagram, defined as
\eqn{FourptExchDef}{
D_4^{\rm exch} &\equiv \!\!\!\!\!\musepic{\figFourExch}\!\!\!\!\! \equiv \iint_{z,w \in {\rm AdS}}\!\!\! \hat{K}_{\Delta_1}(x_1,z) \hat{K}_{\Delta_2}(x_2,z) \hat{G}_{\Delta_0}(z,w) \hat{K}_{\Delta_4}(x_4,w) \hat{K}_{\Delta_5}(x_5,w) \,,
}
which admits the CBD,
\begingroup
\allowdisplaybreaks
\begin{align*}
\musepic{\figFourExch} &= P_{\rm exch}^{(\Delta_0)}  \left(\musepic{\figFourExchCBD}\right)  + \sum_{M_A} P_{\rm exch}^{(12)}(M_A) \left(\musepic{\figFourContCBDa}\right) \\
& \quad + \sum_{M_B} P_{\rm exch}^{(45)}(M_B) \left(\musepic{\figFourContCBDb}\right), \stepcounter{equation}\tag{\theequation}\label{FourptExch}
\end{align*}
\endgroup
with\footnote{For the same reason as the one mentioned in footnote~\ref{fn:notationtypo}, the expressions for the OPE coefficients~\eno{FourptExchOPE} differ from those quoted in Eqn.~(4.17) of Ref.~\cite{Hijano:2015zsa} by particular normalization factors.}
\eqn{FourptExchOPE}{
P_{\rm exch}^{(\Delta_0)} &=\!\!  \left( B(\Delta_{01,2}, \Delta_{02,1})\sum_{M_A} {1\over N_{\Delta_{A}}}{a_{M_A}^{\Delta_1;\Delta_2} \over m_{\Delta_{0}}^2-m_{\Delta_{A}}^2} \right) \!\!
\left( B(\Delta_{04,5}, \Delta_{05,4})\sum_{M_B} {1\over N_{\Delta_{B}}}{a_{M_B}^{\Delta_4;\Delta_5} \over m_{\Delta_{0}}^2-m_{\Delta_{B}}^2} \right)\cr
P_{\rm exch}^{(12)}(M_A) &= \!\! \left( B(\Delta_{A1,2}, \Delta_{A2,1}) {1\over N_{\Delta_{0}}}{a_{M_A}^{\Delta_1;\Delta_2} \over m_{\Delta_{A}}^2-m_{\Delta_{0}}^2} \right) \!\!
\left( B(\Delta_{A4,5}, \Delta_{A5,4})\sum_{M_B} {1\over N_{\Delta_{B}}}{a_{M_B}^{\Delta_4;\Delta_5} \over m_{\Delta_{A}}^2-m_{\Delta_{B}}^2} \right) \cr
P_{\rm exch}^{(45)}(M_B) &= \!\! \left( B(\Delta_{B4,5}, \Delta_{B5,4}) {1\over N_{\Delta_{0}}}{a_{M_B}^{\Delta_4;\Delta_5} \over m_{\Delta_{B}}^2-m_{\Delta_{0}}^2} \right) \!\!
\left( B(\Delta_{B1,2}, \Delta_{B2,1})\sum_{M_A} {1\over N_{\Delta_{A}}}{a_{M_A}^{\Delta_1;\Delta_2} \over m_{\Delta_{B}}^2-m_{\Delta_{A}}^2} \right).
}
Like in the case of the contact diagram, we can in fact evaluate the infinite sums above using the identity~\eno{OPEsumId} to obtain more compact expressions,
\eqn{FourptExchOPEAgain}{
P_{\rm exch}^{(\Delta_0)} &=  C_{\Delta_1\Delta_2\Delta_0} \: C_{\Delta_4\Delta_5\Delta_0} \cr
P_{\rm exch}^{(12)}(M_A) &=  \left(  {1\over N_{\Delta_{0}}}{\alpha_{M_A}^{\Delta_1;\Delta_2} \over m_{\Delta_{A}}^2-m_{\Delta_{0}}^2} \right) C_{\Delta_4\Delta_5\Delta_{A}}\cr
P_{\rm exch}^{(45)}(M_B) &=  \left(  {1\over N_{\Delta_{0}}}{\alpha_{M_B}^{\Delta_4;\Delta_5} \over m_{\Delta_{B}}^2-m_{\Delta_{0}}^2} \right) C_{\Delta_1\Delta_2\Delta_{B}}\,.
}
Note that the four-point contact and exchange diagram OPE coefficients squared are related via
\eqn{4ptOPErelate}{
{P_{\rm exch}^{(12)}(M_A) \over P_{\rm cont}^{(12)}(M_A)} = {1\over N_{\Delta_{0}}}{1 \over m_{\Delta_{A}}^2-m_{\Delta_{0}}^2} \qquad  \qquad
{P_{\rm exch}^{(45)}(M_B) \over P_{\rm cont}^{(45)}(M_B)} = {1\over N_{\Delta_{0}}}{1 \over m_{\Delta_{B}}^2-m_{\Delta_{0}}^2}\,.
}
The normalization factor $N_{\Delta_0}$ above can be absorbed into the definition of the four-point exchange diagram~\eno{FourptExchDef} by using a different normalization for the bulk-to-bulk propagator in~\eno{GNorm}.

In the remainder of this section, we extend the methods of Ref.~\cite{Hijano:2015zsa} to provide a geodesic diagram derivation of the CBD of higher-point bulk diagrams (specifically, five- and six-point tree-level diagrams) without performing any further bulk or contour integrations.
We end the section with a discussion of the CBD of individual diagrams repackaged into a spectral decomposition, which clarifies relations  of the form~\eno{4ptOPErelate} also obeyed by decomposition coefficients of higher-point AdS diagrams.

\subsection{Five-point contact diagram}
\label{5CONTACT}

In this subsection we will perform the CBD of the five-point contact diagram,
\eqn{5ContactIntegral}{
D_5 \equiv \musepic{\figFiveContact} \equiv \int_{z \in {\rm AdS}} \prod_{i=1}^5 \hat{K}_{\Delta_i}(x_i,z)\,.
}
Like for the decomposition of four-point exchange diagrams~\cite{Hijano:2015zsa}, we begin by applying the holographic OPE~\eno{KKexpFig} at the legs $(x_1,x_2)$ and $(x_4,x_5)$, to obtain
\eqn{D5exp}{
D_5 = 4\sum_{M_A,M_B} a_{M_A}^{\Delta_1;\Delta_2} a_{M_B}^{\Delta_4;\Delta_5} \musepic{\figContactA},
}
where $\Delta_A,\Delta_B$ were defined in~\eno{DeltaABDef}. We can now make use of the three-propagator identity~\eno{KGGreal} to perform this bulk integration. The three terms on the RHS of~\eno{KGGreal} admit a physical interpretation, corresponding individually to the contributions coming from the exchange of a pair of a specific combination of multi-trace primaries and their higher-twist cousins (often referred to as multi-twist operators). Indeed, we will show below that
\begingroup
\allowdisplaybreaks
\begin{align*}
\musepic{\figFiveContact} &= \sum_{M_A,M_B} P_{\rm cont}^{(12,45)}(M_A,M_B) \left(\musepic{\figFiveContCBDa} \right) \cr
    &+ \sum_{M_A,M} P_{\rm cont}^{(12,123)}(M_A,M) \left(\musepic{\figFiveContCBDb}\right) \\
    &+ \sum_{M,M_B} P_{\rm cont}^{(345,45)}(M,M_B) \left(\musepic{\figFiveContCBDc}\right), \stepcounter{equation}\tag{\theequation}\label{D5schematic}
\end{align*}
\endgroup
where the coefficients $P^{(s,t)}_{\rm cont}$, to be determined below, are related to the OPE coefficients of the dual CFT, via
\eqn{D5CBDcoeffs}{
P_{\rm cont}^{(12,45)}(M_A,M_B)  &= \alpha_{M_A}^{\Delta_1;\Delta_2} \: \alpha_{M_B}^{\Delta_4;\Delta_5}\: C_{\Delta_3\Delta_A\Delta_B} \cr
P_{\rm cont}^{(345,45)}(M,M_B) &= \alpha_{M}^{\Delta_3;\Delta_B}\:  \alpha_{M_B}^{\Delta_4;\Delta_5}\:
C_{\Delta_1;\Delta_2;2\Delta_{3B,}+2M} \cr
P_{\rm cont}^{(12,123)}(M_A,M) &= \alpha_{M_A}^{\Delta_1;\Delta_2}\:  \alpha_{M}^{\Delta_3;\Delta_A}\:
C_{\Delta_4;\Delta_5;2\Delta_{3A,}+2M}\,.
}
The diagrams in parentheses above represent  five-point conformal blocks reviewed in section~\ref{SIXOPEGEODESIC}, with multi-twist primaries (more accurately double-twist, or even double-twist of double-twist primaries), color-coded for visual aid, exchanged in the intermediate channels. We are using the notation~\eno{doubleTrace} to denote schematically the multi-twist operators.

We now present the computational details. Explicitly, employing~\eno{KGGreal} to evaluate the bulk integration in~\eno{D5exp}, we find
\eqn{D5KGG}{
D_5 &=   4\sum_{M_A,M_B}  a_{M_A}^{\Delta_1;\Delta_2} a_{M_B}^{\Delta_4;\Delta_5} \Bigg[ C_{\Delta_3\Delta_A\Delta_B} \sum_{k_A,k_B=0}^\infty c_{k_A;k_B}^{\Delta_{A};\Delta_3;\Delta_{B}} \musepic{\figContactB}   \cr
 &\qquad + \bigg( \sum_{k_A,k_B=0}^\infty d_{k_A;k_B}^{\Delta_A;\Delta_3;\Delta_B}   \musepic{\figContactC} + ``(A \leftrightarrow B)" \bigg) \Bigg],
}
where $c_{k_A;k_B}^{\Delta_{A};\Delta_3;\Delta_{B}}$ and $d_{k_A;k_B}^{\Delta_A;\Delta_3;\Delta_B}$ are defined in~\eno{c2Def}-\eno{d2Def}. The last term written as ``$(A \leftrightarrow B)$'' is interpreted in the sense explained below~\eno{KGGreal}.

The first term in~\eno{D5KGG} is already in its desired form. Making use of~\eno{calW5}-\eno{calW5confwave} we recognize it to be the first term in the CBD~\eno{D5schematic}, with the CBD coefficient given by
\eqn{P1245cont}{
P_{\rm cont}^{(12,45)}(M_A,M_B)  = \alpha_{M_A}^{\Delta_1;\Delta_2} \: \alpha_{M_B}^{\Delta_4;\Delta_5}\: C_{\Delta_3\Delta_A\Delta_B} \,,
}
where $\alpha_M^{\Delta,\Delta'}$ was defined in~\eno{alphaDef}.

To bring the second term in~\eno{D5KGG} to a form which makes the contribution from the appropriate conformal block manifest, we make use of the following non-trivial identity proven in appendix~\ref{KEY1PROOF},\footnote{This equation may be thought of as the real analog of the simpler identity \eno{BetaToC} that was used in the parallel $p$-adic computation.}
\eqn{key1}{
\sum_{M_A=0}^\infty
a_{M_A}^{\Delta_1;\Delta_2}
d^{2\Delta_{12,}+2M_A;\Delta_3;\Delta_0}_{k_A;k_B}
=
\sum_{M=0}^\infty
{\:
\alpha_{M}^{\Delta_3;\Delta_0}
C_{\Delta_1\,\Delta_2\,2\Delta_{03,}+2M}\,
\over B(\Delta_{031,2}+M,\Delta_{032,1}+M) }   \: c_{k_{AB,}-M;k_B}^{2\Delta_{03,}+2M;\Delta_3;\Delta_{0}}\,.
}
 Recognizing $2\Delta_{12,}+2M_A=\Delta_A$ and setting $\Delta_0 = \Delta_B = 2\Delta_{45,}+2M_B$, we can re-express the sum over $M_A$ in the second term of~\eno{D5KGG} to obtain the following equivalent form:
\eqn{D5KGG2nd}{
 4 \sum_{M,M_B}    { a_{M_B}^{\Delta_4;\Delta_5}
\alpha_{M}^{\Delta_3;\Delta_B}
C_{\Delta_1\,\Delta_2\,2\Delta_{3B,}+2M}\,
\over B(\Delta_{3B1,2}+M,\Delta_{3B2,1}+M) }     \sum_{k_A,k_B=0}^\infty   c_{k_{AB,}-M;k_B}^{2\Delta_{3B,}+2M;\Delta_3;\Delta_{B}} \musepic{\figContactC}.
 }
The final manipulation we need is to send the variable $k_A \to k_A -k_B+M$, which also changes the lower limit of the $k_A$ summation to $k_B-M$. However, it is clear from the explicit form of $c^{2\Delta_{B3,}+2M;\Delta_3;\Delta_B}_{k_A;k_B}$ (refer to the definition~\eno{c2Def}) that it vanishes for negative integer values of $k_A$, as well as for values of $k_A$ from the set $\{0, 1, \ldots, k_B-M\}$ for $k_B>M$. Hence we can change the lower limit of the sum over the transformed variable $k_A$ back to $0$ without affecting the sum. Then~\eno{D5KGG2nd} becomes
\eqn{D5KGG2ndAgain}{
 4 \sum_{M,M_B}    { a_{M_B}^{\Delta_4;\Delta_5}
\alpha_{M}^{\Delta_3;\Delta_B}
C_{\Delta_1\,\Delta_2\,2\Delta_{3B,}+2M}\,
\over B(\Delta_{3B1,2}+M,\Delta_{3B2,1}+M) }     \sum_{k_A,k_B=0}^\infty   c_{k_{A};k_B}^{2\Delta_{3B,}+2M;\Delta_3;\Delta_{B}}\!\!\!\!\!\!\!\!\!\! \musepic{\figContactCMod}\!\!.
 }
The inner sum over $k_A,k_B$ is immediately recognized, with the help of~\eno{calW5}-\eno{calW5confwave} as the conformal block $W_{2\Delta_{3B,}+2M;\Delta_B}^{\Delta_1,\ldots,\Delta_5}(x_i)$ up to an overall factor. This is precisely the third term of~\eno{D5schematic}, with the CBD coefficient taking the form
\eqn{P34545cont}{
P_{\rm cont}^{(345,45)}(M,M_B) = \alpha_{M}^{\Delta_3;\Delta_B}\:  \alpha_{M_B}^{\Delta_4;\Delta_5}\:
C_{\Delta_1\,\Delta_2\,2\Delta_{3B,}+2M}\,.
}
An analogous analysis for the third term in~\eno{D5KGG} reproduces the second term of the CBD~\eno{D5schematic} with the coefficient
\eqn{P12123cont}{
P_{\rm cont}^{(12,123)}(M_A,M) = \alpha_{M_A}^{\Delta_1;\Delta_2}\:  \alpha_{M}^{\Delta_3;\Delta_A}\:
C_{\Delta_4\,\Delta_5\,2\Delta_{3A,}+2M}\,.
}
This completes the CBD of the five-point contact diagram. The reader may note the similarity with the CBD coefficients of the four-point contact diagram.

Moreover, this five-point decomposition is very similar to the one computed in the $p$-adic setup for the corresponding contact diagram in section~\ref{CBDBTTREE}. At the same time, the functional form of the coefficients is preserved.
 Essentially for any tree-level diagram, the $p$-adic CBDs will turn out to be identical to the real decompositions, except that there will be no descendant contributions (corresponding to triviality of $p$-adic conformal blocks), and only the lowest-twist states will contribute; thus all infinite sums such as those in~\eno{D4schematic},~\eno{FourptExch} and~\eno{D5schematic} will collapse to the leading terms in the sums.

\subsection{Five-point exchange diagrams}
\label{5EXCHANGE}

We now turn to five-point exchange diagrams. Up to a relabelling of external operators, the most general five-point tree-level bulk exchange diagram involving the exchange of a single bulk scalar of conformal dimension $\Delta_0$ is
\eqn{D5OneExch}{
D_5^{\rm 1-exch} \equiv \musepic{\figFiveOneExch}.
}
Here, we show that in the basis of the conformal block of figure~\ref{fig:5block} it admits the CBD,
\eqn{5ExchCBDFinal}{
D_5^{\rm 1-exch} &=
\sum_{M_A,M_B=0}^\infty\!  {P}_{\rm 1-exch}^{(12,45)}(M_A,M_B)  \: W_{\Delta_{A};\Delta_{B}}^{\Delta_1,\ldots,\Delta_5} +
\sum_{M_B=0}^\infty {P}_{\rm 1-exch}^{(0,45)}(M_B) \: W_{\Delta_{0};\Delta_{B}}^{\Delta_1,\ldots,\Delta_5} \cr
& +  \sum_{M, M_A=0}^\infty\! {P}_{\rm 1-exch}^{(12,123)}(M_A,M)   \: W_{\Delta_{A};2\Delta_{3A,}+2M}^{\Delta_1,\ldots,\Delta_5}
+  \sum_{M=0}^\infty {P}_{\rm 1-exch}^{(0,03)}(M)   \: W_{\Delta_{0};2\Delta_{03,}+2M}^{\Delta_1,\ldots,\Delta_5} \cr
& +
\sum_{M, M_B=0}^\infty\!  {P}_{\rm 1-exch}^{(345,45)}(M,M_B)  \: W_{2\Delta_{3B,}+2M;\Delta_{B}}^{\Delta_1,\ldots,\Delta_5}
\,,
}
where the CBD coefficients are given by
\eqn{D1ExchCBDcoeffs}{
 P_{\rm 1-exch}^{(12,45)}(M_A,M_B) &= {P_{\rm cont}^{(12,45)}(M_A,M_B) \over N_{\Delta_0}(m_{\Delta_A}^2-m_{\Delta_0}^2)} \cr
  P_{\rm 1-exch}^{(0,45)}(M_B) &=   \alpha_{M_B}^{\Delta_4;\Delta_5}\: C_{\Delta_3\Delta_0\Delta_B}\: C_{\Delta_1\Delta_2\Delta_0}  \cr
P_{\rm 1-exch}^{(12,123)}(M_A,M) &= {P_{\rm cont}^{(12,123)}(M_A,M) \over N_{\Delta_0}(m_{\Delta_A}^2-m_{\Delta_0}^2)} \cr
 P_{\rm 1-exch}^{(0,03)}(M) &= \alpha_M^{\Delta_0;\Delta_3}\: C_{\Delta_4\,\Delta_5\,2\Delta_{03,}+2M} \: C_{\Delta_1\Delta_2\Delta_0} \cr
P^{(345,45)}_{\rm 1-exch}(M,M_B) &= {P^{(345,45)}_{\rm cont}(M,M_B) \over    N_{\Delta_0} (m_{2\Delta_{3B,}+2M}^2-m_{\Delta_0}^2) }
 \,,
}
with the contact diagram coefficients $P_{\rm cont}^{(s,t)}$ given in~\eno{D5CBDcoeffs}.  Like in the case of the contact diagram, the CBD in~\eno{5ExchCBDFinal} is interpreted in terms of the exchange of multi-twist primaries in the intermediate channels, so that we may rewrite it as
\eqn{D5Exch1schematic}{
& D_5^{\rm 1-exch} \cr
&= \sum_{M_A,M_B} P_{\rm 1-exch}^{(12,45)} \left(\musepic{\figFiveContCBDa} \right)
 + \sum_{M_B} P_{\rm 1-exch}^{(0,45)} \left(\musepic{\figFiveCBDa} \right) \cr
    &+ \sum_{M_A,M} P_{\rm 1-exch}^{(12,123)} \left(\musepic{\figFiveContCBDb}\right)
     + \sum_{M} P_{\rm 1-exch}^{(0,03)} \left(\musepic{\figFiveCBDb}\right) \cr
    &+ \sum_{M,M_B} P^{(345,45)}_{\rm 1-exch} \left( \musepic{\figFiveContCBDc}\right).
    }

 To show the decomposition, we begin with an application of the holographic OPE identity~\eno{KKexpFig} on the pairs of legs $(x_1,x_2)$ and $(x_4,x_5)$, followed by an application of the propagator identity~\eno{GGidFig} to evaluate one of the two full AdS integrations, to obtain
\eqn{D5OneExchExp}{
D_5^{\rm 1-exch} = 4 \sum_{M_A,M_B} {a_{M_A}^{\Delta_1;\Delta_2} a_{M_B}^{\Delta_4;\Delta_5} \over m_{\Delta_A}^2-m_{\Delta_0}^2} \left( {1\over N_{\Delta_0}} \musepic{\figContactA} -  {1\over N_{\Delta_A}} \musepic{\figContactO} \right).
}
At this point we can use the propagator identity~\eno{KGGreal} to trade the remaining cubic AdS integration for an expression involving sums over specific products of bulk-to-bulk and bulk-to-boundary propagators:
\begingroup
\allowdisplaybreaks
\begin{align*}
&
D_5^{\rm 1-exch} = 4 \sum_{M_A,M_B} {a_{M_A}^{\Delta_1;\Delta_2} a_{M_B}^{\Delta_4;\Delta_5} \over m_{\Delta_A}^2-m_{\Delta_0}^2} \sum_{k_A,k_B=0}^\infty\left( \frac{C_{\Delta_A\Delta_B\Delta_3}}{N_{\Delta_0}}
c_{k_A;k_B}^{\Delta_A;\Delta_3;\Delta_B}
\musepic{\figContactB} \right. \\
&
+\frac{d_{k_A;k_B}^{\Delta_B;\Delta_3;\Delta_A}}{N_{\Delta_0}}
\musepic{\figContactE}
 +
\frac{d^{\Delta_A;\Delta_3;\Delta_B}_{k_A;k_B}}{N_{\Delta_0}}
\musepic{\figContactC}
\\
&-\frac{C_{\Delta_0\Delta_B\Delta_3}}{N_{\Delta_A}}
c_{k_A;k_B}^{\Delta_0;\Delta_3;\Delta_B}
\musepic{\figContactBtilde}
-\frac{d_{k_A;k_B}^{\Delta_B;\Delta_3;\Delta_0}
}{N_{\Delta_A}}
\musepic{\figContactF}
 \\
&\left.  -\frac{d^{\Delta_0;\Delta_3;\Delta_B}_{k_A;k_B}}{N_{\Delta_A}}
\musepic{\figContactC}\right).  \stepcounter{equation}\tag{\theequation}\label{1ExAllTerms}
\end{align*}
\endgroup
The first and fourth terms in this equation are already written directly as a sum over conformal blocks  $W_{\Delta_{A};\Delta_{B}}^{\Delta_1,\ldots,\Delta_5}$ and $W_{\Delta_{0};\Delta_{B}}^{\Delta_1,\ldots,\Delta_5}$ and their higher-twist analogs,  so we can directly read off the CBD coefficients in~\eno{5ExchCBDFinal} using~\eno{calW5}-\eno{calW5confwave}:
\eqn{}
{
 P_{\rm 1-exch}^{(12,45)}(M_A,M_B) &= {\alpha_{M_A}^{\Delta_1;\Delta_2}\: \alpha_{M_B}^{\Delta_4;\Delta_5} \over N_{\Delta_0}(m_{\Delta_A}^2-m_{\Delta_0}^2)}\:C_{\Delta_A\Delta_B\Delta_3} \cr
  P_{\rm 1-exch}^{(0,45)}(M_B) &=   \alpha_{M_B}^{\Delta_4;\Delta_5} C_{\Delta_0\Delta_B\Delta_3}
  \sum_{M_A=0}^\infty
  \frac{B(\Delta_{01,2},\Delta_{02,1})a_{M_A}^{\Delta_1;\Delta_2}}{N_{\Delta_A}(m^2_{\Delta_0}-m^2_{\Delta_A})} \cr
  &= \alpha_{M_B}^{\Delta_4;\Delta_5}\: C_{\Delta_0\Delta_B\Delta_3}\:C_{\Delta_0\Delta_1\Delta_2}
  \,,
}
where we made use of \eno{OPEsumId} in the last equation. This reproduces the first two terms of \eno{5ExchCBDFinal}. As for the third and fourth terms of \eno{5ExchCBDFinal}, these are equal to the second and fifth terms in \eno{1ExAllTerms}, respectively. This can be seen by changing the order of summation in \eno{1ExAllTerms}, applying equation \eno{key1}, and performing a change of summation variables like the one  used in going from~\eno{D5KGG2nd} to~\eno{D5KGG2ndAgain}, to explicitly rewrite the second and fifth terms as sums over conformal blocks $W_{\Delta_a;2\Delta_{3A,}+2M}^{\Delta_1,...,\Delta_5}$ and $W_{\Delta_0;2\Delta_{03,}+2M}^{\Delta_1,...,\Delta_5}$ respectively, with CBD coefficients
\eqn{}
{
P_{\rm 1-exch}^{(12,123)}(M_A,M) &=
\frac{\alpha_{M_A}^{\Delta_1;\Delta_2}\:\alpha_{M}^{\Delta_A;\Delta_3}}{N_{\Delta_0}(m_{\Delta_A}^2-m_{\Delta_0}^2)}\:C_{\Delta_4\,\Delta_5\,2\Delta_{A3,}+2M}
\cr
P_{\rm 1-exch}^{(0,03)}(M) &=\alpha_{M}^{\Delta_0;\Delta_3}C_{\Delta_4\,\Delta_5\,2\Delta_{03,}+2M}\sum_{M_A}\frac{B(\Delta_{01,2},\Delta_{02,1})a_{M_A}^{\Delta_1;\Delta_2}}{N_{\Delta_A}(m_{\Delta_0}^2-m_{\Delta_A}^2)} \cr
&=
\alpha_{M}^{\Delta_0;\Delta_3}\:C_{\Delta_4\,\Delta_5\,2\Delta_{03,}+2M}\:C_{\Delta_0\Delta_1\Delta_2} \,.
}
The remaining two terms in~\eno{1ExAllTerms} combine since they are proportional to the same geodesic bulk diagram. Together, they can be recast as a weighted sum over conformal blocks. To do that, we need to use  a variant of identity~\eno{key1}, proven in appendix~\ref{KEY2PROOF}, namely
\footnote{This equation is a special case of the identity \eno{key4} that we use below to work out the CBD of six-point exchange diagrams. These two equations can be thought of as analogs of the simpler identity \eno{Elephant} used in the parallel $p$-adic computations.}
\eqn{key2}{
&
\sum_{M_A=0}^\infty
\frac{a_{M_A}^{\Delta_1;\Delta_2}}{m_{\Delta_{1}+\Delta_2+2M_A}^2-m_{\Delta_0}^2}
\bigg[
\frac{d_{k_A;k_B}^{\Delta_1+\Delta_2+2M_A;\Delta_3;\Delta_B}}
{N_{\Delta_0}}
-
\frac{d_{k_A;k_B}^{\Delta_0;\Delta_3;\Delta_B}}
{N_{\Delta_{1}+\Delta_2+2M_A}}
\bigg]
\cr
&
=
\sum_{M=0}^\infty
\frac{1}{N_{\Delta_0}}
\frac{
C_{\Delta_1\,\Delta_2\,2\Delta_{3B,}+2M}
}{m_{2\Delta_{3B,}+2M}^2-m_{\Delta_0}^2}
\frac{\alpha_M^{\Delta_{3};\Delta_{B}}}{B(\Delta_{3B2,1}+M,\Delta_{3B1,2}+M)}
c_{k_{AB,}-M;k_B}^{2\Delta_{3B,}+2M;\Delta_3;\Delta_B}\,.
}
Changing the order of summation between $M_A$ and $k_A,k_B$, applying \eno{key2} to the sum of third and sixth terms in \eno{1ExAllTerms}, and performing a change of variables like the one between~\eno{D5KGG2nd} and~\eno{D5KGG2ndAgain} turns these terms into a sum over conformal blocks $W_{2\Delta_{3B,}+2M;\Delta_B}^{\Delta_1,...,\Delta_5}$, with coefficients given by
\eqn{}
{
P_{\rm 1-exch}^{(345,45)}(M,M_B)=\alpha_{M_B}^{\Delta_4;\Delta_5}
\frac{1}{N_{\Delta_0}}
\frac{
C_{\Delta_1\,\Delta_2\,2\Delta_{3B,}+2M}
}{m_{2\Delta_{3B,}+2M}^2-m_{\Delta_0}^2}
\alpha_M^{\Delta_{3};\Delta_{B}}   \,.
}
Thus the third and sixth terms in \eno{1ExAllTerms} together reproduce the last term in \eno{5ExchCBDFinal}.

Like in the case of the contact diagram, the coefficients~\eno{D1ExchCBDcoeffs} share strong structural similarities with the four-point exchange coefficients~\eno{FourptExchOPEAgain}; importantly the five-point exchange coefficients corresponding to the exchange of multi-twist primaries admit very simple relations with the five-point contact coefficients. Indeed  in~\eno{D1ExchCBDcoeffs}, we have written some of the CBD coefficients directly in terms of the five-point contact diagram CBD coefficients to highlight the simple algebraic relation between the two. These relations take the same form as the ones for four-point coefficients in~\eno{4ptOPErelate}.

\vspace{.5em}
Finally, turning to the five-point exchange diagram involving three cubic contact interaction vertices, its CBD can be obtained using the same method, tools and techniques described above; in particular, the manipulations~\eno{key1} and~\eno{key2} are used again, and no further identities are necessary.
Thus we omit the long but un-illuminating computational details, and only show the final result for the direct channel decomposition:
\begingroup
\allowdisplaybreaks
\begin{align*}
& \musepic{\figFiveTwoExch}
\\ &= P_{\rm 2-exch}^{(0,0^\prime)}\! \left(\!\musepic{\figFiveCBD}\!\right)\! + \!\sum_{M_A,M_B}\!\! P_{\rm 2-exch}^{(12,45)} \!\left(\!\musepic{\figFiveContCBDa} \!\right) \\
&+ \sum_{M_B} P_{\rm 2-exch}^{(0,45)} \left(\musepic{\figFiveCBDa} \right) + \sum_{M} P_{\rm 2-exch}^{(0,03)} \left(\musepic{\figFiveCBDb}\right)  \\
&+  \sum_{M_A} P_{\rm 2-exch}^{(12,0^\prime)} \left(\musepic{\figFiveCBDc} \right) + \sum_{M} P_{\rm 2-exch}^{(0^\prime 3,0^\prime)} \left(\musepic{\figFiveCBDd}\right)  \\
&+\!\!  \sum_{M_A,M}\!\!\! P_{\rm 2-exch}^{(12,123)} \left(\!\!\musepic{\figFiveContCBDb}\!\!\right)\!\! +\!\! \sum_{M,M_B}\!\!\! P_{\rm 2-exch}^{(345,45)} \left(\!\!\musepic{\figFiveContCBDc}\!\!\right) \stepcounter{equation}\tag{\theequation}\label{5TwoExchCBD}
\end{align*}
\endgroup
with the coefficients (whose arguments we suppressed above) given by
\eqn{Fivept2ExchOPE}{
{P}_{\rm 2-exch}^{(0,0^\prime)} =  C_{\Delta_1 \Delta_2 \Delta_0}\:  C_{\Delta_0 \Delta_3 \Delta_{0^\prime}}\:  C_{ \Delta_4 \Delta_5 \Delta_{0^\prime}} \qquad
{P}_{\rm 2-exch}^{(12,45)}(M_A,M_B)  =  { {P}_{\rm 1-exch}^{(12,45)}(M_A,M_B) \over  N_{\Delta_{0^\prime}} (m^2_{\Delta_{B}} - m_{\Delta_{0^\prime}}^2)}  \cr
{P}_{\rm 2-exch}^{(0,45)}(M_B)  =   {P_{\rm 1-exch}^{(0,45)}(M_B) \over  N_{\Delta_{0^\prime}}(m^2_{\Delta_{B}} - m_{\Delta_{0^\prime}}^2)} \qquad
{P}_{\rm 2-exch}^{(0,03)}(M) =   { P_{\rm 1-exch}^{(0,03)}(M) \over  N_{\Delta_{0^\prime}}(m^2_{2\Delta_{03,}+2M} - m_{\Delta_{0^\prime}}^2)}  \cr
{P}_{\rm 2-exch}^{(12,123)}(M_A,M) =   { {P}_{\rm 1-exch}^{(12,123)}(M_A,M) \over N_{\Delta_{0^\prime}}(m^2_{2\Delta_{3A,}+2M} - m_{\Delta_{0^\prime}}^2)} \qquad
{P}_{\rm 2-exch}^{(345,45)}(M,M_B) =   {{P}_{\rm 1-exch}^{(345,45)}(M,M_B) \over N_{\Delta_{0^\prime}}(m^2_{\Delta_{B}} - m_{\Delta_{0^\prime}}^2)}
 \,,
 }
and the remaining two coefficients are obtained by symmetry (or equivalently in terms of the CBD coefficients of an asymmetric scalar exchange diagram obtained from~\eno{D5OneExch} with a simple relabelling):
\eqn{FiveSymOPE1}{
P_{\rm 2-exch}^{(12,0^\prime)}(M_A) &=  {P_{\rm 1-exch}^{(12,0^\prime)}(M_A) \over  N_{\Delta_{0}}(m^2_{\Delta_{A}} - m_{\Delta_{0}}^2)} \cr
P_{\rm 2-exch}^{(0^\prime 3,0^\prime)}(M) &=  { P_{\rm 1-exch}^{(0^\prime 3,0^\prime)}(M) \over  N_{\Delta_{0^\prime}} (m^2_{2\Delta_{0^\prime 3,}+2M} - m_{\Delta_{0}}^2) }
\,,
}
where,  analogous to $P^{(0,45)}_{\rm 1-exch}$ and $P^{(0,03)}_{\rm 1-exch}$ in~\eno{D1ExchCBDcoeffs} for the exchange diagram~\eno{D5OneExch}, we have defined
\eqn{FiveSymOPE2}{
P_{\rm 1-exch}^{(12,0^\prime)}(M_A) &=    \alpha_{M_A}^{\Delta_1;\Delta_2}\: C_{\Delta_3\Delta_{0^\prime} \Delta_A}\: C_{\Delta_4\Delta_5\Delta_{0^\prime}}  \cr
P_{\rm 1-exch}^{(0^\prime 3,0^\prime)}(M) &=   \alpha_M^{\Delta_{0^\prime};\Delta_3}\: C_{\Delta_1\,\Delta_2\,2\Delta_{0^\prime 3,}+2M} \: C_{\Delta_4\Delta_5\Delta_{0^\prime}}
}
for the corresponding exchange diagram obtained from~\eno{D5OneExch} after relabelling.
Finally, we note that the coefficients corresponding to exchanges of solely multi-twist primaries further admit an expression in terms of the contact diagram coefficients:
\eqn{Fivept2ExchOPEAgain}{
{P}_{\rm 2-exch}^{(12,45)}(M_A,M_B)  &= {P_{\rm cont}^{(12,45)}(M_A,M_B) \over N_{\Delta_0}(m_{\Delta_A}^2-m_{\Delta_0}^2) N_{\Delta_{0^\prime}} (m^2_{\Delta_{B}} - m_{\Delta_{0^\prime}}^2)} \cr
{P}_{\rm 2-exch}^{(12,123)}(M_A,M) &=  { {P}_{\rm cont}^{(12,123)}(M_A,M) \over N_{\Delta_0}(m_{\Delta_A}^2-m_{\Delta_0}^2) N_{\Delta_{0^\prime}}(m^2_{2\Delta_{3A,}+2M} - m_{\Delta_{0^\prime}}^2)} \cr
{P}_{\rm 2-exch}^{(345,45)}(M,M_B) &=  {P^{(345,45)}_{\rm cont}(M,M_B) \over    N_{\Delta_0} (m_{2\Delta_{3B,}+2M}^2-m_{\Delta_0}^2)  N_{\Delta_{0^\prime}}(m^2_{\Delta_{B}} - m_{\Delta_{0^\prime}}^2)}\,.
}

\subsection{Six-point diagrams}
\label{SIXCBD}

\begin{figure}[!t]
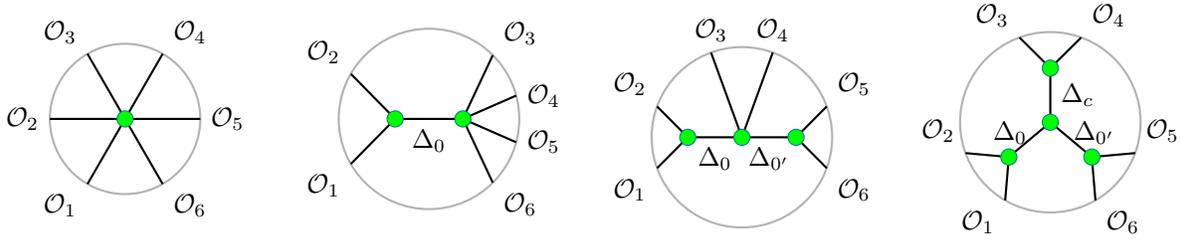

    \begin{subfigure}[]{0.24\textwidth}
    \centering
    \[ \musepic{\figSixContact} \]
    \end{subfigure}
    \begin{subfigure}[]{0.24\textwidth}
    \centering
    \[ \musepic{\figSixExchOneA} \]
    \end{subfigure}
    \begin{subfigure}[]{0.24\textwidth}
    \centering
    \[ \musepic{\figSixExchTwoA} \]
    \end{subfigure}
    \begin{subfigure}[]{0.24\textwidth}
    \centering
    \[ \musepic{\figSixExchThreeA} \]
    \end{subfigure}
    \caption{Inequivalent (up to relabelling) six-point tree-level diagrams which admit a direct channel CBD in the conformal basis of the OPE channel six-point block of section~\ref{SIXOPEGEODESIC}.}
    \label{fig:6OPEdiags}
\end{figure}

\begin{figure}[!t]
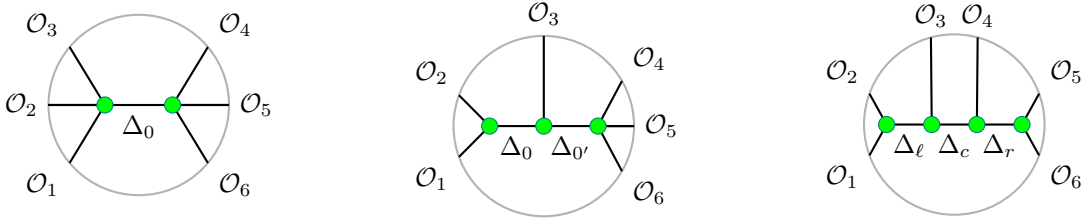

    \begin{subfigure}[]{0.32\textwidth}
    \centering
    \[ \musepic{\figSixExchOneB} \]
    \end{subfigure}
    \begin{subfigure}[]{0.32\textwidth}
    \centering
    \[ \musepic{\figSixExchTwoB} \]
    \end{subfigure}
    \begin{subfigure}[]{0.32\textwidth}
    \centering
    \[ \musepic{\figSixExchThreeB} \]
    \end{subfigure}
    \caption{Inequivalent six-point tree-level diagrams which \emph{do not} admit a direct channel CBD in the conformal basis of the OPE channel six-point block (instead, they \emph{do} so in the comb channel).}
    \label{fig:6combDiags}
\end{figure}

In this section we present the CBD of various tree-level six-point bulk diagrams obtained using geodesic diagram techniques. Figures~\ref{fig:6OPEdiags}-\ref{fig:6combDiags}  catalogue exhaustively all inequivalent (up to relabelling of conformal dimensions and insertion points) tree-level six-point diagrams. In this section we focus on the diagrams shown in figure~\ref{fig:6OPEdiags} since their direct channel CBD can be done in the basis of the OPE channel conformal block discussed in section~\ref{SIXOPEGEODESIC}. The direct channel decomposition of the six-point diagrams of figure~\ref{fig:6combDiags} is expressed in terms of the six-point conformal block in the \emph{comb} channel, for which an explicit form  is currently only known in one and two spacetime dimensions~\cite{Rosenhaus:2018zqn}; the holographic representation is also currently unknown. Thus we will not discuss these diagrams further.\footnote{These diagrams are expected to admit a crossed-channel decomposition in the OPE channel discussed previously, but we will not discuss this point further in this paper.}

The procedure to obtain the CBD for the diagrams in figure~\ref{fig:6OPEdiags} is identical to the one explained in the previous subsection (and the first few steps were already alluded to in section~\ref{SIXOPEGEODESIC}), except we  will require  variants of hypergeometric identities~\eno{key1} and~\eno{key2}, this time involving the expansion coefficients which appear in the propagator identity~\eno{GGGreal}.
Since the computations are fairly straightforward and have been explained in detail before, we refrain from presenting the intermediate steps, but list the new ingredients, in the form of the hypergeometric identities mentioned above.
The key identity required for the CBD of the six-point contact diagram is:
\eqn{key6pt}
{\sum_{M=0}^\infty a_{M}^{\Delta_1;\Delta_2}
d_{k_a;k_b;k_c}^{2\Delta_{12,}+2M;\:\Delta_b;\:\Delta_c}
=
\sum_{M=0}^\infty \frac{
\alpha_M^{\Delta_b;\Delta_c}C_{\Delta_1\,\Delta_2\,2\Delta_{bc,}+2M}
}
{B(\Delta_{bc1,2}+M,\Delta_{bc2,1}+M)}
c_{k_{abc,}-M;\:k_c;\:k_c}^{2\Delta_{bc,}+2M;\:\Delta_b;\:\Delta_c}\,,
}
which is proven in appendix~\ref{KEY3PROOF}.
This leads to the  CBD of the six-point contact diagram in the OPE channel:
\eqn{6ContactCBD}{
& \musepic{\figSixContact} \cr
&= \sum_{M_L,M_C,M_R}\!\!\! {P}_{\rm cont}^{(12,34,56)}\!\! \left( \musepic{\figSixOPECBDa} \right)\!  + \sum_{M_L, M_C, M}\!\!\! {P}_{\rm cont}^{(12,34,1234)}\!\! \left( \musepic{\figSixOPECBDb} \right) \cr
& + \sum_{M_L,M, M_R}\!\!\! {P}_{\rm cont}^{(12,1256,56)}\!\! \left( \musepic{\figSixOPECBDc} \right)\!  + \sum_{M, M_C, M_R}\!\!\! {P}_{\rm cont}^{(3456,34,56)}\!\! \left( \musepic{\figSixOPECBDd} \right) ,
}
where for brevity we have defined the following double-trace higher-twist (i.e.\ double-twist) primaries
\eqn{OLCR}{
{\cal O}_L \equiv [{\cal O}_1 {\cal O}_2]_{M_L}  \qquad {\cal O}_C \equiv [{\cal O}_3 {\cal O}_4]_{M_C} \qquad {\cal O}_R \equiv [{\cal O}_5 {\cal O}_6]_{M_R}\,,
}
with the CBD coefficients
\eqn{6ContactOPE}{
{P}_{\rm cont}^{(12,34,56)}(M_L,M_C,M_R)    &=  \alpha_{M_L}^{\Delta_1;\Delta_2}\:  \alpha_{M_C}^{\Delta_3;\Delta_4} \: \alpha_{M_R}^{\Delta_5;\Delta_6}\:  C_{\Delta_L \Delta_C \Delta_R} \cr
{P}_{\rm cont}^{(12,34,1234)}(M_L,M_C,M)    &=  \alpha_{M_L}^{\Delta_1;\Delta_2}\:  \alpha_{M_C}^{\Delta_3;\Delta_4} \: \alpha_{M}^{\Delta_L;\Delta_C}\:  C_{\Delta_5\, \Delta_6\,2\Delta_{LC,}+2M} \cr
{P}_{\rm cont}^{(12,1256,56)}(M_L,M,M_R)    &=  \alpha_{M_L}^{\Delta_1;\Delta_2}\:  \alpha_{M}^{\Delta_L;\Delta_R} \: \alpha_{M_R}^{\Delta_5;\Delta_6}\:  C_{\Delta_3\, \Delta_4\,2\Delta_{LR,}+2M} \cr
{P}_{\rm cont}^{(3456,34,56)}(M,M_C,M_R)    &=  \alpha_{M}^{\Delta_C;\Delta_R}\:  \alpha_{M_C}^{\Delta_3;\Delta_4} \: \alpha_{M_R}^{\Delta_5;\Delta_6}\:  C_{\Delta_1\, \Delta_2\,2\Delta_{CR,}+2M} \:,
}
where $\Delta_L, \Delta_C$ and $\Delta_R$ were defined in~\eno{DeltaLCRDef}. To guide the eye, we have color-coded the internal lines and the primaries being exchanged in the intermediate channels in~\eno{6ContactCBD}. It is worth pointing out that the CBD of the six-point diagram continues to show strong structural similarities with its four- and five-point contact diagram cousins presented in~\eno{FourptOPEAgain},~\eno{D4schematic} and~\eno{D5schematic}-\eno{D5CBDcoeffs}.

For decomposing the remaining exchange diagrams in figure~\ref{fig:6OPEdiags}, we need the following hypergeometric identity, proven in appendix~\ref{KEY4PROOF}:
\eqn{key4}
{
&
\sum_{M=0}^\infty
\frac{a_{M}^{\Delta_1;\Delta_2}}{m_{2\Delta_{12,}+2M}^2-m_{\Delta_0}^2}
\bigg[
\frac{d_{k_a;k_b;k_c}^{\Delta_1+\Delta_2+2M;\Delta_b;\Delta_c}}
{N_{\Delta_0}}
-
\frac{d_{k_a;k_b;k_c}^{\Delta_0;\Delta_b;\Delta_c}}
{N_{\Delta_{2\Delta_{12,}+2M}}}
\bigg]
\cr
=
\sum_{M=0}^\infty
\frac{1}{N_{\Delta_0}}
&
\frac{
C_{\Delta_1\:\Delta_2\:2\Delta_{bc,}+2M}
}{m_{2\Delta_{bc,}+2M}^2-m_{\Delta_0}^2}
\frac{\alpha_M^{\Delta_{b};\Delta_{c}}}{B(\Delta_{bc2,1}+M,\Delta_{bc1,2}+M)}
c_{k_{abc,}-M;\:k_b;\:k_c}^{2\Delta_{bc,}+2M;\:\Delta_b;\:\Delta_c}\,.
}

Then the exchange diagram built from a cubic and a quintic bulk contact interaction vertex is found to decompose as
\begingroup
\allowdisplaybreaks
\begin{align*}
& \musepic{\figSixExchOneA} = \sum_{M_C,M_R}\! {P}_{\rm 1-exch}^{(0,34,56)}\! \left( \musepic{\figSixOPECBDe} \right) \\
&+ \sum_{M_C, M}\!\! {P}_{\rm 1-exch}^{(0,34,034)}\! \left( \musepic{\figSixOPECBDf} \right) +  \sum_{M, M_R}\!\! {P}_{\rm 1-exch}^{(0,056,56)}\! \left( \musepic{\figSixOPECBDg} \right) \\
&+ \sum_{M_L,M_C,M_R}\!\!\! {P}_{\rm 1-exch}^{(12,34,56)}\!\! \left( \musepic{\figSixOPECBDa} \right)\!  + \sum_{M_L, M_C, M}\!\!\! {P}_{\rm 1-exch}^{(12,34,1234)}\!\! \left( \musepic{\figSixOPECBDb} \right) \\
& + \sum_{M_L,M, M_R}\!\!\! {P}_{\rm 1-exch}^{(12,1256,56)}\!\! \left( \musepic{\figSixOPECBDc} \right)\!  + \sum_{M, M_C, M_R}\!\!\! {P}_{\rm 1-exch}^{(3456,34,56)}\!\! \left( \musepic{\figSixOPECBDd} \right) , \stepcounter{equation}\tag{\theequation}\label{6Exch1CBD}
\end{align*}
\endgroup
where
\eqn{6Exch1OPEa}{
{P}_{\rm 1-exch}^{(0,34,56)}(M_C,M_R) &= \alpha_{M_C}^{\Delta_3;\Delta_4}\: \alpha_{M_R}^{\Delta_5;\Delta_6}\: C_{\Delta_1 \Delta_2 \Delta_0} \: C_{\Delta_0 \Delta_C \Delta_R} \cr
{P}_{\rm 1-exch}^{(0,34,034)}(M_C,M) &= \alpha_{M_C}^{\Delta_3;\Delta_4}\: \alpha_{M}^{\Delta_0;\Delta_C}\: C_{\Delta_1 \Delta_2 \Delta_0} \: C_{\Delta_5\: \Delta_6\: 2\Delta_{0C,}+2M} \cr
{P}_{\rm 1-exch}^{(0,056,56)}(M,M_R) &= \alpha_{M}^{\Delta_0;\Delta_R}\: \alpha_{M_R}^{\Delta_5;\Delta_6}\: C_{\Delta_1 \Delta_2 \Delta_0} \: C_{\Delta_3\: \Delta_4\: 2\Delta_{0R,}+2M} \,,
}
and
\eqn{6Exch1OPEb}{
{P}_{\rm 1-exch}^{(12,34,56)}(M_L,M_C,M_R)    &=  {{P}_{\rm cont}^{(12,34,56)}(M_L,M_C,M_R) \over N_{\Delta_0} (m_{\Delta_L}^2-m_{\Delta_0}^2)} \cr
{P}_{\rm 1-exch}^{(12,34,1234)}(M_L,M_C,M)    &= {{P}_{\rm cont}^{(12,34,1234)}(M_L,M_C,M)  \over N_{\Delta_0} (m_{\Delta_L}^2-m_{\Delta_0}^2)}  \cr
{P}_{\rm 1-exch}^{(12,1256,56)}(M_L,M,M_R)    &= {{P}_{\rm cont}^{(12,1256,56)}(M_L,M,M_R)  \over N_{\Delta_0} (m_{\Delta_L}^2-m_{\Delta_0}^2)} \cr
{P}_{\rm 1-exch}^{(3456,34,56)}(M,M_C,M_R)    &=  {{P}_{\rm cont}^{(3456,34,56)}(M,M_C,M_R) \over  N_{\Delta_0} (m_{2\Delta_{RC,}+2M}^2-m_{\Delta_0}^2)} \,.
}

Moreover, the six-point exchange diagram in figure~\ref{fig:6OPEdiags} with three bulk interaction vertices is decomposed as
\begingroup
\allowdisplaybreaks
\begin{align*}
& \musepic{\figSixExchTwoA} \\
&= \sum_{M_C}\! {P}_{\rm 2-exch}^{(0,34,0^\prime)}\! \left( \musepic{\figSixOPECBDk} \right)\! + \sum_{M}\! {P}_{\rm 2-exch}^{(0,00^\prime,0^\prime)}\! \left( \musepic{\figSixOPECBDl} \right) \\
&+ \sum_{M_C,M_R}\! {P}_{\rm 2-exch}^{(0,34,56)}\! \left( \musepic{\figSixOPECBDe} \right)\! + \sum_{M_L,M_C}\! {P}_{\rm 2-exch}^{(12,34,0^\prime)}\! \left( \musepic{\figSixOPECBDh} \right) \\
&+ \sum_{M_C, M}\!\! {P}_{\rm 2-exch}^{(0,34,034)}\! \left( \musepic{\figSixOPECBDf} \right) +  \sum_{M, M_R}\!\! {P}_{\rm 2-exch}^{(0,056,56)}\! \left( \musepic{\figSixOPECBDg} \right) \\
&+ \sum_{M_C, M}\!\! {P}_{\rm 2-exch}^{(0^\prime34,34,0^\prime)}\! \left( \musepic{\figSixOPECBDi} \right) +  \sum_{M,M_L}\!\! {P}_{\rm 2-exch}^{(12,0^\prime12,0^\prime)}\! \left( \musepic{\figSixOPECBDj} \right) \\
&+ \sum_{M_L,M_C,M_R}\!\!\! {P}_{\rm 2-exch}^{(12,34,56)}\!\! \left( \musepic{\figSixOPECBDa} \right)\!  + \sum_{M_L, M_C, M}\!\!\! {P}_{\rm 2-exch}^{(12,34,1234)}\!\! \left( \musepic{\figSixOPECBDb} \right) \\
& + \sum_{M_L,M, M_R}\!\!\! {P}_{\rm 2-exch}^{(12,1256,56)}\!\! \left( \musepic{\figSixOPECBDc} \right)\!  + \sum_{M, M_C, M_R}\!\!\! {P}_{\rm 2-exch}^{(3456,34,56)}\!\! \left( \musepic{\figSixOPECBDd} \right), \stepcounter{equation}\tag{\theequation}\label{6Exch2CBD}
\end{align*}
\endgroup
with
\eqn{6Exch2OPEa}{
{P}_{\rm 2-exch}^{(0,34,0^\prime)}(M_C) &= \alpha_{M_C}^{\Delta_3;\Delta_4} \: C_{\Delta_1\Delta_2\Delta_0} \: C_{\Delta_0 \Delta_C \Delta_{0^\prime}}\: C_{\Delta_{0^\prime}\Delta_5\Delta_6} \cr
{P}_{\rm 2-exch}^{(0,00^\prime,0^\prime)}(M) &= \alpha_{M}^{\Delta_0;\Delta_{0^\prime}}\:  C_{\Delta_1\Delta_2\Delta_0} \: C_{\Delta_3\: \Delta_4\: 2\Delta_{00^\prime,}+2M}\: C_{\Delta_{0^\prime}\Delta_5\Delta_6}\,,
}
\eqn{6Exch2OPEb}{
{P}_{\rm 2-exch}^{(0,34,56)}(M_C,M_R) &= { {P}_{\rm 1-exch}^{(0,34,56)}(M_C,M_R) \over N_{\Delta_{0^\prime}} (m_{\Delta_R}^2-m_{\Delta_{0^\prime}}^2)}\cr
{P}_{\rm 2-exch}^{(0,34,034)}(M_C,M) &={ {P}_{\rm 1-exch}^{(0,34,034)}(M_C,M) \over N_{\Delta_{0^\prime}} (m_{2\Delta_{0C,}+2M}^2-m_{\Delta_{0^\prime}}^2)} \cr
{P}_{\rm 2-exch}^{(0,056,56)}(M,M_R) &= { {P}_{\rm 1-exch}^{(0,056,56)}(M,M_R) \over N_{\Delta_{0^\prime}} (m_{\Delta_R}^2-m_{\Delta_{0^\prime}}^2)} \,,
}
and
\eqn{6Exch2OPEc}{
{P}_{\rm 2-exch}^{(12,34,56)}(M_L,M_C,M_R)    &=  {{P}_{\rm cont}^{(12,34,56)}(M_L,M_C,M_R) \over N_{\Delta_0} (m_{\Delta_L}^2-m_{\Delta_0}^2) N_{\Delta_{0^\prime}} (m_{\Delta_R}^2-m_{\Delta_{0^\prime}}^2)} \cr
{P}_{\rm 2-exch}^{(12,34,1234)}(M_L,M_C,M)    &= {{P}_{\rm cont}^{(12,34,1234)}(M_L,M_C,M)  \over N_{\Delta_0} (m_{\Delta_L}^2-m_{\Delta_0}^2) N_{\Delta_{0^\prime}} (m_{2 \Delta_{LC,}+2M}^2-m_{\Delta_{0^\prime}}^2)}  \cr
{P}_{\rm 2-exch}^{(12,1256,56)}(M_L,M,M_R)    &= {{P}_{\rm cont}^{(12,1256,56)}(M_L,M,M_R)  \over N_{\Delta_0} (m_{\Delta_L}^2-m_{\Delta_0}^2) N_{\Delta_{0^\prime}} (m_{\Delta_R}^2-m_{\Delta_{0^\prime}}^2)} \cr
{P}_{\rm 2-exch}^{(3456,34,56)}(M,M_C,M_R)    &=  {{P}_{\rm cont}^{(3456,34,56)}(M,M_C,M_R) \over  N_{\Delta_0} (m_{2\Delta_{RC,}+2M}^2-m_{\Delta_0}^2) N_{\Delta_{0^\prime}} (m_{\Delta_R}^2-m_{\Delta_{0^\prime}}^2)} \,.
}
The remaining three CBD coefficients are obtained from~\eno{6Exch2OPEb} by permuting the labels, as demonstrated earlier in a five-point example for the CBD coefficients~\eno{FiveSymOPE1}-\eno{FiveSymOPE2}.

Finally, the six-point diagram in figure~\ref{fig:6OPEdiags} with four cubic interaction vertices admits a direct channel decomposition in a basis of twenty six-point conformal blocks in the OPE channel. Twelve of the blocks are the ones which already appeared in the CBD shown in~\eno{6Exch2CBD}, while the remaining eight are the ones in~\eno{6Exch2CBD}  where the double-trace operator  ${\cal O}_C$ appearing in any of the intermediate channels is replaced with the single-trace primary ${\cal O}_c$.
Displaying only the terms not related to each other by a simple relabelling of indices, the CBD can be expressed as
\begingroup
\allowdisplaybreaks
\begin{align*}
&\musepic{\figSixExchThreeA} \\
&=  P_{\rm 3-exch}^{(0,c,0^\prime)} \left( \musepic{\figSixOPECBDm} \right)
+ \sum_{M_L,M_C,M_R}\!\!\! {P}_{\rm 3-exch}^{(12,34,56)}\!\! \left( \musepic{\figSixOPECBDa} \right)
\\ &
+  \sum_{M_R} {P}_{\rm 3-exch}^{(0,c,56)} \left( \musepic{\figSixOPECBDn} \right)
+ \text{(2 terms obtained by permuting indices)}
\\ &
+  \sum_{ M} {P}_{\rm 3-exch}^{(0^\prime c,c,0^\prime)} \left( \musepic{\figSixOPECBDo} \right)
+ \text{(2 terms obtained by permuting indices)}
\\ &
+  \sum_{M_C,M_R}\! {P}_{\rm 3-exch}^{(0,34,56)}\! \left( \musepic{\figSixOPECBDe} \right)
+ \text{(2 terms obtained by permuting indices)}
\\ &
+ \sum_{M, M_R}\! {P}_{\rm 3-exch}^{(c56,c,56)}\! \left( \musepic{\figSixOPECBDp} \right)
+ \text{(5 terms obtained by permuting indices)}
\\ &
+  \sum_{M, M_C, M_R}\!\!\! {P}_{\rm 3-exch}^{(3456,34,56)}\!\! \left( \musepic{\figSixOPECBDd} \right)
+ \text{(2 terms obtained by permuting indices)} \,, \\[-2em] \stepcounter{equation}\tag{\theequation}\label{6Exch3CBD}
\end{align*}
\endgroup
where by a permutation of indices we mean an element of the permutation group $S_3$ acting on the three ordered sets $(1,2,0)$, $(3,4,c)$, and $(5,6,0')$,\footnote{Under any such permutation, $P_{\rm 3-exch}^{(0,c,0^\prime)}$ gets mapped to itself, and so does ${P}_{\rm 3-exch}^{(12,34,56)}$. But for example, ${P}_{\rm 3-exch}^{(0^\prime c,c,0^\prime)}$ can be mapped to ${P}_{\rm 3-exch}^{(0,00^\prime,0^\prime)}$ and ${P}_{\rm 3-exch}^{(0,c,c0)}$. Likewise ${P}_{\rm 3-exch}^{(c56,c,56)}$ can be mapped to
${P}_{\rm 3-exch}^{(12,c,c12)}$,
${P}_{\rm 3-exch}^{(0,34,034)}$,
${P}_{\rm 3-exch}^{(0,056,56)}$,
${P}_{\rm 3-exch}^{(12,0^\prime 12,0^\prime)}$, and
${P}_{\rm 3-exch}^{(0^\prime 34,34,0^\prime)}$.} and the CBD coefficients are given by
\eqn{6Exch3OPEa}{
P_{\rm 3-exch}^{(0,c,0^\prime)} &=  C_{\Delta_1\Delta_2\Delta_0}\:
 C_{\Delta_3\Delta_4\Delta_c}\:
 C_{\Delta_{0^\prime}\Delta_5\Delta_6}\:
 C_{\Delta_0\Delta_c\Delta_{0^\prime}} \cr
{P}_{\rm 3-exch}^{(0,c,56)}(M_R)    &=   \frac{ \alpha_{M_R}^{\Delta_5;\Delta_6} \:C_{\Delta_1 \Delta_2 \Delta_0}\:C_{\Delta_3 \Delta_4 \Delta_c}\:C_{\Delta_0 \Delta_c \Delta_R} }{N_{\Delta_{0^\prime}} (m^2_{\Delta_R}-m^2_{\Delta_{0^\prime}})}\cr
{P}_{\rm 3-exch}^{(0^\prime c,c,0^\prime)}(M) &=  \frac{\alpha_{M}^{\Delta_{0^\prime};\Delta_c} \:C_{\Delta_1\: \Delta_2\: 2\Delta_{0^\prime c,}+2M}\:C_{\Delta_3 \Delta_4 \Delta_c}\:C_{\Delta_{0^\prime} \Delta_5 \Delta_6} }{N_{\Delta_{0}} (m^2_{2\Delta_{0^\prime c,}+2M}-m^2_{\Delta_{0}})}  \cr
{P}_{\rm 3-exch}^{(c56,c,56)}(M,M_R) &=  \frac{  \alpha_{M}^{\Delta_c;\Delta_R}\:\alpha_{M_R}^{\Delta_{5};\Delta_6}  \:C_{\Delta_1\: \Delta_2\: 2\Delta_{R c,}+2M}\:C_{\Delta_3 \Delta_4 \Delta_c}}{N_{\Delta_{0}} (m^2_{2\Delta_{R c,}+2M}-m^2_{\Delta_{0}}) N_{\Delta_{0^\prime}} (m^2_{\Delta_R}-m^2_{\Delta_{0^\prime}})} \,,
}
and
\eqn{6Exch3OPEb}{
{P}_{\rm 3-exch}^{(0,34,56)}(M_C,M_R) &=  { {P}_{\rm 1-exch}^{(0,34,56)}(M_C,M_R) \over N_{\Delta_{0^\prime}} (m_{\Delta_R}^2-m_{\Delta_{0^\prime}}^2) N_{\Delta_c} (m_{\Delta_C}^2-m_{\Delta_c}^2)}\cr
{P}_{\rm 3-exch}^{(12,34,56)}(M_L,M_C,M_R)    &=  {{P}_{\rm cont}^{(12,34,56)}(M_L,M_C,M_R) \over N_{\Delta_0} (m_{\Delta_L}^2-m_{\Delta_0}^2) N_{\Delta_{0^\prime}} (m_{\Delta_R}^2-m_{\Delta_{0^\prime}}^2) N_{\Delta_c} (m_{\Delta_C}^2-m_{\Delta_c}^2)} \cr
{P}_{\rm 3-exch}^{(3456,34,56)}(M,M_C,M_R) &= {{P}_{\rm cont}^{(3456,34,56)}(M,M_C,M_R) \over  N_{\Delta_0} (m_{2\Delta_{RC,}+2M}^2-m_{\Delta_0}^2) N_{\Delta_{0^\prime}} (m_{\Delta_R}^2-m_{\Delta_{0^\prime}}^2) N_{\Delta_c} (m_{\Delta_C}^2-m_{\Delta_c}^2)}\,.
}
It is worth remarking that the CBD coefficients in~\eno{6Exch3OPEa} may also be rewritten in terms of the CBD coefficients of diagrams with fewer exchanged bulk scalars.

The CBDs for the corresponding six-point diagrams in $p$-adic AdS/CFT are easily obtained from the ones worked out in this section using the mapping between real and $p$-adic results described at the end of section~\ref{5CONTACT} --- all infinite sums should be collapsed to their leading terms, and the real conformal blocks should be replaced with the scaling $p$-adic conformal blocks. The CBD coefficients take the same form, except explicit expressions are obtained by using the $p$-adic versions of the OPE coefficients, mass-dimension relation and normalization factors given in section~\ref{PADIC}.\footnote{Upon setting all integral summation parameters to zero, the $\alpha_{M}^{s;t}$ functions appearing in the real CBD coefficients reduce identically to unity and thus their $p$-adic analogs are simply constant factors of unity.}

\subsection{Algebraic origin of logarithmic singularities}
\label{LOG}

The decomposition of AdS diagrams discussed above had generic external and internal conformal dimensions. For certain combinations of non-generic dimensions, the diagrams are expected to develop logarithmic singularities, corresponding to the contributions from anomalous dimensions of multi-twist operators at tree-level~\cite{Liu:1998th,Freedman:1998bj,DHoker:1999mic}. These are the so-called integrality conditions. For instance, for the four-point contact diagram~\eno{FourptContact}, the condition on external dimensions is $\Delta_1 + \Delta_2 - \Delta_4 - \Delta_5 \in \mathbb{Z}$. These conditions were originally obtained from analyzing directly the divergence of the associated integrals, and are repackaged in Mellin space as double poles of the Mellin amplitude. In Ref.~\cite{Hijano:2015zsa} the appearance of logarithms is associated trivially with certain algebraic conditions. As can be seen from the explicit form of the decomposition~\eno{FourptContact} and the associated coefficients in~\eno{FourptOPE}, logarithms appear in the CBD when $m_{\Delta_A}^2 = m_{\Delta_B}^2$, explicitly, $m_{\Delta_1 + \Delta_2 + 2M_A}^2 = m_{\Delta_4 + \Delta_5+2M_B}^2$ for $M_A, M_B \in \mathbb{Z}^{\ge 0}$.\footnote{The four-point contact diagram admits direct channel decompositions in other channels as well, as long as the boundary insertions satisfy the relevant OPE convergence conditions. In such cases there will be corresponding algebraic conditions in the other channels. The same will be true for higher-point diagrams to be discussed shortly, but this point will be not be explicitly discussed.} These are equivalent to the integrality conditions mentioned above~\cite{Hijano:2015zsa}. In the case of the four-point exchange diagram, the decomposition~\eno{FourptExch} and the associated coefficients~\eno{FourptExchOPE} immediately yield the condition for logarithmic terms; they appear whenever any of $m_{\Delta_A}^2, m_{\Delta_B}^2, m_{\Delta_0}^2$ coincide~\cite{Hijano:2015zsa}.

This continues to hold for higher-point diagrams as well. For example, for the five-point contact diagram~\eno{D5schematic}, one can use the identity~\eno{OPEsumId} to re-express the structure constants appearing in the decomposition coefficients~\eno{D5CBDcoeffs} in their series representation, to make the algebraic origin of the logarithms transparent. While there are several non-unique choices for the series representation due to the totally symmetric nature of the structure constants, given the CBD~\eno{D5schematic} only particular choices of the series representation for each CBD coefficient will make manifest the algebraic conditions; these choices are dictated by the precise operators being exchanged in the intermediate channels in the corresponding conformal block. This immediately leads to the result that logarithmic singularities appear whenever
\eqn{5contactAlg}{
m_{\Delta_1 + \Delta_2 + 2M_A}^2 =  m_{\Delta_3 + \Delta_4 + \Delta_5 + 2M + 2M_B}^2 \quad {\rm or} \quad m_{\Delta_4 + \Delta_5 + 2M_B}^2 = m_{\Delta_1 + \Delta_2 + \Delta_3 + 2M_A + 2M}^2\,,
}
for $M, M_A, M_B \in \mathbb{Z}^{\ge 0}$. The associated integrality conditions are $\Delta_1 + \Delta_2 - \Delta_3 - \Delta_4 - \Delta_5 \in \mathbb{Z}$ or $\Delta_1 + \Delta_2 + \Delta_3 - \Delta_4 -\Delta_5 \in \mathbb{Z}$.
Likewise, for the five-point exchange diagram~\eno{5ExchCBDFinal}, the form of the decomposition coefficients~\eno{D1ExchCBDcoeffs} dictates the algebraic conditions for logarithmic singularities. In addition to the conditions~\eno{5contactAlg}, logarithms will appear whenever any of the following holds:
\eqn{5exch1Alg}{
m_{\Delta_0}^2 = m_{\Delta_1+\Delta_2+2M_A}^2\,, \quad m_{\Delta_0}^2 = m_{\Delta_3+\Delta_4+\Delta_5+2M+2M_B}^2\,, \quad m_{\Delta_0 + \Delta_3 + 2M}^2 = m_{\Delta_4 + \Delta_5 + 2M_B}^2\,.
}
The algebraic conditions for the five-point diagram in~\eno{5TwoExchCBD} also follow trivially from a similar analysis. In addition to the conditions~\eno{5contactAlg} and~\eno{5exch1Alg}, there are a few more possibilities for non-generic conformal dimensions which admit logarithmic terms at tree-level. They are
\eqn{5exch2Alg}{
m_{\Delta_{0^\prime}}^2 = m_{\Delta_4+\Delta_5+2M_B}^2\,, \qquad& m_{\Delta_{0^\prime}}^2 = m_{\Delta_1 + \Delta_2 + \Delta_3 + 2M_A + 2M}^2\,, \qquad m_{\Delta_{0^\prime}+\Delta_3 + 2M}^2 = m_{\Delta_1 + \Delta_2 + 2M_A}^2\,, \cr
m_{\Delta_{0}}^2 &= m_{\Delta_{0^\prime} + \Delta_3 + 2M}^2\,, \qquad m_{\Delta_{0^\prime}}^2 = m_{\Delta_0 + \Delta_3 + 2M}^2\,.
}
We invite the reader to note the agreement between these conditions and those obtained in the $p$-adic framework in section~\ref{PLOGARITHMS}.

One can similarly obtain the algebraic conditions  for the six-point diagrams presented in this paper leading to logarithmic singularities. For example, for the six-point contact diagram decomposed in the OPE channel as in~\eno{6ContactCBD}, logarithms appear at tree-level whenever any of the following conditions are met:
\eqn{6contactAlg}{
m_{\Delta_1 + \Delta_2 + 2M_L}^2 = m_{\Delta_3 + \Delta_4 + \Delta_5 + \Delta_6 + 2M_C + 2M_R}^2 \,, \qquad
m_{\Delta_3 + \Delta_4 + 2M_C}^2 = m_{\Delta_1 + \Delta_2 + \Delta_5 + \Delta_6 + 2M_L + 2M_R}^2\,, \cr
m_{\Delta_5 + \Delta_6 + 2M_R}^2 = m_{\Delta_1 + \Delta_2 + \Delta_3 + \Delta_4 + 2M_L + 2M_C}^2\,, \qquad\qquad\qquad\qquad
}
where $M_L,M_C,M_R \in \mathbb{Z}^{\ge 0}$. Likewise similar algebraic conditions can be read off of the explicit CBD and the associated CBD coefficients of the other exchange six-point diagrams presented in section~\ref{SIXCBD}. As another example, the exchange diagram in~\eno{6Exch1CBD} admits, in addition to~\eno{6contactAlg}, the following conditions:
\eqn{6exch1Alg}{
m_{\Delta_{0}}^2 = m_{\Delta_1 + \Delta_2 + 2M_L}^2\,, \qquad m_{\Delta_{0}}^2 = m_{\Delta_3+\Delta_4+\Delta_5+\Delta_6+2M_C+2M_R}^2\,, \qquad \quad \cr
m_{\Delta_{0}+\Delta_3+\Delta_4 + 2M+2M_C}^2 = m_{\Delta_5+\Delta_6+2M_R}^2\,, \qquad
m_{\Delta_{0}+\Delta_5+\Delta_6 + 2M+2M_R}^2 = m_{\Delta_3+\Delta_4+2M_C}^2\,.
}
It is a trivial exercise to determine similar conditions for the remaining six-point diagrams; we omit stating the somewhat lengthy list of the conditions here.\footnote{The $p$-adic analogs of the six-point conditions mentioned in this section can be obtained simply by setting all integral parameters $M_i$ to zero, and using the $p$-adic analog of the mass-dimension relation~\eno{pMassDim}.}

\subsection{Spectral decomposition of AdS diagrams}
\label{SPECTRAL}

The conformal block decomposition of tree-level diagrams can also be  obtained in the framework of the shadow formalism.
 Using the split representation~\cite{Penedones:2010ue} one can recast all bulk integrations in the diagram into three-point contact integrals which can be readily evaluated.
  The ensuing boundary integrals are recognized as conformal partial waves, corresponding to the exchange of states in the principal series representation of the conformal group.
  This gives the spectral decomposition of AdS diagrams, with the poles of the spectral density function under the contour integral dictating the explicit conformal block decomposition.
  Two detailed examples are provided in appendix~\ref{4SPECTRAL} for illustrative purposes.

  Conformal partial waves  themselves are linear combinations of conformal blocks and their shadow blocks, so one can trade conformal partial waves in the integrand for conformal blocks in the shadow formalism, to make the connection with CBD manifest.
For example, in the case of the four-point diagrams, this computation leads to the following spectral decomposition (see appendix~\ref{4SPECTRAL})
  \eqn{4cont}{
\musepic{\figFourContact} &= \int_{-i\infty}^{i\infty} {dc \over 2\pi i} {\zeta_\infty(d+2c)\over \zeta_\infty(2c)} C_{\Delta_1\: \Delta_2\: {d\over 2}+c}\: C_{{d\over 2}+c\: \Delta_4\: \Delta_5} \left(\musepic{\figFourPrinc}\right)
}
\eqn{4exch}{
\musepic{\figFourExch} &= {1\over N_{\Delta_0}}\int_{-i\infty}^{i\infty} {dc \over 2\pi i} {\zeta_\infty(d+2c)\over \zeta_\infty(2c)} {C_{\Delta_1\:\Delta_2\: {d\over 2}+c}\: C_{{d\over 2}+c\: \Delta_4\: \Delta_5} \over m^2_{{d\over2}+c}-m^2_{\Delta_0}} \!\left(\!\musepic{\figFourPrinc}\!\right)\!,
}
where the local zeta function $\zeta_\infty$ was defined in~\eno{zetainftyDef}, and the OPE coefficients and normalization factor $N_\Delta$ can be found in~\eno{OPEreal} and~\eno{GNorm} respectively.
Evaluating the contour integral using the residue theorem reproduces the CBDs in~\eno{D4schematic} and~\eno{FourptExch} with the right decomposition coefficients.
We note that we have  written the spectral density in the decompositions above in a form which makes the pole structure manifest and admits a direct generalization to higher-point diagrams.

 Likewise higher-point diagrams considered in this section also admit similar spectral decompositions. For example, the five-point diagrams decompose as
 \eqn{5cont}{
\musepic{\figFiveContact} &= \int_{-i\infty}^{i\infty} \left(\prod_{j=A,B}  {dc_j \over 2\pi i} {\zeta_\infty(d+2c_j)\over \zeta_\infty(2c_j)} \right) \left(\musepic{\figFivePrinc}\right) \cr
&\qquad\quad \times C_{\Delta_1\,\Delta_2\,{d\over 2}+c_A} \: C_{{d\over2}+c_A\,\Delta_3\,{d\over2}+c_B} \: C_{{d\over 2}+c_B\,\Delta_4\,\Delta_5}\,,
}
\eqn{5exch1}{
\musepic{\figFiveOneExch} &= {1\over N_{\Delta_0}} \int_{-i\infty}^{i\infty} \left(\prod_{j=A,B}  {dc_j \over 2\pi i} {\zeta_\infty(d+2c_j)\over \zeta_\infty(2c_j)} \right) \left(\musepic{\figFivePrinc}\right)  \cr
&\qquad\qquad \times  {C_{\Delta_1\,\Delta_2\,{d\over 2}+c_A}\: C_{{d\over2}+c_A\,\Delta_3\,{d\over2}+c_B}\: C_{{d\over 2}+c_B\,\Delta_4\,\Delta_5} \over m^2_{{d\over2}+c_A}-m^2_{\Delta_0}}\,,
}
\eqn{5exch2}{
\musepic{\figFiveTwoExch} &= {1\over N_{\Delta_0} N_{\Delta_{0^\prime}}} \int_{-i\infty}^{i\infty} \left(\prod_{j=A,B}  {dc_j \over 2\pi i} {\zeta_\infty(d+2c_j)\over \zeta_\infty(2c_j)} \right) \left(\musepic{\figFivePrinc}\right)  \cr
&\qquad\qquad \times  {C_{\Delta_1\,\Delta_2\,{d\over 2}+c_A}\: C_{{d\over2}+c_A\,\Delta_3\,{d\over2}+c_B}\: C_{{d\over 2}+c_B\,\Delta_4\,\Delta_5} \over (m^2_{{d\over2}+c_A}-m^2_{\Delta_0}) (m^2_{{d\over2}+c_B}-m^2_{\Delta_{0^\prime}})}\,.
}
In the integrands above, the object in parantheses is the global scalar five-point conformal block~\cite{Rosenhaus:2018zqn,Parikh:2019ygo} discussed  briefly in section~\ref{SIXOPEGEODESIC}.
Similarly, the six-point diagrams which admit an OPE channel direct channel decomposition can be written as
\eqn{6cont}{
\musepic{\figSixContact} &= \int_{-i\infty}^{i\infty} \left(\prod_{j=L,C,R}  {dc_j \over 2\pi i} {\zeta_\infty(d+2c_j)\over \zeta_\infty(2c_j)} \right) \left(\musepic{\figSixPrinc}\right) \cr
&\qquad\quad \times C_{\Delta_1\,\Delta_2\,{d\over 2}+c_L} \:C_{{d\over2}+c_L\,{d\over2}+c_C\,{d\over2}+c_R}\: C_{{d\over2}+c_C\,\Delta_3\,\Delta_4} \: C_{{d\over 2}+c_R\,\Delta_5\,\Delta_6}\,,
}
\eqn{6exch1}{
\musepic{\figSixExchOneA} &={1\over N_{\Delta_0}} \int_{-i\infty}^{i\infty} \left(\prod_{j=L,C,R}  {dc_j \over 2\pi i} {\zeta_\infty(d+2c_j)\over \zeta_\infty(2c_j)} \right) \left(\musepic{\figSixPrinc}\right) \cr
&\qquad\quad \times {C_{\Delta_1\,\Delta_2\,{d\over 2}+c_L} \:C_{{d\over2}+c_L\,{d\over2}+c_C\,{d\over2}+c_R}\: C_{{d\over2}+c_C\,\Delta_3\,\Delta_4} \: C_{{d\over 2}+c_R\,\Delta_5\,\Delta_6} \over m^2_{{d\over2}+c_L}-m^2_{\Delta_0}}\,,
}
\eqn{6exch2}{
\musepic{\figSixExchTwoA} &={1\over N_{\Delta_0}N_{\Delta_{0^\prime}}} \int_{-i\infty}^{i\infty} \left(\prod_{j=L,C,R}  {dc_j \over 2\pi i} {\zeta_\infty(d+2c_j)\over \zeta_\infty(2c_j)} \right) \left(\musepic{\figSixPrinc}\right) \cr
&\qquad\quad \times {C_{\Delta_1\,\Delta_2\,{d\over 2}+c_L} \:C_{{d\over2}+c_L\,{d\over2}+c_C\,{d\over2}+c_R}\: C_{{d\over2}+c_C\,\Delta_3\,\Delta_4} \: C_{{d\over 2}+c_R\,\Delta_5\,\Delta_6} \over (m^2_{{d\over2}+c_L}-m^2_{\Delta_0})(m^2_{{d\over2}+c_R}-m^2_{\Delta_{0^\prime}})}\,,
}
\eqn{6exch3}{
\musepic{\figSixExchThreeA} &={1\over N_{\Delta_0}N_{\Delta_c}N_{\Delta_{0^\prime}}} \int_{-i\infty}^{i\infty} \left(\prod_{j=L,C,R}  {dc_j \over 2\pi i} {\zeta_\infty(d+2c_j)\over \zeta_\infty(2c_j)} \right) \left(\musepic{\figSixPrinc}\right) \cr
&\qquad\quad \times {C_{\Delta_1\,\Delta_2\,{d\over 2}+c_L} \:C_{{d\over2}+c_L\,{d\over2}+c_C\,{d\over2}+c_R}\: C_{{d\over2}+c_C\,\Delta_3\,\Delta_4} \: C_{{d\over 2}+c_R\,\Delta_5\,\Delta_6} \over (m^2_{{d\over2}+c_L}-m^2_{\Delta_0})(m^2_{{d\over2}+c_C}-m^2_{\Delta_c})(m^2_{{d\over2}+c_R}-m^2_{\Delta_{0^\prime}})}\,.
}
In the integrands above, the object inside parentheses is the global scalar six-point conformal block in the OPE channel, whose holographic representation was obtained in section~\ref{SIXOPEGEODESIC}.
Evaluating the contour integrals yields explicitly the CBDs obtained earlier using geodesic diagram techniques.
 Moreover, the form of the spectral density function explains the algebraic relations between the decomposition coefficients of contact and exchange diagrams involving more and more interaction vertices highlighted earlier in this section.
 Finally, the generalization to \emph{arbitrary} scalar tree-level AdS diagrams should be clear from the examples considered here.

Before closing this section, we point out the closely related results in the $p$-adic AdS/CFT framework of section~\ref{PADIC}.
The same diagrams evaluated on the Bruhat--Tits tree admit identical spectral decompositions as the ones shown above, except  we  must essentially replace all $\zeta_\infty$ local zeta functions in the formulas with the $\zeta_p$ local zeta function defined in~\eno{zetapDef}. More precisely, in the spectral decomposition one should simply use the formulas for the OPE coefficient~\eno{OPEpadic}, the overall normalization factor~\eno{pGNorm} and bulk scalar mass~\eno{pMassDim} as encountered in the $p$-adic framework, as well as the simpler $p$-adic conformal block.
The lack of higher-twist contributions in the $p$-adic CBD seen in section~\ref{CBDBTTREE} is repackaged into the drastically simpler pole structure of the $\zeta_p$ local zeta function, as compared to its real analog, the $\zeta_\infty$ function defined in~\eno{zetainftyDef}.\footnote{Also, owing to the periodicity of $\zeta_p$ in the imaginary direction, in the $p$-adic case the complex variables $c_j$ are not integrated over a line in the complex plane but along a contour that wraps around a cylindrical manifold with circumference $\pi/\log p$; see Ref.~\cite{Jepsen:2018dqp} where the necessary $p$-adic split representation was first worked out.}

 \section{Discussion}
\label{DISCUSSION}

In this paper we presented new results establishing the holographic duals of global scalar conformal blocks for the five-point block (equations~\eno{calW5CBpadic}-\eno{calW5padic}), and the six-point block in the OPE channel (equations~\eno{calW6padic}-\eno{calW6CBpadic}) in $p$-adic AdS/CFT, and the six-point block in the OPE channel  in conventional (real) AdS$_{d+1}$/CFT$_d$ (equations~\eno{calW6star}-\eno{calW6starConfwave}), following the techniques introduced in Ref.~\cite{Parikh:2019ygo} where the dual of the global five-point block in conventional AdS$_{d+1}$/CFT$_d$ was obtained.
Similar to the holographic representation of the global four-point block~\cite{Hijano:2015zsa}, the holographic duals of the higher-point blocks have an integral representation in terms of geodesic diagrams, \emph{viz.}\ variants of bulk Feynman diagrams involving solely bulk  integrals over boundary anchored geodesics.
In the case of the six-point global conformal block in the OPE channel, to our knowledge the holographic dual provides the only known explicit representation of the associated block.

However, in contrast with the four-point block, whose holographic dual is a a single tree-level four-point AdS exchange diagram except with all AdS integrations replaced with geodesic integrals, the precise holographic representations for higher-point blocks turn out to be more complicated for a number of reasons:

First, the holographic representation of the six-point block in the OPE channel admits an  interpretation as the six-point \emph{one-loop} AdS diagram built out of three quartic interactions vertices with bulk-to-bulk propagators assigned special linear combinations of conformal dimensions, but with all AdS integrations replaced by geodesic integrals.

Second, one must perform a weighted sum over an infinite number of diagrams of this class; conceptually this sums up the contributions from the full conformal families associated with the conformal representations being exchanged in the intermediate channels.  Reassuringly, such infinite sums are missing in the holographic duals of the $p$-adic versions of the same conformal blocks and they are represented as \emph{single} geodesic bulk diagrams, since the putative dual $p$-adic CFT lacks descendants~\cite{Melzer:1988he}.
However, the contrast with the holographic dual of the real four-point block~\cite{Hijano:2015zsa} is only superficial. The four-point holographic dual does indeed admit a representation as an infinite sum over geodesic bulk diagrams~\cite{Parikh:2019ygo}; this representation (described in section~\ref{SIXOPEGEODESIC}) is easily summed up analytically leading to the compact closed-form holographic representation of Ref.~\cite{Hijano:2015zsa}. For both practical and conceptual purposes, it would be useful to determine whether the simplification in the case of the four-point block was accidental or if holographic duals of higher-point blocks should also admit further simplifications that allow them to be written as single geodesic bulk diagrams.

Third, the holographic representation of the global five-point comb channel block in real AdS$_{d+1}$/CFT$_d$~\cite{Parikh:2019ygo} does not lend itself to a direct interpretation in terms of a conventional (tree- or loop-level) AdS diagram, albeit with all AdS integrations replaced with geodesic integrals, as can be seen in~\eno{calW5}. This suggests that a more fundamental interpretation of the holographic representations of global conformal blocks which applies more generally to arbitrary $n$-point blocks in any spacetime dimension in any channel is perhaps more subtle.
At first glance, such seems to be the case also for the five- and six-point blocks in $p$-adic AdS/CFT presented in this paper.
 However, the $p$-adic blocks, owing to their drastically simpler scaling forms, do admit a simpler, alternative holographic interpretation on the Bruhat--Tits tree in terms of geodesic diagrams involving cubic bulk interaction vertices and no full bulk integrations, as discussed at the end of section~\ref{DUALBTTREE}. This interpretation relies on the existence of special bulk points, which may either be interpreted as unique points of intersections of geodesics joining boundary insertion points, or as Fermat-Torricelli points solving a geodesic length minimization problem. Other interpretations may also be possible on the Bruhat--Tits tree, and it is not obvious which one, if any, might carry over to the real setup  (although there may conceivably be a connection with Fermat-Torricelli points and Steiner trees in hyperbolic space; such constructs recently appeared in the context of holographic representations of large-$c$ Virasoro conformal blocks~\cite{Alkalaev:2018nik}).

 In any case, since in some aspects the formulation of $p$-adic AdS/CFT~\cite{Gubser:2016guj,Heydeman:2016ldy} is similar to $d=1$ dimensional (real) AdS/CFT, perhaps there is a possibility that at least low-dimensional AdS/CFT may allow  simpler interpretations of the holographic duals for (real) conformal blocks.
 Further in $d=2$, following the work of Ref.~\cite{Kraus:2017ezw} for the four-point global conformal blocks, it would be interesting to extend the higher-point results of this paper to holographic duals of higher-point global blocks in finite temperature CFTs. Moreover in $d=2$ it would be interesting to explore the connections between the higher-point geodesic diagrams of this paper and higher-point Virasoro blocks along the lines of Refs.~\cite{Hijano:2015qja,Banerjee:2016qca} (see also Refs.~\cite{Fitzpatrick:2014vua, Fitzpatrick:2015zha,Hijano:2015rla,Alkalaev:2015wia,Alkalaev:2015lca,Alkalaev:2015fbw,Besken:2016ooo,Alkalaev:2016rjl,Alkalaev:2018nik}).
   In arbitrary spacetime dimensions, it is also natural to consider the generalizations of the holographic duals of the  higher-point scalar blocks of this paper to those involving external and exchanged spinning operators, along the lines of the four-point case~\cite{Nishida:2016vds, Castro:2017hpx, Dyer:2017zef, Chen:2017yia, Tamaoka:2017jce, Nishida:2018opl, Das:2018ajg}.

 \vspace{.5em}

One of the direct applications of the holographic duals of higher-point global conformal blocks was an alternate, direct derivation of the conformal block decomposition of higher-point AdS diagrams.
One of the main technical tools developed in this paper for this purpose was a class of AdS propagator identities involving bulk integration over a common point of intersection of three bulk-to-bulk and/or bulk-to-boundary propagators  (see sections~\ref{PROPBTTREE} and~\ref{PROPIDS}). These identities provide a generalization of the three-point contact diagram, with a subset of boundary points pushed into the bulk.
Indeed, with the knowledge of the holographic duals and various propagator identities which re-express bulk integrations in terms of unintegrated combinations of bulk-to-bulk and bulk-to-boundary propagators, we were able to obtain the explicit direct channel CBD of a number of higher-point tree-level scalar AdS diagrams involving scalar contact interactions.
With various AdS propagator and hypergeometric identities in hand, the procedure to obtain the CBD involved only simple algebraic operations, and no bulk integrations.
Notably, in section~\ref{CBD} we presented the explicit decomposition of all five-point scalar diagrams and the class of all six-point diagrams which admit a direct-channel CBD in the basis of OPE channel six-point blocks.\footnote{The resulting decompositions are presented in~\eno{D5schematic}-\eno{D5CBDcoeffs}, \eno{D1ExchCBDcoeffs}-\eno{D5Exch1schematic}, and \eno{5TwoExchCBD}-\eno{Fivept2ExchOPEAgain} for the five-point diagrams, and~\eno{6ContactCBD}-\eno{6ContactOPE}, \eno{6Exch1CBD}-\eno{6Exch1OPEb}, \eno{6Exch2CBD}-\eno{6Exch2OPEc} and \eno{6Exch3CBD}-\eno{6Exch3OPEb} for the six-point diagrams.}
This procedure  provides a higher-point generalization of the  direct-channel CBD of four-point AdS diagrams using geodesic diagram techniques.
As described in section~\ref{LOG} (as well as section~\ref{PLOGARITHMS}), the conditions for the presence of logarithmic singularities in tree-level AdS diagrams also fall out trivially as simple algebraic relations.
It would be useful to find generalizations of the higher-point method that incorporate spinning AdS diagrams, derivative and spin exchanges.
Progress along this direction may also aid the technically challenging task of the holographic reconstruction of the classical bulk action for higher spin gravity theories beyond quartic interaction vertices~\cite{Giombi:2009wh,Bekaert:2014cea,Bekaert:2015tva}.
The rewriting of spectral decomposition of AdS diagrams in terms of conformal blocks as presented in section~\ref{SPECTRAL} may also turn out to be useful in this regard.

A class of four-point loop diagrams (such as the bubble diagram), which admit a rewriting as a sum over infinitely many tree-level exchange diagrams~\cite{Penedones:2010ue}, can in principle be decomposed in the direct channel using the techniques of Ref.~\cite{Hijano:2015zsa} (see also Ref.~\cite{Fitzpatrick:2011dm}). However, a detailed analysis of the structure and properties of the resulting decomposition coefficients remains insufficiently addressed.
The new propagator identities of this paper provide yet another method to obtain the CBD of such diagrams using only elementary operations. It would be interesting to investigate if these new tools provide new insights into the decomposition of such loop amplitudes, and more ambitiously into the decomposition of arbitrary loop amplitudes.
The evaluation of certain loop diagrams may involve generalizations of AdS propagator identities derived in this paper to products of four or more bulk-to-bulk and/or bulk-to-boundary propagators.
 These would also be helpful in obtaining the decomposition of seven- and higher-point AdS diagrams via geodesic diagram techniques.
We are also hopeful methods presented in this paper may  help inform the  discussion on the CBD of AdS diagrams and conformal partial waves in the crossed channel, which has been the subject of much recent interest --- see e.g.\ Refs.~\cite{Sleight:2018epi,Sleight:2018ryu,Cardona:2018dov,Liu:2018jhs,Gopakumar:2018xqi,Zhou:2018sfz,Li:2019dix,Chen:2019gka} --- especially because the $p$-adic analog of these methods yields, promisingly, a closed-form expression for the crossed channel decomposition of the four-point exchange diagram on the Bruhat--Tits tree~\cite{Gubser:2017tsi}.

We hope to see progress in these directions in the near future.

\subsection*{Acknowledgments}

C.~B.~J. is grateful to S.~S.~Gubser for imparting insight. S.~P.~thanks D.~Meltzer and E.~Perlmutter for valuable discussions. The work of C.~B.~J. was supported in part by the Department of Energy under Grant No.~DE-FG02-91ER40671, and by the Simons Foundation, Grant 511167 (SSG).

\newpage

\appendix

\section{Spectral decomposition: Four-point examples}
\label{4SPECTRAL}

In this appendix, we will derive~\eno{4cont}-\eno{4exch}.

Our starting point is the integral representation of the four-point conformal partial wave associated with the conformal multiplet of weight $(\Delta,J)$, given by~\cite{Ferrara:1972uq,Ferrara:1972ay,Ferrara:1972xe,Ferrara:1973vz,SimmonsDuffin:2012uy}
\eqn{4PartialWaveInt}{
\Psi_{\Delta_,J}^{\Delta_1,\ldots,\Delta_4}(x_i) \equiv  \int_{y \in \partial {\rm AdS}} \langle\langle {\cal O}_1(x_1) {\cal O}_2(x_2) {\cal O}_{\Delta,J}^{\mu_1\ldots \mu_J}(y) \rangle\rangle \langle \langle \widetilde{ {\cal O}}_{\Delta,J,\mu_1\ldots \mu_J}(y) {\cal O}_3(x_3) {\cal O}_4(x_4) \rangle \rangle\,,
}
where $\langle\langle \cdot \rangle\rangle$ is the purely kinematic part of the three-point function, i.e.\ devoid of the OPE coefficient,
and $\widetilde{ {\cal O}}_{\Delta,J,\mu_1\ldots \mu_J}$ is the shadow operator of ${\cal O}_{\Delta,J,\mu_1\ldots\mu_J}$,
\eqn{ShadowOp}{
\widetilde{\cal O}(x) \equiv \int_{y \in \partial{\rm AdS}} {1\over (x-y)^{2(d-\Delta)}} {\cal O}(x)\,.
}
Conformal partial waves are single-valued functions of coordinates and, for integer spins and unphysical complex dimensions $\Delta= d/2 + i \nu$ ($\nu \in \mathbb{R}$)  corresponding to the principal series representation of the conformal group, form a complete set of functions obeying orthogonality relations~\cite{Dobrev:1977qv}.\footnote{In odd $d$ one must also include a discrete series representation; see e.g.\ Ref.~\cite{Gadde:2017sjg}.}
The four-point partial wave transforms like the four-point conformal correlator and in fact is an eigenfunction of the multi-point Casimir constructed out of positions $x_1$ and $x_2$, with the same eigenvalue as the one associated with the four-point conformal block.
However, the integral~\eno{4PartialWaveInt} is \emph{not} the conformal block.
Instead, the precise relation between the four-point conformal partial wave and the global conformal block is~\cite{Dobrev:1977qv,Dolan:2000ut}
\eqn{4PartialWaveBlock}{
\Psi_{\Delta,J}^{\Delta_1,\ldots,\Delta_4}(x_i) = K_{\widetilde{\Delta},J}^{\Delta_3,\Delta_4} W_{\Delta,J}^{\Delta_1,\ldots,\Delta_4} (x_i) +  K_{\Delta,J}^{\Delta_1,\Delta_2} W_{\widetilde{\Delta},J}^{\Delta_1,\ldots,\Delta_4} (x_i)\,,
}
where $W_{\Delta,J}^{\Delta_1,\ldots,\Delta_4}$ is the four-point conformal block which is the contribution to the four-point correlator coming from the exchange of an operator of dimension $\Delta$ and spin $J$ and all its descendants in the $(12,34)$-channel, we have defined $\widetilde{\Delta} = d-\Delta$, and
\eqn{Kcoeffs}{
K_{\Delta,J}^{\Delta_1,\Delta_2} = \left(-{1\over 2}\right)^J { \pi^{d\over 2}\: \Gamma\left(\Delta -{d \over 2}\right) \Gamma(\Delta+J-1) \: \Gamma\left({\widetilde{\Delta}+\Delta_1-\Delta_2+J \over 2}\right) \Gamma\left({\widetilde{\Delta}+\Delta_2-\Delta_1+J\over 2}\right) \over \Gamma(\Delta-1)\: \Gamma(\widetilde{\Delta}+J)\: \Gamma\left({\Delta+\Delta_1-\Delta_2+J \over 2}\right) \Gamma\left({\Delta+\Delta_2-\Delta_1+J \over 2}\right)}\,.
}
The second term in~\eno{4PartialWaveBlock} represents the four-point shadow block.
Since we are focusing on external and exchanged scalars in this paper, we specialize to $J=0$ and drop the spin label altogether; see Refs.~\cite{Costa:2014kfa,Sleight:2017fpc} for a  generalization to higher derivative contact interactions, exchange of spinning operators, and spinning external operators for four-point diagrams.

Consider the four-point contact and exchange diagrams~\eno{FourptContact} and~\eno{FourptExchDef}.
Using the split representations~\cite{Penedones:2010ue,Costa:2014kfa,Bekaert:2015tva} of respectively, the delta function on the contact vertex, and the bulk-to-bulk propagator in the exchange diagram,
\eqn{deltaGSplit}{
\delta^{d+1}(z,w) &=  \int_{y \in \partial {\rm AdS}} \int_{-i\infty}^{i\infty} {dc \over 2\pi i} \: {\rho}(c) \hat{K}_{{d\over 2}+c}(y,z) \hat{K}_{{d\over 2}-c}(y,w) \cr
\hat{G}_\Delta(z,w) &=  \int_{y \in \partial {\rm AdS}} \int_{-i\infty}^{i\infty} {dc \over 2\pi i} \: \rho_{\Delta}(c) \hat{K}_{{d\over 2}+c}(y,z) \hat{K}_{{d\over 2}-c}(y,w),
}
where
\eqn{rhoDef}{
{\rho}(c) &\equiv   { \zeta_\infty(d+2c) \over  2\zeta_\infty(2c)}  {\zeta_\infty(d-2c) \over 2\zeta_\infty(-2c)} \cr
\rho_\Delta(c) &\equiv {-1\over N_{\Delta}}  { {\rho}(c) \over m_\Delta^2-m_{{d\over 2}+c}^2}
}
and the normalization $N_\Delta$ is defined in~\eno{GNorm}, one can recast all bulk integrations in the diagrams into (unphysical) three-point contact diagrams which can be readily evaluated, to give
\eqn{4ContExchSplit}{
D_4 &= \int_{-i\infty}^{i\infty} {dc \over 2\pi i} \:  \rho^{\rm cont}(c) \int_{y \in \partial{\rm AdS}}  \langle\langle {\cal O}_1(x_1) {\cal O}_2(x_2) {\cal O}_{{d\over 2}+c}(y) \rangle\rangle \langle \langle \widetilde{ {\cal O}}_{{d\over 2}+c}(y) {\cal O}_4(x_4) {\cal O}_5(x_5) \rangle \rangle \cr
D_4^{\rm exch} &= \int_{-i\infty}^{i\infty} {dc \over 2\pi i} \:  \rho^{\rm exch}_{\Delta_0}(c) \int_{y \in \partial{\rm AdS}}  \langle\langle {\cal O}_1(x_1) {\cal O}_2(x_2) {\cal O}_{{d\over 2}+c}(y) \rangle\rangle \langle \langle \widetilde{ {\cal O}}_{{d\over 2}+c}(y) {\cal O}_4(x_4) {\cal O}_5(x_5) \rangle \rangle\,,
}
where
\eqn{rhoOPEDef}{
\rho^{\rm cont}(c) &\equiv {\rho}(c) \: C_{\Delta_1\: \Delta_2\: {d\over 2}+c} C_{\Delta_4\: \Delta_5\: {d\over 2}-c} \cr
\rho^{\rm exch}_{\Delta_0}(c) &\equiv {\rho}_{\Delta_0}(c) \: C_{\Delta_1\: \Delta_2\: {d\over 2}+c} C_{\Delta_4\: \Delta_5\: {d\over 2}-c}\,,
}
and $C_{\Delta_i \Delta_j \Delta_k}$ are the OPE coefficients~\eno{OPEreal}.
Identifying the integral representation of the (spin 0) four-point conformal partial wave, one obtains a spectral integral over the principal series
\eqn{4ContAgain}{
D_4 &= \int_{-i\infty}^{i\infty} {dc \over 2\pi i} \:  \rho^{\rm cont}(c)\: \Psi_{{d\over 2}+c}^{\Delta_1,\ldots,\Delta_4}(x_i) \cr
 &= \int_{-i\infty}^{i\infty} {dc \over 2\pi i} \:  2 K_{{d\over 2}-c}^{\Delta_3,\Delta_4}\: \rho^{\rm cont}(c)\: W_{{d\over 2}+c}^{\Delta_1,\ldots,\Delta_4}(x_i)\,,
}
and
\eqn{4ExchAgain}{
D_4^{\rm exch} &= \int_{-i\infty}^{i\infty} {dc \over 2\pi i} \:  \rho^{\rm exch}_{\Delta_0}(c) \Psi_{{d\over 2}+c}^{\Delta_1,\ldots,\Delta_4}(x_i) \cr
 &=  \int_{-i\infty}^{i\infty} {dc \over 2\pi i} \:  2 K_{{d\over 2}-c}^{\Delta_3,\Delta_4}\: \rho^{\rm exch}_{\Delta_0}(c)\: W_{{d\over 2}+c}^{\Delta_1,\ldots,\Delta_4}(x_i) \,,
}
where in the second lines of~\eno{4ContAgain}-\eno{4ExchAgain} we have used~\eno{4PartialWaveBlock} to re-express the spectral decomposition in terms of the four-point conformal blocks (see e.g.\ Ref.~\cite{Zhou:2018sfz}). As noted previously, we are suppressing all spin $J$ subscripts, so for example we have defined $K_{\Delta}^{\Delta_a,\Delta_b} \equiv K_{\Delta,0}^{\Delta_a,\Delta_b}$.
Closing the contour on the right and picking up the poles one obtains the well-known direct channel conformal block decomposition of the four-point contact and exchange diagrams.

As currently written, the pole structure of the spectral density function for the conformal block decompositions is not obvious. However, the spectral density simplifies to a more accessible form,
\eqn{SpecDensitySimplify}{
2 K_{{d\over 2}-c}^{\Delta_3,\Delta_4}\: \rho^{\rm cont}(c) &=    { \zeta_\infty(d+2c) \over  2\zeta_\infty(2c)} \:  C_{\Delta_1\: \Delta_2\: {d\over 2}+c} \:  C_{\Delta_3\: \Delta_4\: {d\over 2}+c} \cr
2 K_{{d\over 2}-c}^{\Delta_3,\Delta_4}\: \rho^{\rm exch}_{\Delta_0}(c) &=    {-1\over N_{\Delta_0}} { \zeta_\infty(d+2c) \over  2\zeta_\infty(2c)} \:  {C_{\Delta_1\: \Delta_2\: {d\over 2}+c}\:   C_{\Delta_3\: \Delta_4\: {d\over 2}+c}  \over m_{\Delta_0}^2-m_{{d\over 2}+c}^2}\,,
}
which makes the pole structure, and thus the spectrum of operators being exchanged in the intermediate channels in the conformal block decomposition, manifest.
For example, the CBD of the contact diagram will be a linear combination of two conformal blocks, one associated with the exchange of multi-twist primaries of dimensions ${d\over 2} + c = \Delta_1+\Delta_2+2M$ for $M \in \mathbb{Z}^{\geq 0}$ and their entire conformal families, and the other associated with the exchange of multi-twist primaries with conformal dimensions ${d\over 2}+c = \Delta_3+\Delta_4+2M$ and the contribution from their conformal families. These poles arise from the OPE coefficients in~\eno{SpecDensitySimplify}. For the exachange diagram, in addition to the poles above, there is a pole at ${d \over 2}+c=\Delta_0$ coming from the zero of the mass-squared expression in the denominator, corresponding to the exchange of the single-trace operator with dimension $\Delta_0$ and its conformal family.

\section{Proofs of important identities}
\label{PROOFIDS}

\subsection{Propagator identities}
\label{PROOFPROPAG}
In this appendix we provide derivations of the three real propagator identities \eno{KKGreal}, \eno{KGGreal}, and \eno{GGGreal}.\footnote{The three $p$-adic propagator identities  \eno{KKGpadic}, \eno{KGGpadic}, and \eno{GGGpadic}, originally given in Ref.~\cite{Gubser:2017tsi} and found by direct computation on the Bruhat-Tits tree, can also be derived in a manner  parallel to the computations over the reals shown in this appendix  using the $p$-adic Schwinger-parametrization and Mellin representation developed in Ref.~\cite{Jepsen:2018dqp} (though various infinite series  encountered in the following calculations get collapsed  to just the leading term of the series).}
For ease of computation, we will use the embedding space formalism. Here, boundary points $x^i \in \mathbb{R}^d$ are described in terms of the space of light rays in $\mathbb{R}^{d+1,1}$ that pass through the origin,
\eqn{}{
-(P^0)^2+(P^1)^2+...+(P^{d+1})^2=0\,,
}
according to
\eqn{}{
&
P^0=\frac{1+x^2}{2}\
\cr
&
P^i=x^i\
\cr
&
P^{d+1}=\frac{1-x^2}{2}\,.
}
Bulk points $z=(z_0,z^i) \in \mathbb{R}^+ \times\mathbb{R}^d$ are described in terms of the hyperboloid in $\mathbb{R}^{+}\times \mathbb{R}^{d+1}$,
\eqn{}{
-(Z^0)^2+(Z^1)^2+...+(Z^{d+1})^2=-1\,,
}
according to
\eqn{}{
&
Z^0=\frac{1+z_0^2+z^2}{2z_0}
\cr
&
Z^i=\frac{z^i}{z_0}
\cr
&
Z^{d+1}=\frac{1-z_0^2-z^2}{2z_0}\,.
}
In embedding space coordinates, the quantity closely related to chordal distance in~\eno{Gform} is written as
\eqn{xiEmbedding}{
{\xi(W,Z) \over 2} = {1 \over (-2W\cdot Z)}\,,
}
the bulk-to-bulk propagator in~\eno{Gform} is given by
\eqn{GEmbedding}{
\hat{G}_{\Delta}(W,Z) =  \left({\xi(W,Z) \over 2}\right)^\Delta\: {}_2F_1\left[{\Delta\over 2},{\Delta+1 \over 2};\Delta-{d\over 2}+1;\xi(W,Z)^2\right],
}
and the bulk-to-boundary propagator~\eno{Kdef} takes the form
\eqn{KEmbedding}{
\hat{K}_{\Delta}(P,Z) = {1 \over (-2 P \cdot Z)^\Delta}\,.
}

\subsubsection{$\int \hat{K}\hat{K}\hat{G}$}
\label{KKGPROOF}

In this subsection we derive~\eno{KKGreal}, explicitly,
\eqn{KKGreal2}{
& \int_{z \in {\rm AdS}} \hat{K}_{\Delta_2}(x_2,z) \hat{K}_{\Delta_3}(x_3,z)  \hat{G}_{\Delta_a}(w_a,z) \cr
&= C_{\Delta_a\Delta_2\Delta_3} \sum_{k_a=0}^\infty c_{k_a}^{\Delta_2;\Delta_a;\Delta_3}
 { \hat{K}_{\Delta_{a2,3}+k_a}(x_2,w_a)  \hat{K}_{\Delta_{a3,2}+k_a}(x_3,w_a) \over (x_{23}^2)^{\Delta_{23,a}-k_a}} \cr
 &\quad + {1 \over m_{2\Delta_{23,}}^2-m_{\Delta_a}^2} {1 \over N_{\Delta_a}} \sum_{k_a=0}^\infty d_{k_a}^{\Delta_2;\Delta_a;\Delta_3} {\hat{K}_{\Delta_2+k_a}(x_2,w_a) \hat{K}_{\Delta_3+k_a}(x_3,w_a) \over (x_{23}^2)^{-k_a}}\,,
}
where
\eqn{c1Def}{
c_{k_a}^{\Delta_2;\Delta_a;\Delta_3} &=  {(-1)^{k_a} \over k_a!}
\left( \Delta_{a2,3} \right)_{k_a}
\left(\Delta_{a3,2}\right)_{k_a}
\left(\Delta_{23,a}\right)_{-k_a}
F_A^{(1)}\Big[\Delta_{a23,}-h;\big\{-k_a\big\};\big\{\Delta_a-h +1\big\};1\Big] \cr
 &= {1 \over k_a!} { (\Delta_{a2,3})_{k_a} (\Delta_{a3,2})_{k_a} \over (\Delta_a-h+1)_{k_a}} \,,
}
and
\eqn{d1Def}{
d_{k_a}^{\Delta_2;\Delta_a;\Delta_3} ={(\Delta_2)_{k_a} (\Delta_3)_{k_a} \over (\Delta_{23,a}+1)_{k_a} (\Delta_{a23,}-h+1)_{k_a} }\,,
}
with
\eqn{hDef}{
h\equiv \frac{d}{2}\,.
}
In writing the $c$-coefficient we have defined the Lauricella function $F_A^{(\ell)}$ of $\ell$ variables
\eqn{LauricellaDef}
{
F_A^{(\ell)}\Big[g;\{a_1,\ldots,a_\ell\};\{b_1,\ldots,b_\ell\};x_1,\ldots,x_\ell\Big]
\equiv\left[\prod_{i=1}^\ell\sum_{n_i=0}^\infty \right](g)_{\sum_{i=1}^\ell n_i}\prod_{i=1}^\ell \frac{(a_i)_{n_i}}{(b_i)_{n_i}}\frac{x_i^{n_i}}{n_i!}\,.
}
Note that the notation $(a)_n \equiv \Gamma(a+n)/\Gamma(a)$ stands for the Pochhammer symbol.

 We therefore need to evaluate the following integral:
\eqn{J}{
J
\equiv
&
\int_{z\in\text{AdS}}\,\hat{K}_{\Delta_2}(x_2,z)\hat{K}_{\Delta_3}(x_3,z)\hat{G}_{\Delta_a}(w_a,z)
\cr
=&
\sum_{\ell=0}^\infty
\frac{4^{\ell}\left(\frac{\Delta_a}{2}\right)_\ell\left(\frac{\Delta_a+1}{2}\right)_\ell}{\ell!\left(\Delta_a-h+1\right)_\ell}
\int_{\text{AdS}}dZ\,
\frac{1}{(-2P_2\cdot Z)^{\Delta_2}}
\frac{1}{(-2P_3\cdot Z)^{\Delta_3}}
\frac{1}{(-2W_a\cdot Z)^{\Delta_a+2\ell}}\,,
}
where in the second line above we have employed the embedding space formalism.
Applying the Schwinger parametrization
\eqn{SchwingerReal}{
\frac{1}{x^\Delta}=\frac{1}{\Gamma(\Delta)}\int_0^\infty
\frac{ds}{s}s^\Delta e^{-2sx}
}
to the three power law factors and using the AdS integral identity~\cite{Penedones:2010ue}
\eqn{}{
\int_{\text{AdS}}dZ\, e^{2Z\cdot Q}=\pi^h\int_0^\infty\frac{dz}{z} z^{-h}e^{-z+\frac{Q^2}{z}}\,,
}
one obtains the equation
\eqn{}{
J=&
\pi^h
\sum_{\ell=0}^\infty
\frac{4^{\ell}\left(\frac{\Delta_a}{2}\right)_\ell\left(\frac{\Delta_a+1}{2}\right)_\ell}{\ell!\left(\Delta_a-h+1\right)_\ell}
\int_0^\infty \frac{ds_2ds_3ds_a}{s_2s_3s_a}
\frac{s_2^{\Delta_2}s_3^{\Delta_3}s_a^{\Delta_a+2\ell}}{\Gamma(\Delta_2)\Gamma(\Delta_3)\Gamma(\Delta_a+2\ell)}
\cr
&
\times
\int_0^\infty
\frac{dz}{{z^{1+h}}}
e^{-z+{1\over z}\left( 2s_2s_3P_2\cdot P_3+2s_2s_aP_2\cdot W_a+2s_3s_aP_3\cdot W_a-s_a^2\right)}\,.
}
Interchanging the order of integrals and changing from variables $s_2$, $s_3$, and $s_a$ to $t_a\equiv\frac{s_2s_3}{z}$, $t_2\equiv\frac{s_as_3}{z}$, and $t_3\equiv\frac{s_as_2}{z}$, one rewrites
\eqn{}{
J=&
\frac{\pi^h}{2\Gamma(\Delta_1)\Gamma(\Delta_2)}
\sum_{\ell=0}^\infty
\frac{4^{\ell}\left(\frac{\Delta_a}{2}\right)_\ell\left(\frac{\Delta_a+1}{2}\right)_\ell}{\ell!\left(\Delta_a-h+1\right)_\ell\Gamma(\Delta_a+2\ell)}
\int_0^\infty
\frac{dz}{z}
z^{\frac{\Delta_{1}+\Delta_{2}+\Delta_{a}-d}{2}+\ell}
e^{-z}
\,
I_t\,
}
where we have introduced the definition
\eqn{}{
I_t\equiv
\int_0^\infty \frac{dt_2dt_3dt_a}{t_2t_3t_a}
t_2^{\Delta_{3a,2}+\ell}
t_3^{\Delta_{2a,3}+\ell}
t_a^{\Delta_{23,a}-\ell}
e^{2t_aP_2\cdot P_3+2t_3P_2\cdot W_a+2t_2P_3\cdot W_a-\frac{t_2t_3}{t_a}}\,.
}
Carrying out the $z$ integral using~\eno{SchwingerReal} and using the fact that
\eqn{}{
\left(\frac{\Delta}{2}\right)_\ell\left(\frac{\Delta+1}{2}\right)_\ell=\frac{1}{4^\ell}(\Delta)_{2\ell}\,,
}
one arrives at
\eqn{I}{
J=&
\frac{\pi^h}{2\Gamma(\Delta_2)\Gamma(\Delta_3)\Gamma(\Delta_a)}
\sum_{\ell=0}^\infty
\frac{\Gamma\left(\Delta_{a23,}-h+\ell\right)}{\ell!\left(\Delta_a-h+1\right)_\ell}
\,
I_t\,.
}

Now let's turn to evaluating $I_t$.
Using the Mellin representation,
\eqn{MellinRep}{
e^{-x}=\int_{\epsilon-i\infty}^{\epsilon+i\infty}\frac{dc}{2\pi i}\frac{\Gamma(c)}{x^c}\,,
}
for $\exp(-t_2 t_3/t_a)$, where $\epsilon$ is a small positive number, followed by changing the order of integration, and subsequently carrying out the $t_i$ integrals ($i=a,2,3$) leads to
\eqn{}{
I_t=
\int_{\epsilon-i\infty}^{\epsilon+i\infty}
\frac{dc}{2\pi i}\Gamma(c)
\frac{\Gamma\left(\Delta_{23,a}-\ell+c\right)
\Gamma\left(\Delta_{2a,3}+\ell-c\right)
\Gamma\left(\Delta_{3a,2}+\ell-c\right)}{
(-2P_2\cdot P_3)^{\Delta_{23,a}-\ell+c}
(-2P_2\cdot W_a)^{\Delta_{2a,3}+\ell-c}
(-2P_3\cdot W_a)^{\Delta_{3a,2}+\ell-c}
}
\,.
}
The remaining $c$ contour integral is straightforward to evaluate.
Closing the contour to the left and summing up the residues at the enclosed poles, at $c=-m$ and $c=\ell-\Delta_{12,a}-m$ where $m\in \mathbb{N}_0$ (i.e.\ the set of natural numbers including zero), we obtain
\eqn{}{
J=J_1+J_2
}
where we have split $J$ into two parts,
\eqn{}{
J_1 &\equiv
\frac{\pi^h}{2\Gamma(\Delta_2)\Gamma(\Delta_3)\Gamma(\Delta_a)}
\sum_{\ell=0}^\infty
\frac{\Gamma\left(\Delta_{23a,}-h+\ell\right)}{\ell!\left(\Delta_a-h+1\right)_\ell}
\,
2\pi i\sum_{m=0}^\infty  \,\underset{c=-m}{\text{Res}}
I_t\
\cr
J_2 &\equiv
\frac{\pi^h}{2\Gamma(\Delta_2)\Gamma(\Delta_3)\Gamma(\Delta_a)}
\sum_{\ell=0}^\infty
\frac{\Gamma\left(\Delta_{23a,}-h+\ell\right)}{\ell!\left(\Delta_a-h+1\right)_\ell}
\,
2\pi i\sum_{m=0}^\infty  \,\underset{c=\ell-\Delta_{12,a}-m}{\text{Res}}
I_t\,.
}
For the first part of $J$, we note that
\eqn{}{
2\pi i  \,\underset{c=-m}{\text{Res}} I_t
=
\frac{(-1)^m}{m!}
\frac{\Gamma\left(\Delta_{23,a}-\ell-m\right)
\Gamma\left(\Delta_{2a,3}+\ell+m\right)
\Gamma\left(\Delta_{3a,2}+\ell+m\right)}{
(-2P_2\cdot P_3)^{\Delta_{23,a}-\ell-m}
(-2P_2\cdot W_a)^{\Delta_{2a,3}+\ell+m}
(-2P_3\cdot W_a)^{\Delta_{3a,2}+\ell+m}
}\,.
}
Now change the summation variable in the expression for $J_1$ from $m$ to $S\equiv m+\ell$ and then change the order of $s$ and $\ell$ summations.
Accordingly, one must change the limits of the individual sums, so that now the $S$ sum runs from zero to infinity and the $\ell$ sum runs from zero to $S$.
But the terms with $\ell>S$ each vanish due to the factor of $(S-\ell)!$ in the denominator. Thus we can freely extend the upper limit of the $S$ sum to infinity, to write
\eqn{}{
J_1=\,
&
\frac{\pi^h}{2\Gamma(\Delta_2)\Gamma(\Delta_3)\Gamma(\Delta_a)}
\sum_{S=0}^\infty
\frac{(-1)^S\Gamma\left(\Delta_{23,a}-S\right)
\Gamma\left(\Delta_{2a,3}+S\right)
\Gamma\left(\Delta_{3a,2}+S\right)}{
(-2P_2\cdot P_3)^{\Delta_{23,a}-S}
(-2P_2\cdot W_a)^{\Delta_{2a,3}+S}
(-2P_3\cdot W_a)^{\Delta_{3a,2}+S}
}
\cr
&
\times
\sum_{\ell=0}^\infty
\frac{(-1)^{\ell}}{(S-\ell)!}
\frac{\Gamma\left(\Delta_{a23,}-h+\ell\right)}{\ell!\left(\Delta_a-h+1\right)_\ell}\,.
}
Using
\eqn{}{
(-1)^S\Gamma(\Delta-S)
=\frac{\Gamma(\Delta)}{(1+\Delta)_S}\,,
}
as well as the result
\eqn{}{
\sum_{\ell=0}^\infty
\frac{(-1)^\ell\Gamma(\Delta_{a23,}-h+\ell)}{(S-\ell)!\,\ell!\,(\Delta_a-h+1)_\ell}
&=
\frac{\Gamma(\Delta_{a23,}-h)}{S!}
\sum_{\ell=0}^\infty
\frac{(-S)_\ell(\Delta_{a23,}-h)_\ell}{\ell!(\Delta_a-h+1)_\ell}
\cr
&=
\frac{\Gamma(\Delta_{a23,}-h)}{S!}
\frac{(1+\Delta_{a,23})_S}{(\Delta_a-h+1)_S}\,,
}
one simplifies $J_1$ to
\eqn{J1}{
J_1 &=
\frac{C_{\Delta_a\Delta_2\Delta_3} }
{(-2P_2\cdot P_3)^{\Delta_{23,a}}
(-2P_2\cdot W_a)^{\Delta_{2a,3}}
(-2P_3\cdot W_a)^{\Delta_{3a,2}}}
\cr
&\quad \times
\sum_{S=0}^\infty
\frac{
(\Delta_{2a,3})_{S}
(\Delta_{3a,2})_{S}}{
S!(\Delta_a-h+1)_S
}
\left[\frac{(-2P_2\cdot P_3)}{(-2P_2\cdot W_a)
(-2P_3\cdot W_a)}
\right]^{S},
}
where the OPE coefficient $C_{ijk}$ was written in~\eno{OPEreal}.

For the second part of $J$, $J_2$, we note that
\eqn{}{
2\pi i \,\underset{c=\ell-\Delta_{12,a}-m}{\text{Res}}
I_t
=
\frac{(-1)^m}{m!}
\frac{\Gamma(\Delta_{a,23}+\ell-m)
\Gamma\left(\Delta_2+m\right)
\Gamma\left(\Delta_3+m\right)}{
(-2P_2\cdot P_3)^{-m}
(-2P_2\cdot W_a)^{\Delta_2+m}
(-2P_3\cdot W_a)^{\Delta_3+m}
}
\,.
}
Changing the order of summation in $J_2$ and carrying out the sum over $\ell$ first, one finds that
\eqn{J2}{
J_2
=\,&
\frac{\pi^h}{2}
\frac{\Gamma(\Delta_{a23,}-h)\Gamma(\Delta_{a,23})\Gamma(\Delta_a-h+1)}{\Gamma(\Delta_a)\Gamma(\Delta_{a,23}+1)
\Gamma(\Delta_{a23,}-h+1)
}
\frac{1}{(-2P_2\cdot W_a)^{\Delta_2}
(-2P_3\cdot W_a)^{\Delta_3}}
\cr
&
\times \sum_{m=0}^\infty
\frac{\left(\Delta_2\right)_m
\left(\Delta_3\right)_m}
{\left(1-\Delta_{a,23}\right)_m
(\Delta_{a23,}-h+1)_m
}
\left[
\frac{
(-2P_2\cdot P_3)}{
(-2P_2\cdot W_a)
(-2P_3\cdot W_a)
}\right]^m.
}
Part of the prefactor in~\eno{J2} may be reexpressed in terms of the mass of the bulk scalar
\eqn{}{
&\frac{\pi^h}{2}
\frac{\Gamma(\Delta_{a23,}-h)\Gamma(\Delta_{a,23})\Gamma(\Delta_a-h+1)}{\Gamma(\Delta_a)\Gamma(\Delta_{a,23}+1)
\Gamma(\Delta_{a23,}-h+1)
}
=
\frac{1}{N_{\Delta_a}(m_{\Delta_{2}+\Delta_3}^2-m^2_{\Delta_a})}\,.
}
Further, the position dependent power law factors in~\eno{J1} and~\eno{J2} can be written in terms of the bulk-to-boundary propagator using~\eno{KEmbedding}.
With these replacements, adding up~\eno{J1} and~\eno{J2} and substituting Poincar\'e coordinates one recovers~\eno{KKGreal2}, completing the proof.

Finally, we note that in contour integral form, this identity takes the form
\eqn{}{
\int_{z\in {\rm AdS}} &\hat{G}_{\Delta_a}(w_a,z)  \hat{K}_{\Delta_2}(x_2,z) \hat{K}_{\Delta_3}(x_3,z) \cr
&=
C_{\Delta_a\Delta_2\Delta_3}
\sum_{k_a=0}^\infty
\frac{\left(\Delta_{a23,}-h\right)_{k_a}}{k_a!\left(\Delta_a-h+1\right)_{k_a}} \int_{\epsilon-i\infty}^{\epsilon+i\infty}
\frac{dc}{2\pi i}\;\Gamma(c) \cr
&\quad \times
\left(\Delta_{a2,3}\right)_{k_a-c}
\left(\Delta_{a3,2}\right)_{k_a-c}
 \left(\Delta_{23,a}\right)_{-k_a+c}
\frac{ \hat{K}_{\Delta_{a2,3}+k_a-c}(x_2,w_a)
\hat{K}_{\Delta_{a3,2}+k_a-c}(x_3,w_a)}{
(x_{23}^2)^{\Delta_{23,a}-k_a+c}}\,.
}


\subsubsection{$\int \hat{K}\hat{G}\hat{G}$}
\label{KGGPROOF}

In this subsection we derive \eno{KGGreal}, reproduced below,
\eqn{KGGreal2}{
& \int_{z \in {\rm AdS}} \hat{K}_{\Delta_3}(x_3,z)\hat{G}_{\Delta_a}(w_a,z)\hat{G}_{\Delta_b}(w_b,z) \cr
&= C_{\Delta_3\Delta_a\Delta_b} \sum_{k_a,k_b=0}^\infty c_{k_a;k_b}^{\Delta_a;\Delta_3;\Delta_b}
 \hat{K}_{\Delta_{3a,b}+k_a-k_b}(x_3,w_a)  \hat{K}_{\Delta_{3b,a}-k_a+k_b}(x_3,w_b)
 \left({\xi(w_a,w_b) \over 2}\right)^{\Delta_{ab,3}+k_a+k_b} \cr
 & + \left(
\sum_{k_a,k_b=0}^\infty
d_{k_a;k_b}^{\Delta_a;\Delta_3;\Delta_b}
 \hat{K}_{\Delta_{3}+k_a}(x_3,w_a)  \hat{K}_{-k_a}(x_3,w_b)
 \left({\xi(w_a,w_b) \over 2}\right)^{\Delta_{b}+2k_b+k_a} + (a\leftrightarrow b)\right),
}
where\footnote{The coefficient $c_{k_a;k_b}^{\Delta_a;\Delta_3;\Delta_b}$ was originally written in Ref.~\cite{Parikh:2019ygo} in terms of a hypergeometric ${}_3F_2$ function (see equation~(2.14) of Ref.~\cite{Parikh:2019ygo}), but using \cite[Equation 4.29]{Paulos:2011ie} we have re-expressed it in terms of the Lauricella function $F_A$ of two variables in~\eno{c2Def}, since this has natural analogs in the case of $\int \hat{K}\hat{K}\hat{G}$ and $\int \hat{G}\hat{G}\hat{G}$ identities in terms of Lauricella functions of one and three variables, respectively.}
\eqn{c2Def}{
c_{k_a;k_b}^{\Delta_a;\Delta_3;\Delta_b} &\equiv  {(-1)^{k_a+k_b} \over k_a!k_b!}
\left( \Delta_{3a,b} \right)_{k_a-k_b}
\left(\Delta_{ab,3}\right)_{k_a+k_b}
\left(\Delta_{3b,a}\right)_{-k_a+k_b} \cr
&\quad  \times  F_A^{(2)}\Big[\Delta_{ab3,}-h;\big\{-k_a,-k_b\big\};\big\{\Delta_a-h +1,\Delta_b-h +1\big\};1,1\Big],
 }
and
\eqn{d2Def}
{
d^{\Delta_a;\Delta_3;\Delta_b}_{k_a;k_b}=&
\frac{\pi^h\Gamma\left(\Delta_{3ab,}-h\right)}{2\Gamma(\Delta_a)}
\frac{(-1)^{k_a+k_b}}{k_a!k_b!}
\left(\Delta_3\right)_{k_a}
\left(\Delta_b\right)_{2k_b+k_a}
\Gamma(\Delta_{a,b3}-k_b-k_a)
\cr
&  \times
F_A^{(2)}\left[
\Delta_{3ab,}-h;\left\{\Delta_{a,b3}-k_b-k_a,-k_b\right\};\left\{\Delta_a-h+1,\Delta_b-h+1\right\};1,1
\right].
}
The Lauricella function $F_A^{(\ell)}$ was defined in~\eno{LauricellaDef}.
Name the left hand side of the identity to be proven~\eno{KGGreal2} ${\cal J}$,
\eqn{JJ}{
\mathcal{J}
\equiv
&
\int_{z\in\text{AdS}}\,\hat{K}_{\Delta_3}(x_3,z)\hat{G}_{\Delta_a}(w_a,z)\hat{G}_{\Delta_b}(w_b,z)\,.
}
The same intermediate steps that lead us from \eno{J} to \eno{I} allow us to recast \eno{JJ} as
\eqn{}{
\mathcal{J}=
\frac{\pi^h}{2\Gamma(\Delta_3)\Gamma(\Delta_a)\Gamma(\Delta_b)}
\sum_{\ell_a,\ell_b=0}^\infty
\frac{\Gamma\left(\Delta_{ab3,}-h+\ell_a+\ell_b\right)}{\ell_a!\left(\Delta_a-h+1\right)_{\ell_a}\ell_b!\left(\Delta_b-h+1\right)_{\ell_b}}
\mathcal{I}_t\,,
}
where we have introduced the definition
\eqn{}{
\mathcal{I}_t \equiv
\int_0^\infty \frac{dt_3dt_adt_b}{t_3t_at_b}
t_3^{\Delta_{ab,3}+\ell_a+\ell_b}
t_a^{\Delta_{b3,a}-\ell_a+\ell_b}
t_b^{\Delta_{a3,b}+\ell_a-\ell_b}
e^{2t_bP_3\cdot W_a+2t_aP_3\cdot W_b+2t_3W_a\cdot W_b-\frac{t_3t_b}{t_a}-\frac{t_3t_a}{t_b}}\,.
}
Applying the Mellin representation~\eno{MellinRep} for the factors $\exp(-t_3 t_b/t_a)$ and $\exp(-t_3 t_a/t_b)$, and carrying out the $t_i$ integrals (for $i=3,a,b$) we obtain
\eqn{}{
\mathcal{I}_t=&
\int_{\epsilon-i\infty}^{\epsilon+i\infty} \frac{dc_a}{2\pi i}
\,\Gamma(c_a)
\int_{\epsilon-i\infty}^{\epsilon+i\infty} \frac{dc_b}{2\pi i}
\,\Gamma(c_b)
\cr
&
\times
\frac{
\Gamma(\Delta_{a3,b}+\ell_a-\ell_b-c_a+c_b)
\,
\Gamma(\Delta_{b3,a}-\ell_a+\ell_b+c_a-c_b)
\,
\Gamma(\Delta_{ab,3}+\ell_a+\ell_b-c_a-c_b)}
{
(-2P_3\cdot W_a)^{\Delta_{a3,b}+\ell_a-\ell_b-c_a+c_b}
(-2P_3\cdot W_b)^{\Delta_{b3,a}-\ell_a+\ell_b+c_a-c_b}
(-2W_a\cdot W_b)^{\Delta_{ab,3}+\ell_a+\ell_b-c_a-c_b}
}
\cr
\equiv&
\int_{\epsilon-i\infty}^{\epsilon+i\infty} \frac{dc_a}{2\pi i}
\int_{\epsilon-i\infty}^{\epsilon+i\infty} \frac{dc_b}{2\pi i}
\,\widetilde{\mathcal{I}}_t
\,.
}
Focusing on the $c_a$ integral first, we close the contour to the left and pick up the residues of the poles at $c_a=-m_a$ and $c_a=\Delta_{a,b3}+\ell_a-\ell_b+c_b-m_a$ with $m_a \in \mathbb{N}_0$:
\eqn{II}{
&
\mathcal{I}_t=
\int_{\epsilon-i\infty}^{\epsilon+i\infty} \frac{dc_b}{2\pi i}
\,\Gamma(c_b)
\frac{(-1)^{m_a}}{m_a!}
\cr
&
\times\Bigg[
\frac{
\Gamma(\Delta_{a3,b}+\ell_a-\ell_b+m_a+c_b)
\,
\Gamma(\Delta_{b3,a}-\ell_a+\ell_b-m_a-c_b)
\,
\Gamma(\Delta_{ab,3}+l_a+l_b+m_a-c_b)}
{
(-2P_3\cdot W_a)^{\Delta_{a3,b}+\ell_a-\ell_b+m_a+c_b}
(-2P_3\cdot W_b)^{\Delta_{b3,a}-\ell_a+\ell_b-m_a-c_b}
(-2W_a\cdot W_b)^{\Delta_{ab,3}+\ell_a+\ell_b+m_a-c_b}
}
\cr
&
\hspace{6mm}
+
\frac{
\Gamma(\Delta_3+m_A)
\,
\Gamma(\Delta_{a,b3}+\ell_a-\ell_b-m_a+c_b)
\Gamma(\Delta_b+2\ell_b+m_a-2c_b)}
{
(-2P_3\cdot W_a)^{\Delta_3+m_a}
(-2P_3\cdot W_b)^{-m_a}
(-2W_a\cdot W_b)^{\Delta_b+2\ell_b+m_a-2c_b}
}
\Bigg]
\,.
}
We can carry out the $c_b$ integral by once again closing the contour to the left. In that case we must sum over the series of residues at $c_b=-m_b$ and $c_b=\Delta_{b,a3}-\ell_a+\ell_b-m_a-m_b$ in the first term in~\eno{II} for $m_b\in\mathbb{N}_0$. In the second term, besides the pole at $c_b = -m_b$, there are also poles at $c_b=\Delta_{b3,a}-\ell_a+\ell_b+m_a-m_b$ coming from the second gamma function in the second term, with residues
\eqn{residues1}{
\underset{\substack{c_a=\Delta_{a,b3}+\ell_a-\ell_b+c_b-m_a,\\c_b=\Delta_{b3,a}-\ell_a+\ell_b+m_a-m_b}}{\text{Res}}
\widetilde{\mathcal{I}}_t
&=
\frac{(-1)^{m_a+m_b}}{m_a!m_b!} \cr
& \times \frac{
\Gamma(\Delta_{b3,a}-\ell_a+\ell_b+m_a-m_b)
\,
\Gamma(\Delta_3+m_a)
\,
\Gamma(\Delta_{a,3}+2\ell_a-m_a+2m_b)}
{
(-2P_3\cdot W_a)^{\Delta_3+m_a}
(-2P_3\cdot W_b)^{-m_a}
(-2W_a\cdot W_b)^{2\Delta_{a,3}+2\ell_a-m_a+2m_b}
}.
}
But in the first term there is a semi-infinite sequence of poles precisely at $c_b=\Delta_{b3,a}-\ell_a+\ell_b-m_a+m_b$, which can lie on the left-half plane for appropriate values of $m_a,m_b$ and thus be enclosed inside the contour, with residues
\eqn{residues2}{
\underset{\substack{c_a=-m_a,\\c_b=\Delta_{b3,a}-\ell_a+\ell_b-m_a+m_b}}{\text{Res}}
\widetilde{\mathcal{I}}_t &=
-
\frac{(-1)^{m_a+m_b}}{m_a!m_b!} \cr
& \times \frac{
\Gamma(\Delta_{b3,a}-\ell_a+\ell_b-m_a+m_b)
\Gamma(\Delta_3+m_b)
\Gamma(\Delta_{a,3}+2\ell_a+2m_a-m_b)}
{
(-2P_3\cdot W_a)^{\Delta_3+m_b}
(-2P_3\cdot W_b)^{-m_b}
(-2W_a\cdot W_b)^{2\Delta_{a,3}+2\ell_a+2m_a-m_b}
}\,.
}
We notice that the above two types of residues occur at the same values of $c_b$ and are equal to minus one another, except with $m_a$ and $m_b$ interchanged.
Hence, on summing over all values of $m_a$ and $m_b$ the contributions to $\mathcal{J}$ from these two types of residues cancel.
With this observation, we decompose ${\cal J}$ into three parts,
\eqn{J123}
{
\mathcal{J}=\mathcal{J}_1+\mathcal{J}_2+\mathcal{J}_3
}
where $\mathcal{J}_i$ is defined to be
\eqn{}{
&\mathcal{J}_i
\equiv
\frac{\pi^h}{2\Gamma(\Delta_3)\Gamma(\Delta_a)\Gamma(\Delta_b)}
\sum_{\ell_a,\ell_b=0}^\infty
\frac{\Gamma\left(\Delta_{ab3,}-h+\ell_a+\ell_b\right)}{\ell_a!\left(\Delta_a-h+1\right)_{\ell_a}\ell_b!\left(\Delta_b-h+1\right)_{\ell_b}}
\sum_{m_a,m_b=0}^\infty  \,\mathcal{R}_i(m_a,m_b)
\,,
}
and we have further introduced the definitions
\eqn{}
{
&
\mathcal{R}_1(m_a,m_b)\equiv
\underset{\substack{c_a=-m_a,\\c_b=-m_b}}{\text{Res}}
\widetilde{\mathcal{I}}_t
\cr
&
\mathcal{R}_2(m_a,m_b)\equiv
\underset{\substack{c_a=\Delta_{a,b3}+\ell_a-\ell_b+c_b-m_a,\\c_b=-m_b}}{\text{Res}}
\widetilde{\mathcal{I}}_t
\cr
&
\mathcal{R}_3(m_a,m_b)\equiv
\underset{\substack{c_a=-m_a,\\c_b=\Delta_{b,a3}-\ell_a+\ell_b -m_a -m_b}}{\text{Res}}
\widetilde{\mathcal{I}}_t\,.
}
Together, this accounts for the total contribution from the residues at all poles picked up upon closing the two contours.
Evaluating ${\cal R}_1$, we get
\eqn{}{
\mathcal{R}_1(m_a,m_b) &=
\frac{(-1)^{m_a+m_b}}{m_a!m_b!} \cr
&\times \frac{
\Gamma(\Delta_{a3,b}+S_a-S_b)
\,\Gamma(\Delta_{b3,a}-S_a+S_b)
\,\Gamma(\Delta_{ab,3}+S_a+S_b)}
{
(-2P_3\cdot W_a)^{\Delta_{a3,b}+S_a-S_b}
(-2P_3\cdot W_b)^{\Delta_{b3,a}-S_a+S_b}
(-2W_a\cdot W_b)^{\Delta_{ab,3}+S_a+S_b}
}\,,
}
where we have defined $S_a\equiv m_a + \ell_a$ and $S_b \equiv m_b+\ell_b$.
Changing summation variables from $m_a$ and $m_b$ to $S_a$ and $S_b$ and changing the order of $\ell_i$ and $S_i$ summations, one finds that
\eqn{}{
\mathcal{J}_1
=\,&
\frac{\pi^h}{2\Gamma(\Delta_3)\Gamma(\Delta_a)\Gamma(\Delta_b)}
\cr
&
\times
\sum_{S_a,S_b=0}^\infty
\frac{
(-1)^{S_a+S_b}
\Gamma(\Delta_{a3,b}+S_a-S_b)
\,\Gamma(\Delta_{b3,a}-S_a+S_b)
\,\Gamma(\Delta_{ab,3}+S_a+S_b)}
{
(-2P_3\cdot W_a)^{\Delta_{a3,b}+S_a-S_b}
(-2P_3\cdot W_b)^{\Delta_{b3,a}-S_a+S_b}
(-2W_a\cdot W_b)^{\Delta_{ab,3}+S_a+S_b}
}
\cr
&
\times
\sum_{\ell_a,\ell_b=0}^\infty
\frac{(-1)^{\ell_a+\ell_b}}{(S_a-\ell_a)!(S_b-\ell_b)!}
\frac{\Gamma\left(\Delta_{ab3,}-h+\ell_a+\ell_b\right)}{\ell_a!\left(\Delta_a-h+1\right)_{\ell_a}\ell_b!\left(\Delta_b-h+1\right)_{\ell_b}}
\,.
}
Twice applying the trivial identity
\eqn{}{
\frac{(-1)^\ell}{(S-\ell)!}=\frac{(-S)_\ell}{S!}\,,
}
the sum over $\ell_a$ and $\ell_b$ is seen to be precisely in the series representation of the Lauricella function $F_A^{(2)}$,
\eqn{}{
F_A^{(2)}\Big[a;\{b,c\};\{d,e\};1,1\Big]=
\sum_{\ell,\ell'=0}^\infty \frac{(a)_{\ell+\ell'}}{\ell!\,\ell'!}\frac{(b)_\ell(c)_{\ell'}}{(d)_{\ell}(e)_{\ell'}}\,.
}
Consequently, we can write
\eqn{JJ1}{
\mathcal{J}_1=&\frac{\pi^h}{2}\frac{\Gamma(\Delta_{ab3,}-h)}{\Gamma(\Delta_3)\Gamma(\Delta_a)\Gamma(\Delta_b)}
\sum_{S_a,S_b=0}^\infty\frac{(-1)^{S_a+S_b}}{S_a!S_b!}
\cr
&
\times
F^{(2)}_A\Big[\Delta_{ab3,}-h;\{-S_a,-S_b\};\left\{\Delta_a-h+1,\Delta_b-h+1\right\};1,1\Big]
\cr
&
\times
\frac{
\Gamma(\Delta_{a3,b}+S_a-S_b)
\,\Gamma(\Delta_{b3,a}-S_a+S_b)
\,\Gamma(\Delta_{ab,3}+S_a+S_b)}
{
(-2P_3\cdot W_a)^{\Delta_{a3,b}+S_a-S_b}
(-2P_3\cdot W_b)^{\Delta_{b3,a}-S_a+S_b}
(-2W_a\cdot W_b)^{\Delta_{ab,3}+S_a+S_b}
}\,.
}
For the second part of $\mathcal{J}$, we note that
\eqn{}{
\mathcal{R}_2(m_a,m_b)=
\frac{(-1)^{m_a+m_b}}{m_a!m_b!}
\frac{
\Gamma(\Delta_3+m_b)
\Gamma(\Delta_{a,b3}-S_b-m_a+\ell_a)
\Gamma(\Delta_b+2S_b+m_a)
}
{
(-2P_3\cdot W_a)^{\Delta_3+m_a}
(-2P_3\cdot W_b)^{-m_a}
(-2W_a\cdot W_b)^{\Delta_b+2S_b+m_a}
}\,.
}
Therefore, changing summation variable from $m_b$ to $S_b$ and changing the order of summation like we did for ${\cal J}_1$, we have that
\eqn{}{
\mathcal{J}_2=\, &
\frac{\pi^h}{2\Gamma(\Delta_3)\Gamma(\Delta_a)\Gamma(\Delta_b)} \cr
& \times \sum_{m_a,S_b=0}^\infty
\frac{(-1)^{m_a+S_b}}{m_a!S_b!}
\frac{
\Gamma(\Delta_3+m_a)
\Gamma(\Delta_b+2S_b+m_a)}
{
(-2P_3\cdot W_a)^{\Delta_3+m_a}
(-2P_3\cdot W_b)^{-m_a}
(-2W_a\cdot W_b)^{\Delta_b+2S_b+m_a}
}
\cr
&
\times
\sum_{\ell_a,\ell_b=0}^\infty
\frac{(-S_b)_{\ell_b}
\,\Gamma(\Delta_{a,b3}-S_b-m_a+\ell_a)
\,\Gamma\left(\Delta_{ab3,}-h+\ell_a+\ell_b\right)}
{\ell_a!\ell_b!\left(\Delta_a-h+1\right)_{\ell_a}\left(\Delta_b-h+1\right)_{\ell_b}}\,.
}
We can recast this expression for $\mathcal{J}_2$ into the following:
\eqn{JJ2}{
\mathcal{J}_2=\,&
\frac{\pi^h\Gamma(\Delta_{ab3,}-h)}{2\Gamma(\Delta_a)} \cr
&\times \sum_{m_a,S_b=0}^\infty
\frac{(-1)^{m_a+S_b}}{m_a!S_b!}
\frac{
\left(\Delta_3\right)_{m_a}
\left(\Delta_b\right)_{2S_b+m_a}
\Gamma(\Delta_{a,b3}-S_b-m_a)
}
{
(-2P_3\cdot W_a)^{\Delta_3+m_a}
(-2P_3\cdot W_b)^{-m_a}
(-2W_a\cdot W_b)^{\Delta_b+2S_b+m_a}
}
\cr
&
\times
F_A^{(2)}\Big[
\Delta_{ab3,}-h;\left\{\Delta_{a,b1}-S_b-m_a,-S_b\right\};\left\{\Delta_a-h+1,\Delta_b-h+1\right\};1,1
\Big].
}
As for the third term in \eno{J123}, $\mathcal{J}_3$, it  differs from $\mathcal{J}_2$ only by a simple relabeling:
\eqn{}{
\mathcal{J}_3
=\,
\mathcal{J}_2\big|_{a\leftrightarrow b}
\,.
}
Consequently, using \eno{JJ1} and \eno{JJ2} and adding together $\mathcal{J}_1$, $\mathcal{J}_2$, and $\mathcal{J}_3$, we obtain \eno{KGGreal2}.


\subsubsection{$\int \hat{G}\hat{G}\hat{G}$}
\label{GGGPROOF}

In this section we derive \eno{GGGreal}, reproduced below,
\eqn{GGGreal2}{
& \int_{z \in {\rm AdS}} \hat{G}_{\Delta_a}(w_a,z)\hat{G}_{\Delta_b}(w_b,z)\hat{G}_{\Delta_c}(w_c,z) \cr
&= C_{\Delta_a\Delta_b\Delta_c} \!\!\! \sum_{k_a,k_b,k_c=0}^\infty c^{\Delta_a;\Delta_b;\Delta_c}_{k_a;k_b;k_c} \!
 \left({\xi(w_a,w_c)\over 2}\right)^{\Delta_{ac,b}+k_{ac,b}}  \!
 \left({\xi(w_a,w_b) \over 2}\right)^{\Delta_{ab,c}+k_{ab,c}} \!
 \left({\xi(w_b,w_c) \over 2}\right)^{\Delta_{bc,a}+k_{bc,a}} \cr
 &\quad + \bigg(
 \sum_{k_a,k_b,k_c=0}^\infty d^{\Delta_a;\Delta_b;\Delta_c}_{k_a;k_b;k_c}
 \left({\xi(w_a,w_c)\over 2}\right)^{\Delta_c+2k_c+k_a}
 \left({\xi(w_a,w_b) \over 2}\right)^{\Delta_b+2k_b+k_a}
 \left({\xi(w_b,w_c) \over 2}\right)^{-k_a}
 \cr
 &\hspace{15mm}+(a\leftrightarrow b)+(a\leftrightarrow c)\bigg),
}
where
\eqn{c3Def}{
c^{\Delta_a;\Delta_b;\Delta_c}_{k_a;k_b;k_c} &=   {(-1)^{k_a+k_b+k_c} \over k_a!k_b!k_c!}
\left( \Delta_{ac,b} \right)_{k_{ac,b}}
\left(\Delta_{ab,c}\right)_{k_{ab,c}}
\left(\Delta_{bc,a}\right)_{k_{bc,a}}  \cr
 &   \times F_A^{(3)}\Big[\Delta_{abc,}-h;\big\{-k_a,-k_b,-k_c\big\};\big\{\Delta_a-h +1,\Delta_b-h +1,\Delta_c-h +1\big\};1,1,1\Big],
}
and
\eqn{d3Def}
{d_{k_a;k_b;k_c}^{\Delta_a;\Delta_b;\Delta_c}
&=
\frac{(-1)^{k_{abc,}}}{k_a!k_b!k_c!}
(\Delta_b)_{2k_b+k_a}
(\Delta_c)_{2k_c+k_a}
\Gamma(\Delta_{a,bc}-k_{abc,})
\frac{\pi^h}{2}
\frac{\Gamma(\Delta_{abc,}-h)}{\Gamma(\Delta_a)}
\cr
\times
F_A^{(3)}&\Big[
\Delta_{abc,}-h;
\big\{\Delta_{a,bc}-k_{abc,},-k_b,-k_c\big\};
\big\{\Delta_a-h+1,\Delta_b-h+1,\Delta_c-h+1\big\};
1,1,1
\Big].
}
Here in writing $k_{ab,c}$ etc., we are using the shorthand~\eno{kabc} --- note the factor of two difference with the corresponding notation for conformal dimensions~\eno{DeltaijkDef}.

We therefore turn our attention to the integral
\eqn{}
{
\mathfrak{J}
\equiv
&
\int_{Z\in\text{AdS}}\,\hat{G}_{\Delta_a}(W_a,Z)\hat{G}_{\Delta_b}(W_b,Z)\hat{G}_{\Delta_c}(W_c,Z)\,,
}
in embedding space coordinates.
Using the same steps which led us from \eno{J} to \eno{I}, we arrive this time at
\eqn{}{
\mathfrak{J}=&
\frac{\pi^h}{2\Gamma(\Delta_a)\Gamma(\Delta_b)\Gamma(\Delta_c)}
\sum_{\ell_a,\ell_b,\ell_c=0}^\infty
\frac{\Gamma(\Delta_{abc,}-h+\ell_a+\ell_b+\ell_c)}{
\ell_a!\left(\Delta_a-h+1\right)_{\ell_a}
\ell_b!\left(\Delta_b-h+1\right)_{\ell_b}
\ell_c!\left(\Delta_c-h+1\right)_{\ell_c}
}
\,
\mathfrak{I}_t\,,
}
where we have defined
\eqn{}{
\mathfrak{I}_t\equiv
\int_0^\infty \frac{dt_adt_bdt_c}{t_at_bt_c}
&
t_a^{\Delta_{bc,a}-\ell_a+\ell_b+\ell_c}
t_b^{\Delta_{ac,b}+\ell_a-\ell_b+\ell_c}
t_c^{\Delta_{ab,c}+\ell_a+\ell_b-\ell_c}
\cr
&
\times e^{2t_a W_b\cdot W_c+2t_bW_a\cdot W_c+2t_cW_a\cdot W_b-\frac{t_bt_c}{t_a}-\frac{t_at_c}{t_b}-\frac{t_at_b}{t_c}}\,.
}
Applying the Mellin representation three times, and carrying out the $t_i$ integrals for $i=a,b,c$, we get
\eqn{}{
\mathfrak{I}_t=&
\int_{\epsilon-i\infty}^{\epsilon+i\infty} \frac{dc_a}{2\pi i}
\,\Gamma(c_a)
\!\! \int_{\epsilon-i\infty}^{\epsilon+i\infty} \frac{dc_b}{2\pi i}
\,\Gamma(c_b)
\!\! \int_{\epsilon-i\infty}^{\epsilon+i\infty} \frac{dc_c}{2\pi i}
\,\Gamma(c_c)
\frac{
\Gamma(\Delta_{ab,c}+\ell_a+\ell_b-\ell_c-c_a-c_b+c_c)
}
{(-2W_a\cdot W_b)^{\Delta_{ab,c}+\ell_a+\ell_b-\ell_c-c_a-c_b+c_c}}
\cr
&\hspace{83.5mm}
\times \frac{
\Gamma(\Delta_{ac,b}+\ell_a-\ell_b+\ell_c-c_a+c_b-c_c)
}
{(-2W_a\cdot W_c)^{\Delta_{ac,b}+\ell_a-\ell_b+\ell_c-c_a+c_b-c_c}}
\cr
&\hspace{83.5mm}
\times \frac{
\Gamma(\Delta_{bc,a}-\ell_a+\ell_b+\ell_c+c_a-c_b-c_c)
}
{(-2W_b\cdot W_c)^{\Delta_{bc,a}-\ell_a+\ell_b+\ell_c+c_a-c_b-c_c}}
\cr
\equiv
&
\int_{\epsilon-i\infty}^{\epsilon+i\infty} \frac{dc_a}{2\pi i}
\int_{\epsilon-i\infty}^{\epsilon+i\infty} \frac{dc_b}{2\pi i}
\int_{\epsilon-i\infty}^{\epsilon+i\infty} \frac{dc_c}{2\pi i}
\,
\widetilde{\mathfrak{I}}_t
\,.
}
Performing the $c_a$ integral by closing the contour on the left and summing over the semi-infinite sequence of residues at $c_a=-m_a$ and $c_a=\Delta_{a,bc}+\ell_a-\ell_b-\ell_c+c_b+c_c-m_a$ (for $m_a \in \mathbb{N}_0$), and then subsequently carrying out the $c_b$ integral by closing the contour on the left and summing over the sequence of residues in the $c_b$-plane at $c_b=-m_b$ and $c_b=\Delta_{bc,a}-\ell_a+\ell_b+\ell_c+m_a-m_b$ (for $m_b \in \mathbb{N}_0$), one finds that
\eqn{ccIntegral}{
&
\mathfrak{I}_t=
\int_{\epsilon-i\infty}^{\epsilon+i\infty} \frac{dc_c}{2\pi i}
\Gamma(c_c)
\frac{(-1)^{m_a+m_b}}{m_a!m_a!}
\cr
&
\times
\Bigg[
\frac{
\Gamma(\Delta_{ac,b}+S_a-S_b+\ell_c-c_c)
\Gamma(\Delta_{bc,a}-S_a+S_b+\ell_c-c_c)
\Gamma(\Delta_{ab,c}+S_a+S_b-\ell_c+c_c)
}
{
(-2W_c\cdot W_a)^{\Delta_{ac,b}+S_a-S_b+\ell_c-c_c}
(-2W_c\cdot W_b)^{\Delta_{bc,a}-S_a+S_b+\ell_c-c_c}
(-2W_a\cdot W_b)^{\Delta_{ab,c}+S_a+S_b-\ell_c+c_c}
}
\cr
&
\qquad
+\frac{
\Gamma(\Delta_c+2\ell_c-2c_c+m_a)
\Gamma(\Delta_{a,bc}-S_b-m_a+\ell_a-\ell_c+c_c)
\Gamma(\Delta_b+2S_b+m_a)
}
{
(-2W_c\cdot W_a)^{\Delta_c+2\ell_c-2c_c+m_a}
(-2W_c\cdot W_b)^{-m_a}
(-2W_a\cdot W_b)^{\Delta_b+2S_b+m_a}
}
\cr
&
\qquad
+\frac{
\Gamma(\Delta_c+2\ell_c-2c_c+m_b)
\Gamma(\Delta_{b,ac}-S_a-m_b+\ell_b-\ell_c+c_c)
\Gamma(\Delta_a+2S_a+m_b)
}
{
(-2W_c\cdot W_b)^{\Delta_c+2\ell_c-2c_c+m_b}
(-2W_c\cdot W_a)^{-m_b}
(-2W_a\cdot W_b)^{\Delta_a+2S_a+m_b}
}
\Bigg]
\,.
}
Here like before, we are using the definitions $S_a = m_a + \ell_a$ and $S_b = m_b  + \ell_b$.
The last contour integral can be carried out by closing the contour on the left as well. In the first term, we sum over the semi-infinite sequence of residues at $c_c=-m_c$ and $c_c=\Delta_{c,ab}-S_a-S_b+\ell_c-m_c$ with $m_c\in \mathbb{N}_0$. We will return to the spurious poles in the first term shortly. The second term has poles at $c_c = -m_c$ and
$c_c=\Delta_{bc,a}+S_b+m_a-\ell_a+\ell_c-m_c$, with the residue in the latter case given by
\eqn{}
{
&
\underset{\substack{c_a=\Delta_{a,bc}+\ell_a-\ell_b-\ell_c+c_b+c_c-m_a,\\c_b=-m_b,\\c_c=\Delta_{bc,a}+S_b+m_a-\ell_a+\ell_c-m_c}}{\text{Res}}\widetilde{\mathfrak{I}}_t
=\frac{(-1)^{m_a+m_b+m_c}}{m_a!m_b!m_c!}
\cr
&
\times
\frac{
\Gamma(2\Delta_{a,b}-2S_b-m_a+2\ell_a+2m_c)
\Gamma(\Delta_{bc,a}+S_b+m_a-\ell_a+\ell_b)
\Gamma(\Delta_b+2S_b+m_a)
}
{
(-2W_c\cdot W_a)^{2\Delta_{a,b}-2S_b-m_a+2\ell_a+2m_c}
(-2W_c\cdot W_b)^{-m_a}
(-2W_a\cdot W_b)^{\Delta_b+2S_b+m_a}
}\,.
}
But there are also residues at $c_c=\Delta_{bc,a}-S_a+S_b+\ell_c+m_c$ on account of the second factor of gamma function in the first term in~\eno{ccIntegral}:
\eqn{}
{
&
\underset{\substack{c_a=-m_a,\\c_b=-m_b,\\c_c=\Delta_{bc,a}-S_a+S_b+\ell_c+m_c}}{\text{Res}}\widetilde{\mathfrak{I}}_t
=
-
\frac{(-1)^{m_a+m_b+m_c}}{m_a!m_b!m_c!}
\cr
&
\hspace{20mm}
\times
\frac{
\Gamma(2\Delta_{a,b}+2S_a-2S_b-m_c)
\Gamma(\Delta_{bc,a}-S_a+S_b+\ell_c+m_c)
\Gamma(\Delta_{b}+2S_b+m_c)
}
{
(-2W_c\cdot W_a)^{2\Delta_{a,b}+2S_a-2S_b-m_c}
(-2W_c\cdot W_b)^{-m_c}
(-2W_a\cdot W_b)^{\Delta_{ab,c}+S_a+S_b-\ell_c+c_c}
}\,.
}
The above two types of residues occur at the same values of $c_c$ and are equal to minus one another, except with $m_a$ and $m_c$ interchanged. Hence, on summing over $m_a$ and $m_c$, these residues cancel. Similarly, the residues at $c_c=\Delta_{ac,b}+S_a+m_b-\ell_b+\ell_c-m_c$ due to the second gamma function in the third term on the RHS of \eno{ccIntegral} cancel with the residues at $c_c=\Delta_{ac,b}+S_a-S_b+\ell_c+m_c$ due to the first gamma function in the first term. Thus accounting for the cancellations of residues from spurious poles, we decompose $\mathfrak{J}$ as follows:
\eqn{J1234}
{
\mathfrak{J}=\mathfrak{J}_1+\mathfrak{J}_2+\mathfrak{J}_3+\mathfrak{J}_4\,,
}
where the four parts $\mathfrak{J}_i$ are given by
\eqn{}{
\mathfrak{J}_i &=
\frac{\pi^h}{2\Gamma(\Delta_a)\Gamma(\Delta_b)\Gamma(\Delta_c)}
\sum_{\ell_a,\ell_b,\ell_c=0}^\infty
\frac{\Gamma(\Delta_{abc,}-h+\ell_a+\ell_b+\ell_c)}{
\ell_a!\left(\Delta_a-h+1\right)_{\ell_a}
\ell_b!\left(\Delta_b-h+1\right)_{\ell_b}
\ell_c!\left(\Delta_c-h+1\right)_{\ell_c}
}
\cr
& \quad \times \sum_{m_a,m_b,m_c=0}^\infty
\mathfrak{R}_i(m_a,m_b,m_c)
\,,
}
and we have introduced the definitions
\eqn{}
{
&
\mathfrak{R}_1(m_a,m_b,m_c)\equiv
\underset{\substack{c_a=-m_a,\\c_b=-m_b,\\c_c=-m_c\textcolor{white}{,}}}{\text{Res}}
\widetilde{\mathfrak{I}}_t
\cr
&
\mathfrak{R}_2(m_a,m_b,m_c)\equiv
\underset{\substack{c_a = \Delta_{a,bc}+\ell_a-\ell_b-\ell_c+c_b+c_c-m_a,\\c_b=-m_b,\\c_c=-m_c}}{\text{Res}}
\widetilde{\mathfrak{I}}_t
\cr
&
\mathfrak{R}_3(m_a,m_b,m_c)\equiv
\underset{\substack{c_a=-m_a,\\c_b=\Delta_{b,ac}-\ell_a+\ell_b-\ell_c+c_c-m_a-m_b,\\c_c=-m_c}}{\text{Res}}
\widetilde{\mathfrak{I}}_t
\cr
&
\mathfrak{R}_4(m_a,m_b,m_c)\equiv
\underset{\substack{c_a=-m_a,\\c_b=-m_b,\\c_c=\Delta_{c,ab}-\ell_a-\ell_b+\ell_c-m_a-m_b-m_c}}{\text{Res}}
\widetilde{\mathfrak{I}}_t\,.
}
Defining $S_c\equiv \ell_c+m_c$, we have
\eqn{}
{
\mathfrak{R}_1=\,&
\frac{(-1)^{m_a+m_b+m_c}}{m_a!m_b!m_c!}
\cr
&\times
\frac{
\Gamma(\Delta_{ac,b}+S_a-S_b+S_C)
\Gamma(\Delta_{bc,a}-S_a+S_b+S_c)
\Gamma(\Delta_{ab,c}+S_a+S_b-S_c)}
{
(-2W_c\cdot W_a)^{\Delta_{ac,b}+S_a-S_b+S_c}
(-2W_c\cdot W_b)^{\Delta_{bc,a}-S_a+S_b+S_c}
(-2W_a\cdot W_b)^{\Delta_{ab,c}+S_a+S_b-S_c}
}
\,.
}
Changing summation variables from $m_a$, $m_b$, and $m_c$ to $S_a$, $S_b$, and $S_c$, changing order of summation of $S_i$ and $\ell_i$ sums, and recognizing that the sum over $\ell_a$, $\ell_b$, and $\ell_b$ constitutes the Lauricella function  $F_A^{(3)}$, we obtain
\eqn{JJJ1}
{
\mathfrak{J}_1=\,
&\frac{\pi^h}{2}\frac{
\Gamma\left(\Delta_{abc,}-h\right)}{\Gamma(\Delta_a)\Gamma(\Delta_b)\Gamma(\Delta_c)}
\sum_{S_a,S_b,S_c=0}^\infty\frac{(-1)^{S_a+S_b+S_c}}{S_a!S_b!S_c!}
\cr
&
\times
F^{(3)}_A\Big[\Delta_{abc,}-h;\{-S_a,-S_b,-S_c\};\{\Delta_a-h+1,\Delta_b-h+1,\Delta_c-h+1\};1,1,1\Big]
\cr
&
\times
\frac{
\Gamma(\Delta_{bc,a}-S_a+S_b+S_c)
\Gamma(\Delta_{ac,b}+S_a-S_b+S_c)
\Gamma(\Delta_{ab,c}+S_a+S_b-S_c)
}
{
(-2W_c\cdot W_b)^{\Delta_{bc,a}-S_a+S_b+S_c}
(-2W_c\cdot W_a)^{\Delta_{ac,b}+S_a-S_b+S_c}
(-2W_a\cdot W_b)^{\Delta_{ab,c}+S_a+S_b-S_c}
}\,.
}
For the second part of $\mathfrak{J}$, we start with the observation that
\eqn{}{
\mathfrak{R}_2=\frac{(-1)^{m_a+m_b+m_c}}{m_a!m_b!m_c!}
\frac{
\Gamma(\Delta_c+2S_c+m_a)
\Gamma(\Delta_{a,bc}-S_b-m_a+\ell_a-s_c)
\Gamma(\Delta_b+2S_b+m_a)}
{
(-2W_c\cdot W_b)^{\Delta_c+2S_c+m_a}
(-2W_c\cdot W_b)^{-m_a}
(-2W_a\cdot W_b)^{\Delta_b+2S_b+m_a}
}\,.
}
Changing summation variable from $m_b$ and $m_c$ to $S_b$ and $S_c$, changing order of summation, and re-casting the sum over $\ell_a$, $\ell_b$, $\ell_c$ as a Lauricella function, one finds that the second part of $\mathfrak{J}$ is given by
\eqn{JJJ2}
{
\mathfrak{J}_2=\,&
\frac{\pi^h\Gamma\left(\Delta_{abc,}-h\right)}{2\Gamma(\Delta_a)}
\sum_{m_a,S_b,S_c=0}^\infty
\frac{(-1)^{m_a+S_b+S_c}}{m_a!S_b!S_c!}
\cr
&
\times
\frac{
\left(\Delta_c\right)_{2S_c+m_a}
\left(\Delta_b\right)_{2S_b+m_a}
\Gamma(\Delta_{a,bc}-S_b-S_c-m_a)
}
{
(-2W_c\cdot W_b)^{-m_a}
(-2W_c\cdot W_a)^{\Delta_c+2S_c+m_a}
(-2W_a\cdot W_b)^{\Delta_b+2S_b+m_a}
}
\cr
&
\times
F_A^{(3)}\Big[\Delta_{abc,}-
h;\{\Delta_{a,bc}-S_b-S_c-m_a,-S_b,-S_c\};
\cr
&
\hspace{32.5mm}
\left\{\Delta_a-h+1,\Delta_b-\frac{n}{2}+1,\Delta_c-h+1\right\};1,1,1
\Big].
}
The remaining two terms in \eno{J1234} are obtained by a mere relabeling of indices:
\eqn{}{
\mathfrak{J}_3
=\,
\mathfrak{J}_2\big|_{a\leftrightarrow b}
\,,
\hspace{20mm}
\mathfrak{J}_4
=\,
\mathfrak{J}_2\big|_{a\leftrightarrow c}
\,.
}
Finally, using \eno{JJJ1} and \eno{JJJ2} to add up $\mathfrak{J}_1$, $\mathfrak{J}_2$, $\mathfrak{J}_3$, and $\mathfrak{J}_4$, it is straightforward to reproduce~\eno{GGGreal2}.


\subsection{Hypergeometric identities}

\subsubsection{Proof of equation \eno{OPEsumId}}
\label{OPEsumIdProof}

In this appendix we provide a derivation of equation \eno{OPEsumId},
\eqn{OPEsumToProve}
{
\sum_{M=0}^\infty
\frac{1}{N_{\Delta_{1}+\Delta_{2}+2M}}
\frac{a_M^{\Delta_1;\Delta_2}}{m^2_{\Delta_0}-m^2_{\Delta_{1}+\Delta_{2}+2M}}
=\frac{C_{\Delta_0\Delta_1\Delta_2}}{B(\Delta_{01,2}
\,,\,\Delta_{02,1})}\,.
}

Using the definitions
\eno{CasimirEigenvalue},
\eno{GNorm}, and
\eno{aDef} for $m^2_\Delta$, $N_\Delta$, and $a_M^{\Delta_1;\Delta_2}$, we can explicitly write out the LHS of \eno{OPEsumToProve} as

\eqn{}
{
LHS=&
-2\pi^h\sum_{M=0}^\infty
\frac{\Gamma(2\Delta_{12,}+2M-h+1)}{\Gamma(2\Delta_{12,}+2M)}
\frac{(-1)^M}{M!}
\frac{\Gamma(2\Delta_{12,}+2M)}{\Gamma(\Delta_1)\Gamma(\Delta_2)}
\frac{\Gamma(2\Delta_{12,}+M-h)}{\Gamma(2\Delta_{12,}+2M-h)}
\cr
&
\hspace{20mm}
\times
\frac{1}{\Delta_0(\Delta_0-2h)-4(\Delta_{12,}+M)(\Delta_{12,}+M-h)}\,.
}
The sum above may be recast in terms of the  hypergeometric ${}_4F_3$ function:
\eqn{jc}
{
LHS=\,&
\frac{\pi^h\Gamma(2\Delta_{12,}-h)}{\Gamma(\Delta_1)\Gamma(\Delta_2)}
\frac{
\Gamma(\Delta_{12,}-\frac{h}{2}+1)
\Gamma(\Delta_{12,}-\frac{\Delta_0}{2})
\Gamma(\Delta_{12,}-h+\frac{\Delta_0}{2})
}{
\Gamma(\Delta_{12,}-\frac{h}{2})
\Gamma(\Delta_{12,}-\frac{\Delta_0}{2}+1)
\Gamma(\Delta_{12,}-h+\frac{\Delta_0}{2}+1)
}
\cr
& \times
{}_4F_3\bigg[
\left\{
2\Delta_{12,}-h\,\,,\,\,
\Delta_{12,}-h+\frac{\Delta_0}{2}\,\,,\,\,
\Delta_{12,}-\frac{\Delta_0}{2}\,\,,\,\,
\Delta_{12,}-\frac{h}{2}+1
\right\};
\cr
&
\hspace{14mm}
\left\{
\Delta_{12,}-\frac{\Delta_0}{2}+1\,\,,\,\,
\Delta_{12,}-h+\frac{\Delta_0}{2}+1\,\,,\,\,
\Delta_{12,}-\frac{h}{2}
\right\}
;-1\bigg].
}

The arguments of the above ${}_4F_3$ hypergeometric functions are not all independent. In fact, hypergeometric functions whose arguments exhibit the precise linear dependence of the arguments of the ${}_4F_3$ function in~\eno{jc} are known in the mathematics literature as ``very-well poised hypergeometric functions'' and in the present case simplify to a ratio of gamma functions according to the identity~\cite{wolfram4F3}:
\eqn{wellPoised4F3}
{
{}_4F_3\bigg[
\left\{
a,b,c,\frac{a}{2}+1\right\};
\left\{
\frac{a}{2},
a-b+1,
a-c+1
\right\};-1
\bigg]
=
\frac{\Gamma(a-b+1)\Gamma(a-c+1)}{\Gamma(a+1)\Gamma(a-b-c+1)}\,.
}
Applying \eno{wellPoised4F3} to \eno{jc}, cancelling factors in numerator and denominator, and recalling the definition \eno{OPEreal}, we recover~\eno{OPEsumIdProof}:

\eqn{}
{
LHS=&\frac{\pi^h}{2}
\frac{\Gamma(\Delta_{12,0})
\Gamma(\Delta_{012,}-h)
}{\Gamma(\Delta_1)\Gamma(\Delta_2)}
=
\frac{C_{\Delta_0\Delta_1\Delta_2}}{B(\Delta_{01,2}\,,\,\Delta_{02,1})}\,.
}

\subsubsection{Proof of equation \eno{key1}}
\label{KEY1PROOF}

In this appendix, we provide a derivation of equation \eno{key1}. That is, we show that
\eqn{key1show}{
&\sum_{M=0}^\infty
a_{M}^{\Delta_1;\Delta_2}
d^{2\Delta_{12,}+2M;\Delta_3;\Delta_0}_{k_a;k_b}
=
\sum_{M=0}^\infty
f(M) \: c_{k_{ab,}-M;k_b}^{2\Delta_{03,}+2M;\Delta_3;\Delta_{0}}\,,
}
where the function $f(M)$ is given by
\eqn{fM}
{
f(M)
=&
{\:
\alpha_{M}^{\Delta_3;\Delta_0}
C_{\Delta_1\,\Delta_2\,2\Delta_{03,}+2M}\,
\over \beta_\infty(2\Delta_{013,2}+2M\,,\,2\Delta_{023,1}+2M) }
\cr
=&
\frac{\pi^h}{2}
\frac{(-1)^M}{M!}
\frac{(\Delta_0)_M(\Delta_3)_M}{\Gamma(\Delta_1)\Gamma(\Delta_2)}
\frac{\Gamma(2\Delta_{03,}+M-h)\Gamma(\Delta_{0123,}+M-h)
\Gamma(\Delta_{12,03}-M)}{\Gamma(2\Delta_{03,}+2M-h)}
\,,
}
and the $c$ and $d$ symbols were defined in~\eno{c2Def}-\eno{d2Def}.
We first recall that the $F_2^{(A)}$ Lauricella function, which appears in expressions for both $c$ and $d$ symbols, is defined via a double-sum, one of which can be carried out to yield a ratio of gamma functions:
\eqn{LauriRe}
{
F_2^{(A)}
\Big[e,\{a,b\},\{A,B\};1,1\Big]
=&
\sum_{\ell_b=0}^\infty
\frac{(e)_{\ell_b}(b)_{\ell_b}}{\ell_b!(B)_{\ell_b}}
\sum_{\ell_a=0}^\infty
\frac
{(e+\ell_b)_{\ell_a}(a)_{\ell_a}}
{\ell_a!(A)_{\ell_a}}
\cr
=&
\sum_{\ell_b=0}^\infty
\frac{(e)_{\ell_b}(b)_{\ell_b}}{\ell_b!(B)_{\ell_b}}
\frac{\Gamma(A)\Gamma(A-a-e-\ell_b)}{\Gamma(A-a)\Gamma(A-e-\ell_b)}\,.
}
Applying this identity to the Lauricella functions appearing in the symbols $d^{2\Delta_{12}+2M;\Delta_3;\Delta_0}_{k_a;k_b}$ and
$c_{k_{ab}-M;k_b}^{2\Delta_{03}+2M;\Delta_3;\Delta_{0}}$, we can recast the left- and right-hand sides of \eno{key1show} as sums over $\ell_b$ as follows:
\eqn{}
{
LHS=
\frac{\Gamma(\Delta_3+k_a)
\Gamma(\Delta_0+2k_b+k_a)
}
{k_a!k_b!}
\sum_{\ell_b=0}^{k_b}
\frac{
(-k_b)_{\ell_b}
\Gamma(k_{ab,}-\ell_b+1)
}{
\ell_b!(\Delta_0-h+1)_{\ell_b}}
L_{\ell_b}
}
\eqn{}
{
RHS=
\frac{\Gamma(\Delta_3+k_a)
\Gamma(\Delta_0+2k_b+k_a)
}
{k_a!k_b!}
\sum_{\ell_b=0}^{k_b}
\frac{
(-k_b)_{\ell_b}
\Gamma(k_{ab,}-\ell_b+1)
}{
\ell_b!(\Delta_0-h+1)_{\ell_b}}
R_{\ell_b}\,,
}
where $L_{\ell_b}, R_{\ell_b}$ in the summands are given by
\eqn{Lb}
{
L_{\ell_b}=\,&
\frac{\pi^h}{2}
\frac{(-1)^{k_{ab,}}}
{\Gamma(\Delta_0)\Gamma(\Delta_3)}
\sum_{M=0}^\infty
\frac{a_M^{\Delta_1;\Delta_2}}{\Gamma(2\Delta_{12,}+2M)}
\Gamma(\Delta_{12,03}+M-k_{ab,})
\cr
& \times
\frac{
\Gamma(\Delta_{0123,}+M-h+\ell_b)
\Gamma(2\Delta_{12,}+2M-h+1)
}
{\Gamma(\Delta_{0123,}+M-h+k_{ab,}+1)
\Gamma(\Delta_{12,03}+M-\ell_b+1)
}
}
\eqn{Rb}
{
R_{\ell_b}=\,&
\sum_{M=0}^\infty  f(M)
{M! \over (k_{ab,}-M)!\Gamma(\Delta_3+M)\Gamma(\Delta_0+M)}
\cr
& \times
\frac{
\Gamma(2\Delta_{03,}+M-h+\ell_b)
\Gamma(2\Delta_{03,}+2M-h+1)
}{
\Gamma(2\Delta_{03,}+M-h)
\Gamma(2\Delta_{03,}+M-h+1+k_{ab,})\Gamma(M+1-\ell_b)}\,.
}
Now, the identity \eno{key1show} follows from the fact that the summands above are identical, that is \eno{Lb} and \eno{Rb} are equal, which can be seen by first noting that the sums over $M$ may be expressed in terms of ${}_4F_3$ hypergeometric functions:
\eqn{LLb}
{
L_{\ell_b}=
&
\frac{\pi^h}{2}
\frac{(-1)^{k_{ab,}}}
{\Gamma(\Delta_0)\Gamma(\Delta_1)\Gamma(\Delta_2)\Gamma(\Delta_3)}
\frac{
\Gamma(2\Delta_{12,}-h+1)
\Gamma(\Delta_{12,03}-k_{ab,})
\Gamma(\Delta_{0123,}-h+\ell_b)
}
{\Gamma(\Delta_{0123,}-h+1+k_{ab,})
\Gamma(\Delta_{12,03}-\ell_b+1)
}
\cr
&
{}_4F_3\bigg[\left\{2\Delta_{12,}-h\,\,,\,\,
\Delta_{12,03}-k_{ab,}\,\,,\,\,
\Delta_{0123,}-h+\ell_b\,\,,\,\,
\Delta_{12,}-\frac{h}{2}+1\right\};
\cr
&
\hspace{9mm}
\left\{\Delta_{12,}-\frac{h}{2}\,\,,\,\,
\Delta_{0123,}-h+1+k_{ab,}\,\,,\,\,
\Delta_{12,03}-\ell_b+1\right\};-1\bigg]
}
\eqn{RRb}
{
R_{\ell_b}=&
\frac{\pi^h}{2}
\frac{(-1)^{\ell_b}}
{\Gamma(\Delta_0)\Gamma(\Delta_1)\Gamma(\Delta_2)\Gamma(\Delta_3)}
\frac{
\Gamma(2\Delta_{03,}-h+2\ell_b+1)
\Gamma(\Delta_{0123,}+\ell_b-h)
\Gamma(\Delta_{12,03}-\ell_b)
}{
\Gamma(2\Delta_{03,}+\ell_b-h+1+k_{ab,})
\Gamma(k_{ab,}-\ell_b+1)
}
\cr
&
{}_4F_3
\bigg[
\left\{
2\Delta_{03,}-h+2\ell_b\,\,,\,\,
\ell_b-k_{ab,}\,\,,\,\,
\Delta_{0123,}+\ell_b-h\,\,,\,\,
\Delta_{03,}-\frac{h}{2}+\ell_b+1
\right\};
\cr
&
\hspace{9mm}
\left\{
2\Delta_{03,}-h+\ell_b+k_{ab,}+1\,\,,\,\,
\Delta_{03,12}+\ell_b+1\,\,,\,\,
\Delta_{03,}-\frac{h}{2}+\ell_b
\right\}
;-1
\bigg].
}
The ${}_4F_3$ hypergeometric functions in~\eno{LLb}-\eno{RRb} are well-poised as well. Applying~\eno{wellPoised4F3} to these, we find that $L_{\ell_b}$ and $R_{\ell_b}$ can be simplified to the same expression,
\eqn{LbeqRb}
{
L_{\ell_b}=
&
\frac{\pi^h}{2}
\frac{(-1)^{k_{ab,}}}
{\Gamma(\Delta_0)\Gamma(\Delta_1)\Gamma(\Delta_2)\Gamma(\Delta_3)}
\frac{
\Gamma(\Delta_{12,03}-k_{ab,})
\Gamma(\Delta_{0123,}-h+\ell_b)
}
{
\Gamma(k_{ab,}-\ell_b+1)
}
=R_{\ell_b},
}
which establishes \eno{key1show}.

\subsubsection{Proof of equation \eno{key2}}
\label{KEY2PROOF}

In this appendix we prove the identity
\eqn{key2show}{
\sum_{M=0}^\infty
\frac{a_{M}^{\Delta_1;\Delta_2}}{m_{2\Delta_{12,}+2M}^2-m_0^2}
\bigg[
\frac{d_{k_a,k_b}^{2\Delta_{12,}+2M;\Delta_3;\Delta_b}}
{N_{\Delta_0}}
-
\frac{d_{k_a;k_b}^{\Delta_0;\Delta_3;\Delta_b}}
{N_{\Delta_{2\Delta_{12,}+2M}}}
\bigg]
=
\sum_{M=0}^\infty
g(M)
c_{k_{ab,}-M;k_b}^{2\Delta_{b3,}+2M;\Delta_3;\Delta_b}\,,
}
where the function $g(M)$ is given by
\eqn{g}
{
g(M)=\,&\frac{1}{N_{\Delta_0}}
\frac{
C_{\Delta_1\,\Delta_2\,2\Delta_{b3,}+2M}
}{m_{2\Delta_{b3,}+2M}^2-m_{\Delta_0}^2}
\frac{\alpha_M^{\Delta_{b};\Delta_{3}}}{B(\Delta_{b23,1}+M\,,\,\Delta_{b13,2}+M)}
\cr
=&
\left(
\frac{\pi^h}{2}
\right)^2
\frac{\Gamma(\Delta_0-h+1)}{\Gamma(\Delta_0)\Gamma(\Delta_1)\Gamma(\Delta_2)}
\frac{(-1)^M}{M!}
\frac{(\Delta_b)_M(\Delta_3)_M}{(\Delta_{b03,}+M-h)(\Delta_{0,b3}-M)}
\cr
& \times
\frac{\Gamma(\Delta_{b123,}+M-h)\Gamma(\Delta_{12,b3}-M)\Gamma(2\Delta_{b3,}+M-h)}{\Gamma(2\Delta_{b3,}+2M-h)}\,,
}
and the $c,d$ symbols can be found in~\eno{c2Def}-\eno{d2Def}.
Using the series expansion~\eno{LauriRe} of Lauricella functions in $c_{k_{ab,}-M;k_b}^{2\Delta_{b3,}+2M;\Delta_3;\Delta_b}$, we may rewrite the RHS of \eno{key2show} as
\eqn{}
{
\mathcal{RHS}=
\frac{\Gamma(\Delta_3+k_a)
\Gamma(\Delta_b+2k_b+k_a)
}
{k_a!k_b!}
\sum_{\ell_b=0}^{k_b}
\frac{
(-k_b)_{\ell_b}
\Gamma(k_{ab,}-\ell_b+1)
}{
\ell_b!(\Delta_b-h+1)_{\ell_b}}
\mathcal{R}_{\ell_b}\,,
}
where $\mathcal{R}_{\ell_b}$ is given by
\eqn{calR}
{
\mathcal{R}_{\ell_b} &=
\sum_{M=0}^\infty  g(M)
{M! \over (k_{ab,}-M)!\Gamma(\Delta_3+M)\Gamma(\Delta_b+M)}
\cr
&
\hspace{17mm}
\times
\frac{
\Gamma(2\Delta_{b3,}+M-h+\ell_b)
\Gamma(2\Delta_{b3,}+2M-h+1)
}{
\Gamma(2\Delta_{b3,}+M-h)
\Gamma(2\Delta_{b3,}+M-h+1+k_{ab,})\Gamma(M+1-\ell_b)}
\cr
&=
\frac{\pi^{2h}}{4}
\frac{
(-1)^{\ell_b+1}\Gamma(\Delta_0-h+1)
\Gamma(\Delta_{12,b3}-\ell_b)
\Gamma(2\Delta_{b3,}-h+2\ell_b+1)
\Gamma(\Delta_{b123,}-h+\ell_b)
}{
\Gamma(\Delta_b)\Gamma(\Delta_0)\Gamma(\Delta_1)\Gamma(\Delta_2)\Gamma(\Delta_3)\Gamma(k_{ab,}-\ell_b+1)
\Gamma(2\Delta_{b3,}-h+1+\ell_b+k_{ab,})
}
\cr
& \quad
\times
\frac{
1}
{
(\Delta_{b03,}-h+\ell_b)
(\Delta_{b3,0}+\ell_b)}
\cr
& \quad \times
{}_6F_5
\bigg[
\bigg\{
2\Delta_{b3,}-h+2\ell_b\,\,,\,\,
\ell_b-k_{ab,}\,\,,\,\,
\Delta_{b123,}-h+\ell_b\,\,,\,\,
\Delta_{b03,}-h+\ell_b\,\,,\,\,
\Delta_{b3,0}+\ell_b\,\,,\,\,  \cr
& \qquad\qquad \quad \left. \Delta_{b3,}-\frac{h}{2}+\ell_b+1
\right\};
\bigg\{
2\Delta_{b3,}-h+\ell_b+k_{ab,}+1\,\,,\,\,
\Delta_{b3,12}+\ell_b+1\,\,,\,\, \cr
& \qquad\qquad\quad \left.
\Delta_{b3,0}+\ell_b+1\,\,,\,\,
\Delta_{b03,}-h+\ell_b+1\,\,,\,\,
\Delta_{b3,}-\frac{h}{2}+\ell_b
\right\};-1
\bigg]\,.
}
The ${}_6F_5$ hypergeometric function is very well-poised as well and can be reduced to a ${}_3F_2$ hypergeometric function with the help of~\cite[Theorem 3.4.6]{andrews1999special} to give
\eqn{}{
\mathcal{R}_{\ell_b}
=\,&
\frac{\pi^{2h}}{4}
\frac{
(-1)^{\ell_b+1}\Gamma(\Delta_0-h+1)
\Gamma(\Delta_{12,b3}-\ell_b)
\Gamma(\Delta_{b123,}-h+\ell_b)
}{
\Gamma(\Delta_b)\Gamma(\Delta_0)\Gamma(\Delta_1)\Gamma(\Delta_2)\Gamma(\Delta_3)\Gamma(k_{ab,}-\ell_b+1)
}
\cr
&
\times
\frac{\Gamma(\Delta_{b3,12}+\ell_b+1)}{\Gamma(\Delta_{b3,12}+k_{ab,}+1)}
\frac{
1}
{
(\Delta_{b03,}-h+\ell_b)
(\Delta_{b3,0}+\ell_b)}
\cr
&
\times
{}_3F_2
\bigg[
\left\{
1\,\,,\,\,
\ell_b-k_{ab,}\,\,,\,\,
\Delta_{b123,}-h+\ell_b
\right\};
\left\{
\Delta_{b3,0}+\ell_b+1\,\,,\,\,
\Delta_{b03,}-h+\ell_b+1
\right\};
1
\bigg].
}
Turning to the LHS in~\eno{key2show}, we can use the identity \eno{OPEsumId} to directly evaluate the sum over $M$ in the second term inside the square brackets. Then, applying~\eno{LauriRe} to the symbols $d_{k_a;k_b}^{\Delta_1+\Delta_2+2M;\Delta_3;\Delta_b}$ and $d_{k_a;k_b}^{\Delta_0;\Delta_3;\Delta_b}$, we get
\eqn{}
{
\mathcal{LHS}=\,&
\sum_{M=0}^\infty
\frac{a_{M}^{\Delta_1;\Delta_2}}{m_{2\Delta_{12,}+2M}^2-m_0^2}
\frac{d_{k_a;k_b}^{2\Delta_{12,}+2M;\Delta_3;\Delta_b}}
{N_{\Delta_0}}
+
\frac{C_{\Delta_0\Delta_1\Delta_2}}{B(\Delta_{01,2}\,,\,\Delta_{02,1})}
d_{k_a;k_b}^{\Delta_0;\Delta_3;\Delta_b}
\cr
=\,&
\frac{\Gamma(\Delta_3+k_a)
\Gamma(\Delta_b+2k_b+k_a)
}
{k_a!k_b!}
\sum_{\ell_b=0}^{k_b}
\frac{
(-k_b)_{\ell_b}
\Gamma(k_{ab,}-\ell_b+1)
}{
\ell_b!(\Delta_b-h+1)_{\ell_b}}
(\mathcal{L}^{(1)}_{\ell_b}+\mathcal{L}^{(2)}_{\ell_b})\,,
}
where the terms $\mathcal{L}^{(2)}_{\ell_b}$ and $\mathcal{L}^{(1)}_{\ell_b}$ are given by
\eqn{calL2}
{
\mathcal{L}_{\ell_b}^{(2)}
=&
\frac{C_{\Delta_0\Delta_1\Delta_2}}{B(\Delta_{01,2}\,,\,\Delta_{02,1})}
\frac{\pi^h}{2}
\frac{(-1)^{k_{ab,}}\Gamma\left(\Delta_{03b,}-h\right)}{\Gamma(\Delta_0)\Gamma(\Delta_3)\Gamma(\Delta_b)}
\frac{
\Gamma(\Delta_{0,b3}-k_{ab,})
(\Delta_{03b,}-h)_{\ell_b}
\Gamma(\Delta_0-h+1)
}{
\Gamma(\Delta_{0,b3}-\ell_b+1)
\Gamma(\Delta_{b03,}-h+k_{ab,}+1)
}
\,,
}
and
\eqn{calL1}
{
\mathcal{L}_{\ell_b}^{(1)}
&=
\frac{\pi^h}{2}
\frac{(-1)^{k_{ab,}}}{\Gamma(\Delta_3)\Gamma(\Delta_b)N_{\Delta_0}}
\sum_{M=0}^\infty
\frac{a_{M}^{\Delta_1;\Delta_2}}{m_{2\Delta_{12,}+2M}^2-m_{\Delta_0}^2}
\cr
&
\times
\frac{\Gamma\left(\Delta_{b123,}+M-h\right)\Gamma(\Delta_{12,b3}+M-k_{ab,})(\Delta_{b123,}+M-h)_{\ell_b}}
{\Gamma(2\Delta_{12,}+2M)\Gamma(\Delta_{12,b3}+M-\ell_b+1)(2\Delta_{12,}+2M-h+1)_{k_{ab,}+\Delta_{b3,12}-M}}
\cr
&=
 \frac{\pi^{2h}}{4}
\frac{\Gamma(\Delta_0-h+1)}{\Delta_{12,0}(\Delta_{012,}-h)}
\frac{(-1)^{k_{ab,}+1}
\Gamma(\Delta_{12,b3}-k_{ab,})
\Gamma(2\Delta_{12,}-h+1)
\Gamma(\Delta_{b123,}-h+\ell_b)
}{\Gamma(\Delta_b)\Gamma(\Delta_0)\Gamma(\Delta_1)\Gamma(\Delta_2)\Gamma(\Delta_3)
\Gamma(\Delta_{b123,}-h+k_{ab,}+1)
\Gamma(\Delta_{12,b3}-\ell_b+1)
}
\cr
&
\times
{}_6
F_5
\bigg[
\left\{
2\Delta_{12}-h\,\,,\,\,
\Delta_{12,b3}-k_{ab,}\,\,,\,\,
\Delta_{b123,}-h+\ell_b\,\,,\,\,
\Delta_{12,0}\,\,,\,\,
\Delta_{012,}-h\,\,,\,\,
\Delta_{12,}-\frac{h}{2}+1
\right\};
\cr
&\hspace{8.2mm}
\left\{
\Delta_{b123,}-h+k_{ab,}+1\,\,,\,\,
\Delta_{12,b3}-\ell_b+1\,\,,\,\,
\Delta_{12,0}+1\,\,,\,\,
\Delta_{012,}-h+1\,\,,\,\,
\Delta_{12,}-\frac{h}{2}
\right\};-1
\bigg].
}
Once again, the ${}_6F_5$ hypergeometric function above is  very well-poised, so it can be reduced to a ${}_3F_2$ function with the help of~\cite[Theorem 3.4.6]{andrews1999special}, to give
\eqn{}{
\mathcal{L}_{\ell_b}^{(1)}
=&
 \frac{\pi^{2h}}{4}
\frac{\Gamma(\Delta_0-h+1)}{\Delta_{12,0}(\Delta_{012,}-h)\Gamma(\Delta_0)}
\frac{(-1)^{k_{ab,}+1}}{\Gamma(\Delta_b)\Gamma(\Delta_1)\Gamma(\Delta_2)\Gamma(\Delta_3)}\frac{
\Gamma(\Delta_{12,b3}-k_{ab,})
\Gamma(\Delta_{b123,}-h+\ell_b)}
{\Gamma(k_{ab,}-\ell_b+1)}
\cr
&
\times
{}_3
F_2
\bigg[
\left\{
1\,\,,\,\,
\Delta_{12,b3}-k_{ab,}\,\,,\,\,
\Delta_{b123,}-h+\ell_b
\right\};
\left\{
\Delta_{12,0}+1\,\,,\,\,
\Delta_{012,}-h+1
\right\};1
\bigg].
}
Finally, using the identity~\cite[Case 28]{Milgram},
\eqn{}
{
&
 \frac{
{}_3F_2\big[\{1,-k,a+b\};\{b+c+1,b-c+1\};1\big]
}{(b+c)(b-c)}
-
\frac{
{}_3F_2\big[\{1,a-b-k,a+b\};\{a+c+1,a-c+1\};1\big]
}{(a-c)(a+c)}
\cr
&  =
(-1)^{k}
\frac{
\Gamma(k+1)
\Gamma(a-c)
\Gamma(a+c)
\Gamma(b-c)
\Gamma(b+c)
}
{
\Gamma(a-b-k)
\Gamma(a+b)
\Gamma(b-c+k+1)
\Gamma(b+c+k+1)
}
\,,
}
where $k \in \mathbb{N}_0$, it can be checked that
\eqn{L1L2R}
{\mathcal{L}_{\ell_b}^{(1)}+\mathcal{L}_{\ell_b}^{(2)}=\mathcal{R}_{\ell_b}}
which establishes \eno{key2show}.

\subsubsection{Proof of equation \eno{key6pt}}
\label{KEY3PROOF}

In this appendix we provide a derivation of equation \eno{key6pt}:
\eqn{key6ptToProve}
{\sum_{M=0}^\infty a_{M}^{\Delta_1;\Delta_2}
d_{k_a;k_b;k_c}^{2\Delta_{12,}+2M;\Delta_b;\Delta_c}
=
\sum_{M=0}^\infty \tilde{f}(M)
c_{k_{abc,}-M;k_c;k_c}^{2\Delta_{bc,}+2M;\Delta_b;\Delta_c}\,,
}
where the function $\tilde{f}(M)$ is given by
\eqn{ftilde}
{
\tilde{f}(M)=&
\frac{
\alpha_M^{\Delta_b;\Delta_c}C_{\Delta_1\,\Delta_2\,2\Delta_{bc,}+2M}
}
{B(\Delta_{bc1,2}+M\,,\,\Delta_{bc2,1}+M)}
\cr
=&
\frac{(-1)^M}{M!}
\frac{\pi^h}{2}
\Gamma(\Delta_b+M)\Gamma(\Delta_c+M)
\frac{\Gamma(\Delta_{12,bc}-M)
\Gamma(\Delta_{bc12,}-h+M)}
{\Gamma(\Delta_b)\Gamma(\Delta_c)\Gamma(\Delta_1)\Gamma(\Delta_2)}
\frac{\Gamma(2\Delta_{bc,}-h+M)}{\Gamma(2\Delta_{bc,}+2M-h)}\,,
}
and the $c,d$ symbols can be found in~\eno{c3Def}-\eno{d3Def}.
The $F_3^{(A)}$ Lauricella function which appears in the $c,d$ symbols is defined via a triple-sum (see~\eno{LauricellaDef}), one of which can be performed to yield a double-sum over a ratio of gamma functions:
\eqn{Lauri3}
{
F_3^{(A)}
\big[e,\{a,b,c\},\{A,B,C\};1,1\big]
=&
\sum_{\ell_b,\ell_c=0}^\infty
\frac{(e)_{\ell_{bc}}(b)_{\ell_b}(c)_{\ell_c}}{\ell_b!\ell_c!(B)_{\ell_b}(C)_{\ell_c}}
\sum_{\ell_a=0}^\infty
\frac
{(e+\ell_{bc})_{\ell_a}(a)_{\ell_a}}
{\ell_a!(A)_{\ell_a}}
\cr
=&
\sum_{\ell_b,\ell_c=0}^\infty
\frac{(e)_{\ell_{bc}}(b)_{\ell_b}(c)_{\ell_c}}{\ell_b!\ell_c!(B)_{\ell_b}(C)_{\ell_c}}
\frac{\Gamma(A)\Gamma(A-a-e-\ell_{bc})}{\Gamma(A-a)\Gamma(A-e-\ell_{bc})}\,,
}
where we have defined $\ell_{bc,}\equiv \ell_b+\ell_c$. Applying this to the $F_3^{(A)}$ Lauricella functions appearing in the expressions for the symbols  $d_{k_a;k_b;k_c}^{2\Delta_{12,}+2M_a;\Delta_b;\Delta_c}$ and $
c_{k_{abc,}-M;k_c;k_c}^{2\Delta_{bc,}+2M;\Delta_b;\Delta_c}$, we can recast the left- and right-hand sides of \eno{key6ptToProve} as follows:
\eqn{}{
\widetilde{LHS}=
\frac{\Gamma(\Delta_c+2k_c+k_a)\Gamma(\Delta_b+2k_b+k_a)}{k_a!k_b!k_c!}
\sum_{\ell_b,\ell_c=0}^\infty
\frac{(-k_b)_{\ell_b}(-k_c)_{\ell_c}
\Gamma(k_{abc,}-\ell_{bc,}+1)
}{\ell_b!\ell_c!(\Delta_b-h+1)_{\ell_b}(\Delta_c-h+1)_{\ell_c}}
\widetilde{L}_{\ell_b,\ell_c}
}
\eqn{}{
\widetilde{RHS}=
\frac{\Gamma(\Delta_c+2k_c+k_a)\Gamma(\Delta_b+2k_b+k_a)}{k_a!k_b!k_c!}
\sum_{\ell_b,\ell_c=0}^\infty
\frac{(-k_b)_{\ell_b}(-k_c)_{\ell_c}
\Gamma(k_{abc,}-\ell_{bc,}+1)
}{\ell_b!\ell_c!(\Delta_b-h+1)_{\ell_b}(\Delta_c-h+1)_{\ell_c}}
\widetilde{R}_{\ell_b,\ell_c}\,,
}
where $\widetilde{L}_{\ell_b,\ell_c},\widetilde{R}_{\ell_b,\ell_c}$ in the summands are given by
\eqn{Ltilde}
{
\widetilde{L}_{\ell_b,\ell_c}=\,&
\frac{\pi^h}{2}
\frac{(-1)^{k_{abc,}}}
{\Gamma(\Delta_b)\Gamma(\Delta_c)}
\sum_{M=0}^\infty
\frac{a_{M}^{\Delta_1;\Delta_2}}
{\Gamma(2\Delta_{12,}+2M)}
\Gamma(\Delta_{12,bc}+M-k_{abc,})
\cr
& \times
\frac{
\Gamma(\Delta_{bc12,}+M-h+\ell_{bc,})
\Gamma(2\Delta_{12,}+2M-h+1)
}{
\Gamma(\Delta_{bc12,}+M-h+k_{abc,}+1)
\Gamma(\Delta_{12,bc}+M-\ell_{bc,}+1)
}\,,
}
and
\eqn{Rtilde}
{
\widetilde{R}_{\ell_b,\ell_c}
=\,&
\sum_{M=0}^{\infty} \tilde{f}(M)
\frac{M!}
{
(k_{abc,}-M)!
\Gamma(\Delta_{b}+M)
\Gamma(\Delta_{c}+M)}
\cr
& \times
(2\Delta_{bc,}+M-h)_{\ell_{bc,}}
\frac{
\Gamma(2\Delta_{bc,}+2M-h+1)
}{
\Gamma(2\Delta_{bc,}+M-h+k_{abc,}+1)
\Gamma(M+1-\ell_{bc,})
}\,.
}
Now, by comparing \eno{ftilde} with \eno{fM}, \eno{Ltilde} with \eno{Lb}, and \eno{Rtilde} with \eno{Rb}, we observe that
\eqn{}
{
\widetilde{L}_{\ell_b,\ell_c}=L_{\ell_b}\bigg|_{\substack{\Delta_0\rightarrow\Delta_b\\\Delta_3\rightarrow\Delta_c\\ \ell_b\rightarrow \ell_{bc,}\\k_{ab,}\rightarrow k_{abc,}}}
\hspace{10mm}
\widetilde{R}_{\ell_b,\ell_c}=R_{\ell_b}\bigg|_{\substack{\Delta_0\rightarrow\Delta_b\\\Delta_3\rightarrow\Delta_c\\ \ell_b\rightarrow \ell_{bc,}\\k_{ab,}\rightarrow k_{abc,}}}
\,.
}
Hence \eno{LbeqRb} implies $\widetilde{L}_{\ell_b,\ell_c}=\widetilde{R}_{\ell_b,\ell_c}$, from which \eno{key6ptToProve} follows.

\subsubsection{Proof of equation \eno{key4}}
\label{KEY4PROOF}

In this appendix we derive equation \eno{key4}:
\eqn{key4toProve}
{
\sum_{M=0}^\infty
\frac{a_{M}^{\Delta_1;\Delta_2}}{m_{2\Delta_{12,}+2M}^2-m_0^2}
\bigg[
\frac{d_{k_a;k_b;k_c}^{2\Delta_{12,}+2M;\Delta_b;\Delta_c}}
{N_{\Delta_0}}
-
\frac{d_{k_a;k_b;k_c}^{\Delta_0;\Delta_b;\Delta_c}}
{N_{\Delta_{2\Delta_{12,}+2M}}}
\bigg]
=
\sum_{M=0}^\infty
\tilde{g}(M)
c_{k_{abc,}-M;k_b;k_c}^{2\Delta_{bc,}+2M;\Delta_b;\Delta_c}\,,
}
where the function $\tilde{g}(M)$ is given by
\eqn{gtilde}
{
\tilde{g}(M)=&\frac{1}{N_{\Delta_0}}
\frac{
C_{\Delta_1\,\Delta_2\,2\Delta_{bc,}+2M}
}{m_{2\Delta_{bc,}+2M}^2-m_{\Delta_0}^2}
\frac{\alpha_M^{\Delta_{b};\Delta_{c}}}{B(\Delta_{bc2,1}+M,\Delta_{bc1,2}+M)}\,,
}
and the $c,d$ symbols are given in~\eno{c3Def}-\eno{d3Def}.
By applying the $F_3^{(A)}$ identity~\eno{Lauri3} to the Lauricella function appearing in the symbol $c_{k_{abc,}-M;k_b;k_c}^{2\Delta_{bc,}+2M;\Delta_b;\Delta_c}$, we may rewrite the RHS of~\eno{key4toProve} as
\eqn{}
{
\widetilde{\mathcal{RHS}}=
\frac{\Gamma(\Delta_c+2k_c+k_a)
\Gamma(\Delta_b+2k_b+k_a)
}
{k_a!k_b!k_c!}
\sum_{\ell_b=0}^{k_b}
\frac{
(-k_b)_{\ell_b}
(-k_c)_{\ell_c}
\Gamma(k_{abc,}-\ell_{bc,}+1)
}{
\ell_b!(\Delta_b-h+1)_{\ell_b}
\ell_c!(\Delta_c-h+1)_{\ell_c}}
\widetilde{\mathcal{R}}_{\ell_b,\ell_c}\,,
}
where $\widetilde{{\cal R}}_{\ell_b,\ell_c}$ in the summand is given by
\eqn{tildeR}
{
\widetilde{\mathcal{R}}_{\ell_b,\ell_c}
=\,
&
\sum_{M=0}^\infty \tilde{g}(M)
 {M! \over (k_{abc,}-M)!\Gamma(\Delta_c+M)\Gamma(\Delta_b+M)}  \cr
 & \times
\frac{
(2\Delta_{bc,}+M-h)_{\ell_{bc,}}
\Gamma(2\Delta_{bc,}+2M-h+1)
}{
\Gamma(2\Delta_{bc,}+M-h+k_{abc,}+1)
\Gamma(M-\ell_{bc,}+1)
}\,.
}
Turning to the left-hand side of \eno{key4toProve}, applying \eno{OPEsumId} and using \eno{Lauri3} we are led to
\eqn{}
{
\widetilde{\mathcal{LHS}}=&
\sum_{M=0}^\infty
\frac{a_{M}^{\Delta_1;\Delta_2}}{m_{2\Delta_{12,}+2M}^2-m_0^2}
\frac{d_{k_a;k_b;k_c}^{2\Delta_{12,}+2M;\Delta_b;\Delta_c}}
{N_{\Delta_0}}
+\frac{C_{\Delta_0\Delta_1\Delta_2}}{B(\Delta_{01,2}\,,\,\Delta_{02,1})}
d_{k_a;k_b;k_c}^{\Delta_0;\Delta_b;\Delta_c}
\cr
=&
\frac{\Gamma(\Delta_c+2k_c+k_a)
\Gamma(\Delta_b+2k_b+k_a)
}
{k_a!k_b!k_c!}
\sum_{\ell_b=0}^{k_b}
\frac{
(-k_b)_{\ell_b}
(-k_c)_{\ell_c}
\Gamma(k_{abc,}-\ell_{bc,}+1)
}{
\ell_b!(\Delta_b-h+1)_{\ell_b}
\ell_c!(\Delta_c-h+1)_{\ell_c}}
(\widetilde{\mathcal{L}}^{(1)}_{\ell_b,\ell_c}+
\widetilde{\mathcal{L}}^{(2)}_{\ell_b,\ell_c}
)\,,
}
where the $\widetilde{\cal L}_{\ell_b,\ell_c}^{(i)}$ functions in the summand are given by
\eqn{tildeL1}
{
\widetilde{\mathcal{L}}^{(1)}_{\ell_b,\ell_c}=&\,
\frac{\pi^h}{2}\frac{(-1)^{k_{abc,}}}{\Gamma(\Delta_b)\Gamma(\Delta_c)}\sum_{M=0}^\infty
\frac{a_{M}^{\Delta_1;\Delta_2}}{N_{\Delta_0}(m_{2\Delta_{12,}+2M}^2-m_0^2)}
\cr
&
\times
\frac{
\Gamma(\Delta_{bc12,}+M-h+\ell_{bc,})
\Gamma(\Delta_{12,bc}+M-k_{abc,})
\Gamma(2\Delta_{12,}+2M-h+1)
}{
\Gamma(2\Delta_{12,}+2M)
\Gamma(\Delta_{12,bc}+M-\ell_{bc,}+1)
\Gamma(\Delta_{bc12,}+M-h+k_{abc,}+1)
}\,,
}
and
\eqn{tildeL2}
{
\widetilde{\mathcal{L}}^{(2)}_{\ell_b,\ell_c}=&
\frac{C_{\Delta_0\Delta_1\Delta_2}}{B(\Delta_{01,2}\,,\,\Delta_{02,1})}
\frac{\pi^h}{2}
\frac{(-1)^{k_{abc,}}\Gamma(\Delta_{bc0,}-h)
}{\Gamma(\Delta_b)\Gamma(\Delta_c)\Gamma(\Delta_0)}
\frac{
\Gamma(\Delta_{0,bc}-k_{abc,})
(\Delta_{bc0,}-h)_{\ell_{bc,}}
\Gamma(\Delta_0-h+1)
}
{
\Gamma(\Delta_{0,bc}-\ell_{bc,}+1)
\Gamma(\Delta_{bc0,}-h+k_{abc,}+1)
}\,.
}
By comparing \eno{gtilde} with \eno{g}, \eno{tildeR} with \eno{calR}, \eno{tildeL1} with \eno{calL1}, and \eno{tildeL2} with \eno{calL2}, we see that
\eqn{}
{
\widetilde{\mathcal{L}}^{(1)}_{\ell_b,\ell_c}=\mathcal{L}^{(1)}_{\ell_b}\bigg|_{\substack{\Delta_3\rightarrow\Delta_c\\ \ell_b\rightarrow \ell_{bc,}\\k_{ab,}\rightarrow k_{abc,}}}
\hspace{10mm}
\widetilde{\mathcal{L}}^{(2)}_{\ell_b,\ell_c}=\mathcal{L}^{(2)}_{\ell_b}\bigg|_{\substack{\Delta_3\rightarrow\Delta_c\\ \ell_b\rightarrow \ell_{bc,}\\k_{ab,}\rightarrow k_{abc,}}}
\hspace{10mm}
\widetilde{\mathcal{R}}_{\ell_b,\ell_c}=\mathcal{R}_{\ell_b}\bigg|_{\substack{\Delta_3\rightarrow\Delta_c\\ \ell_b\rightarrow \ell_{bc,}\\k_{ab,}\rightarrow k_{abc,}}}
\,.
}
Consequently, \eno{L1L2R} implies
$
\widetilde{\mathcal{L}}^{(1)}_{\ell_{b},\ell_{c}}
+
\widetilde{\mathcal{L}}^{(2)}_{\ell_{b},\ell_{c}}
=
\widetilde{\mathcal{R}}_{\ell_{b},\ell_{c}}\,,
$
which proves \eno{key4toProve}.

\bibliographystyle{ssg}
\bibliography{draft}

\begingroup\raggedright\begin{thebibliography}{100}

\bibitem{Maldacena:1997re}
J.~M. Maldacena, ``{The Large N limit of superconformal field theories and
  supergravity},'' {\em Int. J. Theor. Phys.} {\bf 38} (1999) 1113--1133,
  \href{https://doi.org/10.1023/A:1026654312961,
  10.4310/ATMP.1998.v2.n2.a1}{10.1023/A:1026654312961,
  10.4310/ATMP.1998.v2.n2.a1}, \href{http://arxiv.org/abs/hep-th/9711200}{{\tt
  hep-th/9711200}}.

\bibitem{Gubser:1998bc}
S.~S. Gubser, I.~R. Klebanov, and A.~M. Polyakov, ``{Gauge theory correlators
  from noncritical string theory},'' {\em Phys. Lett.} {\bf B428} (1998)
  105--114,
  \href{https://doi.org/10.1016/S0370-2693(98)00377-3}{10.1016/S0370-2693(98)00377-3},
  \href{http://arxiv.org/abs/hep-th/9802109}{{\tt hep-th/9802109}}.

\bibitem{Witten:1998qj}
E.~Witten, ``{Anti-de Sitter space and holography},'' {\em Adv. Theor. Math.
  Phys.} {\bf 2} (1998) 253--291,
  \href{https://doi.org/10.4310/ATMP.1998.v2.n2.a2}{10.4310/ATMP.1998.v2.n2.a2},
  \href{http://arxiv.org/abs/hep-th/9802150}{{\tt hep-th/9802150}}.

\bibitem{Ferrara:1973yt}
S.~Ferrara, A.~F. Grillo, and R.~Gatto, ``{Tensor representations of conformal
  algebra and conformally covariant operator product expansion},'' {\em Annals
  Phys.} {\bf 76} (1973) 161--188,
  \href{https://doi.org/10.1016/0003-4916(73)90446-6}{10.1016/0003-4916(73)90446-6}.

\bibitem{Polyakov:1974gs}
A.~M. Polyakov, ``{Nonhamiltonian approach to conformal quantum field
  theory},'' {\em Zh. Eksp. Teor. Fiz.} {\bf 66} (1974) 23--42.

\bibitem{Rattazzi:2008pe}
R.~Rattazzi, V.~S. Rychkov, E.~Tonni, and A.~Vichi, ``{Bounding scalar operator
  dimensions in 4D CFT},'' {\em JHEP} {\bf 12} (2008) 031,
  \href{https://doi.org/10.1088/1126-6708/2008/12/031}{10.1088/1126-6708/2008/12/031},
  \href{http://arxiv.org/abs/0807.0004}{{\tt 0807.0004}}.

\bibitem{ElShowk:2012ht}
S.~El-Showk, M.~F. Paulos, D.~Poland, S.~Rychkov, D.~Simmons-Duffin, and
  A.~Vichi, ``{Solving the 3D Ising Model with the Conformal Bootstrap},'' {\em
  Phys. Rev.} {\bf D86} (2012) 025022,
  \href{https://doi.org/10.1103/PhysRevD.86.025022}{10.1103/PhysRevD.86.025022},
  \href{http://arxiv.org/abs/1203.6064}{{\tt 1203.6064}}.

\bibitem{Poland:2018epd}
D.~Poland, S.~Rychkov, and A.~Vichi, ``{The Conformal Bootstrap: Theory,
  Numerical Techniques, and Applications},'' {\em Rev. Mod. Phys.} {\bf 91}
  (2019), no.~1 15002,
  \href{https://doi.org/10.1103/RevModPhys.91.015002}{10.1103/RevModPhys.91.015002},
  \href{http://arxiv.org/abs/1805.04405}{{\tt 1805.04405}}.

\bibitem{Heemskerk:2009pn}
I.~Heemskerk, J.~Penedones, J.~Polchinski, and J.~Sully, ``{Holography from
  Conformal Field Theory},'' {\em JHEP} {\bf 10} (2009) 079,
  \href{https://doi.org/10.1088/1126-6708/2009/10/079}{10.1088/1126-6708/2009/10/079},
  \href{http://arxiv.org/abs/0907.0151}{{\tt 0907.0151}}.

\bibitem{Aharony:2016dwx}
O.~Aharony, L.~F. Alday, A.~Bissi, and E.~Perlmutter, ``{Loops in AdS from
  Conformal Field Theory},'' {\em JHEP} {\bf 07} (2017) 036,
  \href{https://doi.org/10.1007/JHEP07(2017)036}{10.1007/JHEP07(2017)036},
  \href{http://arxiv.org/abs/1612.03891}{{\tt 1612.03891}}.

\bibitem{Mack:2009mi}
G.~Mack, ``{D-independent representation of Conformal Field Theories in D
  dimensions via transformation to auxiliary Dual Resonance Models. Scalar
  amplitudes},'' \href{http://arxiv.org/abs/0907.2407}{{\tt 0907.2407}}.

\bibitem{Penedones:2010ue}
J.~Penedones, ``{Writing CFT correlation functions as AdS scattering
  amplitudes},'' {\em JHEP} {\bf 03} (2011) 025,
  \href{https://doi.org/10.1007/JHEP03(2011)025}{10.1007/JHEP03(2011)025},
  \href{http://arxiv.org/abs/1011.1485}{{\tt 1011.1485}}.

\bibitem{Fitzpatrick:2012cg}
A.~L. Fitzpatrick and J.~Kaplan, ``{AdS Field Theory from Conformal Field
  Theory},'' {\em JHEP} {\bf 02} (2013) 054,
  \href{https://doi.org/10.1007/JHEP02(2013)054}{10.1007/JHEP02(2013)054},
  \href{http://arxiv.org/abs/1208.0337}{{\tt 1208.0337}}.

\bibitem{Hijano:2015zsa}
E.~Hijano, P.~Kraus, E.~Perlmutter, and R.~Snively, ``{Witten Diagrams
  Revisited: The AdS Geometry of Conformal Blocks},'' {\em JHEP} {\bf 01}
  (2016) 146,
  \href{https://doi.org/10.1007/JHEP01(2016)146}{10.1007/JHEP01(2016)146},
  \href{http://arxiv.org/abs/1508.00501}{{\tt 1508.00501}}.

\bibitem{Gopakumar:2016wkt}
R.~Gopakumar, A.~Kaviraj, K.~Sen, and A.~Sinha, ``{Conformal Bootstrap in
  Mellin Space},'' {\em Phys. Rev. Lett.} {\bf 118} (2017), no.~8 081601,
  \href{https://doi.org/10.1103/PhysRevLett.118.081601}{10.1103/PhysRevLett.118.081601},
  \href{http://arxiv.org/abs/1609.00572}{{\tt 1609.00572}}.

\bibitem{Gopakumar:2016cpb}
R.~Gopakumar, A.~Kaviraj, K.~Sen, and A.~Sinha, ``{A Mellin space approach to
  the conformal bootstrap},'' {\em JHEP} {\bf 05} (2017) 027,
  \href{https://doi.org/10.1007/JHEP05(2017)027}{10.1007/JHEP05(2017)027},
  \href{http://arxiv.org/abs/1611.08407}{{\tt 1611.08407}}.

\bibitem{Gopakumar:2018xqi}
R.~Gopakumar and A.~Sinha, ``{On the Polyakov-Mellin bootstrap},'' {\em JHEP}
  {\bf 12} (2018) 040,
  \href{https://doi.org/10.1007/JHEP12(2018)040}{10.1007/JHEP12(2018)040},
  \href{http://arxiv.org/abs/1809.10975}{{\tt 1809.10975}}.

\bibitem{Fitzpatrick:2011ia}
A.~L. Fitzpatrick, J.~Kaplan, J.~Penedones, S.~Raju, and B.~C. van Rees, ``{A
  Natural Language for AdS/CFT Correlators},'' {\em JHEP} {\bf 11} (2011) 095,
  \href{https://doi.org/10.1007/JHEP11(2011)095}{10.1007/JHEP11(2011)095},
  \href{http://arxiv.org/abs/1107.1499}{{\tt 1107.1499}}.

\bibitem{Paulos:2011ie}
M.~F. Paulos, ``{Towards Feynman rules for Mellin amplitudes},'' {\em JHEP}
  {\bf 10} (2011) 074,
  \href{https://doi.org/10.1007/JHEP10(2011)074}{10.1007/JHEP10(2011)074},
  \href{http://arxiv.org/abs/1107.1504}{{\tt 1107.1504}}.

\bibitem{Nandan:2011wc}
D.~Nandan, A.~Volovich, and C.~Wen, ``{On Feynman Rules for Mellin Amplitudes
  in AdS/CFT},'' {\em JHEP} {\bf 05} (2012) 129,
  \href{https://doi.org/10.1007/JHEP05(2012)129}{10.1007/JHEP05(2012)129},
  \href{http://arxiv.org/abs/1112.0305}{{\tt 1112.0305}}.

\bibitem{Kharel:2013mka}
S.~Kharel and G.~Siopsis, ``{Tree-level Correlators of scalar and vector fields
  in AdS/CFT},'' {\em JHEP} {\bf 11} (2013) 159,
  \href{https://doi.org/10.1007/JHEP11(2013)159}{10.1007/JHEP11(2013)159},
  \href{http://arxiv.org/abs/1308.2515}{{\tt 1308.2515}}.

\bibitem{Goncalves:2014rfa}
V.~Gonçalves, J.~Penedones, and E.~Trevisani, ``{Factorization of Mellin
  amplitudes},'' {\em JHEP} {\bf 10} (2015) 040,
  \href{https://doi.org/10.1007/JHEP10(2015)040}{10.1007/JHEP10(2015)040},
  \href{http://arxiv.org/abs/1410.4185}{{\tt 1410.4185}}.

\bibitem{Cardona:2017tsw}
C.~Cardona, ``{Mellin-(Schwinger) representation of One-loop Witten diagrams in
  AdS},'' \href{http://arxiv.org/abs/1708.06339}{{\tt 1708.06339}}.

\bibitem{Yuan:2017vgp}
E.~Y. Yuan, ``{Loops in the Bulk},''
  \href{http://arxiv.org/abs/1710.01361}{{\tt 1710.01361}}.

\bibitem{Yuan:2018qva}
E.~Y. Yuan, ``{Simplicity in AdS Perturbative Dynamics},''
  \href{http://arxiv.org/abs/1801.07283}{{\tt 1801.07283}}.

\bibitem{Ghosh:2018bgd}
K.~Ghosh, ``{Polyakov-Mellin Bootstrap for AdS loops},''
  \href{http://arxiv.org/abs/1811.00504}{{\tt 1811.00504}}.

\bibitem{Liu:2018jhs}
J.~Liu, E.~Perlmutter, V.~Rosenhaus, and D.~Simmons-Duffin, ``{$d$-dimensional
  SYK, AdS Loops, and $6j$ Symbols},''
  \href{http://arxiv.org/abs/1808.00612}{{\tt 1808.00612}}.

\bibitem{Albayrak:2018tam}
S.~Albayrak and S.~Kharel, ``{Towards the higher point holographic momentum
  space amplitudes},'' {\em JHEP} {\bf 02} (2019) 040,
  \href{https://doi.org/10.1007/JHEP02(2019)040}{10.1007/JHEP02(2019)040},
  \href{http://arxiv.org/abs/1810.12459}{{\tt 1810.12459}}.

\bibitem{Albayrak:2019asr}
S.~Albayrak, C.~Chowdhury, and S.~Kharel, ``{New relation for AdS
  amplitudes},'' \href{http://arxiv.org/abs/1904.10043}{{\tt 1904.10043}}.

\bibitem{Raju:2010by}
S.~Raju, ``{BCFW for Witten Diagrams},'' {\em Phys. Rev. Lett.} {\bf 106}
  (2011) 091601,
  \href{https://doi.org/10.1103/PhysRevLett.106.091601}{10.1103/PhysRevLett.106.091601},
  \href{http://arxiv.org/abs/1011.0780}{{\tt 1011.0780}}.

\bibitem{Raju:2011mp}
S.~Raju, ``{Recursion Relations for AdS/CFT Correlators},'' {\em Phys. Rev.}
  {\bf D83} (2011) 126002,
  \href{https://doi.org/10.1103/PhysRevD.83.126002}{10.1103/PhysRevD.83.126002},
  \href{http://arxiv.org/abs/1102.4724}{{\tt 1102.4724}}.

\bibitem{Raju:2012zr}
S.~Raju, ``{New Recursion Relations and a Flat Space Limit for AdS/CFT
  Correlators},'' {\em Phys. Rev.} {\bf D85} (2012) 126009,
  \href{https://doi.org/10.1103/PhysRevD.85.126009}{10.1103/PhysRevD.85.126009},
  \href{http://arxiv.org/abs/1201.6449}{{\tt 1201.6449}}.

\bibitem{Raju:2012zs}
S.~Raju, ``{Four Point Functions of the Stress Tensor and Conserved Currents in
  AdS$_4$/CFT$_3$},'' {\em Phys. Rev.} {\bf D85} (2012) 126008,
  \href{https://doi.org/10.1103/PhysRevD.85.126008}{10.1103/PhysRevD.85.126008},
  \href{http://arxiv.org/abs/1201.6452}{{\tt 1201.6452}}.

\bibitem{Giombi:2017hpr}
S.~Giombi, C.~Sleight, and M.~Taronna, ``{Spinning AdS Loop Diagrams: Two Point
  Functions},'' {\em JHEP} {\bf 06} (2018) 030,
  \href{https://doi.org/10.1007/JHEP06(2018)030}{10.1007/JHEP06(2018)030},
  \href{http://arxiv.org/abs/1708.08404}{{\tt 1708.08404}}.

\bibitem{Bertan:2018khc}
I.~Bertan and I.~Sachs, ``{Loops in Anti–de Sitter Space},'' {\em Phys. Rev.
  Lett.} {\bf 121} (2018), no.~10 101601,
  \href{https://doi.org/10.1103/PhysRevLett.121.101601}{10.1103/PhysRevLett.121.101601},
  \href{http://arxiv.org/abs/1804.01880}{{\tt 1804.01880}}.

\bibitem{Bertan:2018afl}
I.~Bertan, I.~Sachs, and E.~D. Skvortsov, ``{Quantum $\phi^4$ Theory in
  AdS${}_4$ and its CFT Dual},'' {\em JHEP} {\bf 02} (2019) 099,
  \href{https://doi.org/10.1007/JHEP02(2019)099}{10.1007/JHEP02(2019)099},
  \href{http://arxiv.org/abs/1810.00907}{{\tt 1810.00907}}.

\bibitem{Hikida:2016wqj}
Y.~Hikida, ``{The masses of higher spin fields on AdS${}_4$ and conformal
  perturbation theory},'' {\em Phys. Rev.} {\bf D94} (2016), no.~2 026004,
  \href{https://doi.org/10.1103/PhysRevD.94.026004}{10.1103/PhysRevD.94.026004},
  \href{http://arxiv.org/abs/1601.01784}{{\tt 1601.01784}}.

\bibitem{Hikida:2017ecj}
Y.~Hikida and T.~Wada, ``{Marginal deformations of 3d supersymmetric U(N) model
  and broken higher spin symmetry},'' {\em JHEP} {\bf 03} (2017) 047,
  \href{https://doi.org/10.1007/JHEP03(2017)047}{10.1007/JHEP03(2017)047},
  \href{http://arxiv.org/abs/1701.03563}{{\tt 1701.03563}}.

\bibitem{Alday:2017xua}
L.~F. Alday and A.~Bissi, ``{Loop Corrections to Supergravity on $AdS_5 \times
  S^5$},'' {\em Phys. Rev. Lett.} {\bf 119} (2017), no.~17 171601,
  \href{https://doi.org/10.1103/PhysRevLett.119.171601}{10.1103/PhysRevLett.119.171601},
  \href{http://arxiv.org/abs/1706.02388}{{\tt 1706.02388}}.

\bibitem{Aprile:2017bgs}
F.~Aprile, J.~M. Drummond, P.~Heslop, and H.~Paul, ``{Quantum Gravity from
  Conformal Field Theory},'' {\em JHEP} {\bf 01} (2018) 035,
  \href{https://doi.org/10.1007/JHEP01(2018)035}{10.1007/JHEP01(2018)035},
  \href{http://arxiv.org/abs/1706.02822}{{\tt 1706.02822}}.

\bibitem{Ponomarev:2019ltz}
D.~Ponomarev, E.~Sezgin, and E.~Skvortsov, ``{On one loop corrections in higher
  spin gravity},'' \href{http://arxiv.org/abs/1904.01042}{{\tt 1904.01042}}.

\bibitem{Goncalves:2019znr}
V.~Gonçalves, R.~Pereira, and X.~Zhou, ``{$20'$ Five-Point Function from
  $AdS_5\times S^5$ Supergravity},''
  \href{http://arxiv.org/abs/1906.05305}{{\tt 1906.05305}}.

\bibitem{Mack:1974jjo}
G.~Mack, ``{Group Theoretical Approach to Conformal Invariant Quantum Field
  Theory},'' {\em NATO Sci. Ser. B} {\bf 5} (1974) 123--157,
  \href{https://doi.org/10.1007/978-1-4615-8909-9\_7}{10.1007/978-1-4615-8909-9\_7}.

\bibitem{Mack:1974sa}
G.~Mack, ``{Osterwalder-Schrader Positivity in Conformal Invariant Quantum
  Field Theory},'' {\em Lect. Notes Phys.} {\bf 37} (1975) 66--91,
  \href{https://doi.org/10.1007/3-540-07160-1\_3}{10.1007/3-540-07160-1\_3}.

\bibitem{Dobrev:1977qv}
V.~K. Dobrev, G.~Mack, V.~B. Petkova, S.~G. Petrova, and I.~T. Todorov,
  ``{Harmonic Analysis on the n-Dimensional Lorentz Group and Its Application
  to Conformal Quantum Field Theory},'' {\em Lect. Notes Phys.} {\bf 63} (1977)
  1--280, \href{https://doi.org/10.1007/BFb0009678}{10.1007/BFb0009678}.

\bibitem{Dobrev:1975ru}
V.~K. Dobrev, V.~B. Petkova, S.~G. Petrova, and I.~T. Todorov, ``{Dynamical
  Derivation of Vacuum Operator Product Expansion in Euclidean Conformal
  Quantum Field Theory},'' {\em Phys. Rev.} {\bf D13} (1976) 887,
  \href{https://doi.org/10.1103/PhysRevD.13.887}{10.1103/PhysRevD.13.887}.

\bibitem{Ferrara:1972uq}
S.~Ferrara, A.~F. Grillo, G.~Parisi, and R.~Gatto, ``{The shadow operator
  formalism for conformal algebra. Vacuum expectation values and operator
  products},'' {\em Lett. Nuovo Cim.} {\bf 4S2} (1972) 115--120,
  \href{https://doi.org/10.1007/BF02907130}{10.1007/BF02907130}.

\bibitem{Ferrara:1972ay}
S.~Ferrara, A.~F. Grillo, and G.~Parisi, ``{Nonequivalence between conformal
  covariant wilson expansion in euclidean and minkowski space},'' {\em Lett.
  Nuovo Cim.} {\bf 5S2} (1972) 147--151,
  \href{https://doi.org/10.1007/BF02815915}{10.1007/BF02815915}.

\bibitem{Ferrara:1972xe}
S.~Ferrara and G.~Parisi, ``{Conformal covariant correlation functions},'' {\em
  Nucl. Phys.} {\bf B42} (1972) 281--290,
  \href{https://doi.org/10.1016/0550-3213(72)90480-4}{10.1016/0550-3213(72)90480-4}.

\bibitem{Ferrara:1973vz}
S.~Ferrara, A.~F. Grillo, G.~Parisi, and R.~Gatto, ``{Covariant expansion of
  the conformal four-point function},'' {\em Nucl. Phys.} {\bf B49} (1972)
  77--98, \href{https://doi.org/10.1016/0550-3213(72)90587-1,
  10.1016/0550-3213(73)90467-7}{10.1016/0550-3213(72)90587-1,
  10.1016/0550-3213(73)90467-7}.

\bibitem{SimmonsDuffin:2012uy}
D.~Simmons-Duffin, ``{Projectors, Shadows, and Conformal Blocks},'' {\em JHEP}
  {\bf 04} (2014) 146,
  \href{https://doi.org/10.1007/JHEP04(2014)146}{10.1007/JHEP04(2014)146},
  \href{http://arxiv.org/abs/1204.3894}{{\tt 1204.3894}}.

\bibitem{Ferrara:1971vh}
S.~Ferrara, A.~F. Grillo, and R.~Gatto, ``{Manifestly conformal covariant
  operator-product expansion},'' {\em Lett. Nuovo Cim.} {\bf 2S2} (1971)
  1363--1369, \href{https://doi.org/10.1007/BF02770435}{10.1007/BF02770435}.

\bibitem{Ferrara:1974ny}
S.~Ferrara, R.~Gatto, and A.~F. Grillo, ``{Properties of Partial Wave
  Amplitudes in Conformal Invariant Field Theories},'' {\em Nuovo Cim.} {\bf
  A26} (1975) 226,
  \href{https://doi.org/10.1007/BF02769009}{10.1007/BF02769009}.

\bibitem{Dolan:2000ut}
F.~A. Dolan and H.~Osborn, ``{Conformal four point functions and the operator
  product expansion},'' {\em Nucl. Phys.} {\bf B599} (2001) 459--496,
  \href{https://doi.org/10.1016/S0550-3213(01)00013-X}{10.1016/S0550-3213(01)00013-X},
  \href{http://arxiv.org/abs/hep-th/0011040}{{\tt hep-th/0011040}}.

\bibitem{Dolan:2003hv}
F.~A. Dolan and H.~Osborn, ``{Conformal partial waves and the operator product
  expansion},'' {\em Nucl. Phys.} {\bf B678} (2004) 491--507,
  \href{https://doi.org/10.1016/j.nuclphysb.2003.11.016}{10.1016/j.nuclphysb.2003.11.016},
  \href{http://arxiv.org/abs/hep-th/0309180}{{\tt hep-th/0309180}}.

\bibitem{Dolan:2011dv}
F.~A. Dolan and H.~Osborn, ``{Conformal Partial Waves: Further Mathematical
  Results},'' \href{http://arxiv.org/abs/1108.6194}{{\tt 1108.6194}}.

\bibitem{Rosenhaus:2018zqn}
V.~Rosenhaus, ``{Multipoint Conformal Blocks in the Comb Channel},'' {\em JHEP}
  {\bf 02} (2019) 142,
  \href{https://doi.org/10.1007/JHEP02(2019)142}{10.1007/JHEP02(2019)142},
  \href{http://arxiv.org/abs/1810.03244}{{\tt 1810.03244}}.

\bibitem{Fortin:2019fvx}
J.-F. Fortin and W.~Skiba, ``{A recipe for conformal blocks},''
  \href{http://arxiv.org/abs/1905.00036}{{\tt 1905.00036}}.

\bibitem{Fortin:2019dnq}
J.-F. Fortin and W.~Skiba, ``{New Methods for Conformal Correlation
  Functions},'' \href{http://arxiv.org/abs/1905.00434}{{\tt 1905.00434}}.

\bibitem{Komargodski:2012ek}
Z.~Komargodski and A.~Zhiboedov, ``{Convexity and Liberation at Large Spin},''
  {\em JHEP} {\bf 11} (2013) 140,
  \href{https://doi.org/10.1007/JHEP11(2013)140}{10.1007/JHEP11(2013)140},
  \href{http://arxiv.org/abs/1212.4103}{{\tt 1212.4103}}.

\bibitem{Fitzpatrick:2012yx}
A.~L. Fitzpatrick, J.~Kaplan, D.~Poland, and D.~Simmons-Duffin, ``{The Analytic
  Bootstrap and AdS Superhorizon Locality},'' {\em JHEP} {\bf 12} (2013) 004,
  \href{https://doi.org/10.1007/JHEP12(2013)004}{10.1007/JHEP12(2013)004},
  \href{http://arxiv.org/abs/1212.3616}{{\tt 1212.3616}}.

\bibitem{Fitzpatrick:2014vua}
A.~L. Fitzpatrick, J.~Kaplan, and M.~T. Walters, ``{Universality of
  Long-Distance AdS Physics from the CFT Bootstrap},'' {\em JHEP} {\bf 08}
  (2014) 145,
  \href{https://doi.org/10.1007/JHEP08(2014)145}{10.1007/JHEP08(2014)145},
  \href{http://arxiv.org/abs/1403.6829}{{\tt 1403.6829}}.

\bibitem{Kaviraj:2015cxa}
A.~Kaviraj, K.~Sen, and A.~Sinha, ``{Analytic bootstrap at large spin},'' {\em
  JHEP} {\bf 11} (2015) 083,
  \href{https://doi.org/10.1007/JHEP11(2015)083}{10.1007/JHEP11(2015)083},
  \href{http://arxiv.org/abs/1502.01437}{{\tt 1502.01437}}.

\bibitem{Kaviraj:2015xsa}
A.~Kaviraj, K.~Sen, and A.~Sinha, ``{Universal anomalous dimensions at large
  spin and large twist},'' {\em JHEP} {\bf 07} (2015) 026,
  \href{https://doi.org/10.1007/JHEP07(2015)026}{10.1007/JHEP07(2015)026},
  \href{http://arxiv.org/abs/1504.00772}{{\tt 1504.00772}}.

\bibitem{Alday:2015ewa}
L.~F. Alday and A.~Zhiboedov, ``{An Algebraic Approach to the Analytic
  Bootstrap},'' {\em JHEP} {\bf 04} (2017) 157,
  \href{https://doi.org/10.1007/JHEP04(2017)157}{10.1007/JHEP04(2017)157},
  \href{http://arxiv.org/abs/1510.08091}{{\tt 1510.08091}}.

\bibitem{Alday:2016njk}
L.~F. Alday, ``{Large Spin Perturbation Theory for Conformal Field Theories},''
  {\em Phys. Rev. Lett.} {\bf 119} (2017), no.~11 111601,
  \href{https://doi.org/10.1103/PhysRevLett.119.111601}{10.1103/PhysRevLett.119.111601},
  \href{http://arxiv.org/abs/1611.01500}{{\tt 1611.01500}}.

\bibitem{Simmons-Duffin:2016wlq}
D.~Simmons-Duffin, ``{The Lightcone Bootstrap and the Spectrum of the 3d Ising
  CFT},'' {\em JHEP} {\bf 03} (2017) 086,
  \href{https://doi.org/10.1007/JHEP03(2017)086}{10.1007/JHEP03(2017)086},
  \href{http://arxiv.org/abs/1612.08471}{{\tt 1612.08471}}.

\bibitem{Caron-Huot:2017vep}
S.~Caron-Huot, ``{Analyticity in Spin in Conformal Theories},'' {\em JHEP} {\bf
  09} (2017) 078,
  \href{https://doi.org/10.1007/JHEP09(2017)078}{10.1007/JHEP09(2017)078},
  \href{http://arxiv.org/abs/1703.00278}{{\tt 1703.00278}}.

\bibitem{Albayrak:2019gnz}
S.~Albayrak, D.~Meltzer, and D.~Poland, ``{More Analytic Bootstrap:
  Nonperturbative Effects and Fermions},''
  \href{http://arxiv.org/abs/1904.00032}{{\tt 1904.00032}}.

\bibitem{Parikh:2019ygo}
S.~Parikh, ``{Holographic dual of the five-point conformal block},'' {\em JHEP}
  {\bf 05} (2019) 051,
  \href{https://doi.org/10.1007/JHEP05(2019)051}{10.1007/JHEP05(2019)051},
  \href{http://arxiv.org/abs/1901.01267}{{\tt 1901.01267}}.

\bibitem{Gubser:2016guj}
S.~S. Gubser, J.~Knaute, S.~Parikh, A.~Samberg, and P.~Witaszczyk, ``{$p$-adic
  AdS/CFT},'' {\em Commun. Math. Phys.} {\bf 352} (2017), no.~3 1019--1059,
  \href{https://doi.org/10.1007/s00220-016-2813-6}{10.1007/s00220-016-2813-6},
  \href{http://arxiv.org/abs/1605.01061}{{\tt 1605.01061}}.

\bibitem{Gubser:2017tsi}
S.~S. Gubser and S.~Parikh, ``{Geodesic bulk diagrams on the Bruhat–Tits
  tree},'' {\em Phys. Rev.} {\bf D96} (2017), no.~6 066024,
  \href{https://doi.org/10.1103/PhysRevD.96.066024}{10.1103/PhysRevD.96.066024},
  \href{http://arxiv.org/abs/1704.01149}{{\tt 1704.01149}}.

\bibitem{Liu:1998th}
H.~Liu, ``{Scattering in anti-de Sitter space and operator product
  expansion},'' {\em Phys. Rev.} {\bf D60} (1999) 106005,
  \href{https://doi.org/10.1103/PhysRevD.60.106005}{10.1103/PhysRevD.60.106005},
  \href{http://arxiv.org/abs/hep-th/9811152}{{\tt hep-th/9811152}}.

\bibitem{Costa:2014kfa}
M.~S. Costa, V.~Gonçalves, and J.~Penedones, ``{Spinning AdS Propagators},''
  {\em JHEP} {\bf 09} (2014) 064,
  \href{https://doi.org/10.1007/JHEP09(2014)064}{10.1007/JHEP09(2014)064},
  \href{http://arxiv.org/abs/1404.5625}{{\tt 1404.5625}}.

\bibitem{Sleight:2017fpc}
C.~Sleight and M.~Taronna, ``{Spinning Witten Diagrams},'' {\em JHEP} {\bf 06}
  (2017) 100,
  \href{https://doi.org/10.1007/JHEP06(2017)100}{10.1007/JHEP06(2017)100},
  \href{http://arxiv.org/abs/1702.08619}{{\tt 1702.08619}}.

\bibitem{Zhou:2018sfz}
X.~Zhou, ``{Recursion Relations in Witten Diagrams and Conformal Partial
  Waves},'' \href{http://arxiv.org/abs/1812.01006}{{\tt 1812.01006}}.

\bibitem{Hoffmann:2000mx}
L.~Hoffmann, A.~C. Petkou, and W.~Ruhl, ``{Aspects of the conformal operator
  product expansion in AdS / CFT correspondence},'' {\em Adv. Theor. Math.
  Phys.} {\bf 4} (2002) 571--615,
  \href{https://doi.org/10.4310/ATMP.2000.v4.n3.a3}{10.4310/ATMP.2000.v4.n3.a3},
  \href{http://arxiv.org/abs/hep-th/0002154}{{\tt hep-th/0002154}}.

\bibitem{Sleight:2018epi}
C.~Sleight and M.~Taronna, ``{Spinning Mellin Bootstrap: Conformal Partial
  Waves, Crossing Kernels and Applications},'' {\em Fortsch. Phys.} {\bf 66}
  (2018), no.~8-9 1800038,
  \href{https://doi.org/10.1002/prop.201800038}{10.1002/prop.201800038},
  \href{http://arxiv.org/abs/1804.09334}{{\tt 1804.09334}}.

\bibitem{Sleight:2018ryu}
C.~Sleight and M.~Taronna, ``{Anomalous Dimensions from Crossing Kernels},''
  {\em JHEP} {\bf 11} (2018) 089,
  \href{https://doi.org/10.1007/JHEP11(2018)089}{10.1007/JHEP11(2018)089},
  \href{http://arxiv.org/abs/1807.05941}{{\tt 1807.05941}}.

\bibitem{Cardona:2018dov}
C.~Cardona and K.~Sen, ``{Anomalous dimensions at finite conformal spin from
  OPE inversion},'' {\em JHEP} {\bf 11} (2018) 052,
  \href{https://doi.org/10.1007/JHEP11(2018)052}{10.1007/JHEP11(2018)052},
  \href{http://arxiv.org/abs/1806.10919}{{\tt 1806.10919}}.

\bibitem{Li:2019dix}
W.~Li, ``{Closed-form expression for cross-channel conformal blocks near the
  lightcone},'' \href{http://arxiv.org/abs/1906.00707}{{\tt 1906.00707}}.

\bibitem{Chen:2019gka}
H.-Y. Chen and H.~Kyono, ``{On Conformal Block, Crossing Kernel and
  Multi-variable Hypergeometric Functions},''
  \href{http://arxiv.org/abs/1906.03135}{{\tt 1906.03135}}.

\bibitem{Nishida:2016vds}
M.~Nishida and K.~Tamaoka, ``{Geodesic Witten diagrams with an external
  spinning field},'' {\em PTEP} {\bf 2017} (2017), no.~5 053B06,
  \href{https://doi.org/10.1093/ptep/ptx055}{10.1093/ptep/ptx055},
  \href{http://arxiv.org/abs/1609.04563}{{\tt 1609.04563}}.

\bibitem{Castro:2017hpx}
A.~Castro, E.~Llabrés, and F.~Rejon-Barrera, ``{Geodesic Diagrams,
  Gravitational Interactions \& OPE Structures},'' {\em JHEP} {\bf 06} (2017)
  099, \href{https://doi.org/10.1007/JHEP06(2017)099}{10.1007/JHEP06(2017)099},
  \href{http://arxiv.org/abs/1702.06128}{{\tt 1702.06128}}.

\bibitem{Dyer:2017zef}
E.~Dyer, D.~Z. Freedman, and J.~Sully, ``{Spinning Geodesic Witten Diagrams},''
  {\em JHEP} {\bf 11} (2017) 060,
  \href{https://doi.org/10.1007/JHEP11(2017)060}{10.1007/JHEP11(2017)060},
  \href{http://arxiv.org/abs/1702.06139}{{\tt 1702.06139}}.

\bibitem{Chen:2017yia}
H.-Y. Chen, E.-J. Kuo, and H.~Kyono, ``{Anatomy of Geodesic Witten Diagrams},''
  {\em JHEP} {\bf 05} (2017) 070,
  \href{https://doi.org/10.1007/JHEP05(2017)070}{10.1007/JHEP05(2017)070},
  \href{http://arxiv.org/abs/1702.08818}{{\tt 1702.08818}}.

\bibitem{Kraus:2017ezw}
P.~Kraus, A.~Maloney, H.~Maxfield, G.~S. Ng, and J.-q. Wu, ``{Witten Diagrams
  for Torus Conformal Blocks},'' {\em JHEP} {\bf 09} (2017) 149,
  \href{https://doi.org/10.1007/JHEP09(2017)149}{10.1007/JHEP09(2017)149},
  \href{http://arxiv.org/abs/1706.00047}{{\tt 1706.00047}}.

\bibitem{Tamaoka:2017jce}
K.~Tamaoka, ``{Geodesic Witten diagrams with antisymmetric tensor exchange},''
  {\em Phys. Rev.} {\bf D96} (2017), no.~8 086007,
  \href{https://doi.org/10.1103/PhysRevD.96.086007}{10.1103/PhysRevD.96.086007},
  \href{http://arxiv.org/abs/1707.07934}{{\tt 1707.07934}}.

\bibitem{Nishida:2018opl}
M.~Nishida and K.~Tamaoka, ``{Fermions in Geodesic Witten Diagrams},'' {\em
  JHEP} {\bf 07} (2018) 149,
  \href{https://doi.org/10.1007/JHEP07(2018)149}{10.1007/JHEP07(2018)149},
  \href{http://arxiv.org/abs/1805.00217}{{\tt 1805.00217}}.

\bibitem{Das:2018ajg}
S.~Das, ``{Comments on spinning OPE blocks in AdS$_{3}$/CFT$_{2}$},''
  \href{https://doi.org/10.1016/j.physletb.2019.03.058}{10.1016/j.physletb.2019.03.058},
  \href{http://arxiv.org/abs/1811.09375}{{\tt 1811.09375}}.

\bibitem{Bhatta:2016hpz}
A.~Bhatta, P.~Raman, and N.~V. Suryanarayana, ``{Holographic Conformal Partial
  Waves as Gravitational Open Wilson Networks},'' {\em JHEP} {\bf 06} (2016)
  119, \href{https://doi.org/10.1007/JHEP06(2016)119}{10.1007/JHEP06(2016)119},
  \href{http://arxiv.org/abs/1602.02962}{{\tt 1602.02962}}.

\bibitem{Bhatta:2018gjb}
A.~Bhatta, P.~Raman, and N.~V. Suryanarayana, ``{Scalar Blocks as Gravitational
  Wilson Networks},'' \href{http://arxiv.org/abs/1806.05475}{{\tt 1806.05475}}.

\bibitem{Hijano:2015rla}
E.~Hijano, P.~Kraus, and R.~Snively, ``{Worldline approach to semi-classical
  conformal blocks},'' {\em JHEP} {\bf 07} (2015) 131,
  \href{https://doi.org/10.1007/JHEP07(2015)131}{10.1007/JHEP07(2015)131},
  \href{http://arxiv.org/abs/1501.02260}{{\tt 1501.02260}}.

\bibitem{Fitzpatrick:2015zha}
A.~L. Fitzpatrick, J.~Kaplan, and M.~T. Walters, ``{Virasoro Conformal Blocks
  and Thermality from Classical Background Fields},'' {\em JHEP} {\bf 11}
  (2015) 200,
  \href{https://doi.org/10.1007/JHEP11(2015)200}{10.1007/JHEP11(2015)200},
  \href{http://arxiv.org/abs/1501.05315}{{\tt 1501.05315}}.

\bibitem{Hijano:2015qja}
E.~Hijano, P.~Kraus, E.~Perlmutter, and R.~Snively, ``{Semiclassical Virasoro
  blocks from AdS$_{3}$ gravity},'' {\em JHEP} {\bf 12} (2015) 077,
  \href{https://doi.org/10.1007/JHEP12(2015)077}{10.1007/JHEP12(2015)077},
  \href{http://arxiv.org/abs/1508.04987}{{\tt 1508.04987}}.

\bibitem{Besken:2016ooo}
M.~Besken, A.~Hegde, E.~Hijano, and P.~Kraus, ``{Holographic conformal blocks
  from interacting Wilson lines},'' {\em JHEP} {\bf 08} (2016) 099,
  \href{https://doi.org/10.1007/JHEP08(2016)099}{10.1007/JHEP08(2016)099},
  \href{http://arxiv.org/abs/1603.07317}{{\tt 1603.07317}}.

\bibitem{Alkalaev:2015wia}
K.~B. Alkalaev and V.~A. Belavin, ``{Classical conformal blocks via AdS/CFT
  correspondence},'' {\em JHEP} {\bf 08} (2015) 049,
  \href{https://doi.org/10.1007/JHEP08(2015)049}{10.1007/JHEP08(2015)049},
  \href{http://arxiv.org/abs/1504.05943}{{\tt 1504.05943}}.

\bibitem{Alkalaev:2015lca}
K.~B. Alkalaev and V.~A. Belavin, ``{Monodromic vs geodesic computation of
  Virasoro classical conformal blocks},'' {\em Nucl. Phys.} {\bf B904} (2016)
  367--385,
  \href{https://doi.org/10.1016/j.nuclphysb.2016.01.019}{10.1016/j.nuclphysb.2016.01.019},
  \href{http://arxiv.org/abs/1510.06685}{{\tt 1510.06685}}.

\bibitem{Alkalaev:2015fbw}
K.~B. Alkalaev and V.~A. Belavin, ``{From global to heavy-light: 5-point
  conformal blocks},'' {\em JHEP} {\bf 03} (2016) 184,
  \href{https://doi.org/10.1007/JHEP03(2016)184}{10.1007/JHEP03(2016)184},
  \href{http://arxiv.org/abs/1512.07627}{{\tt 1512.07627}}.

\bibitem{Banerjee:2016qca}
P.~Banerjee, S.~Datta, and R.~Sinha, ``{Higher-point conformal blocks and
  entanglement entropy in heavy states},'' {\em JHEP} {\bf 05} (2016) 127,
  \href{https://doi.org/10.1007/JHEP05(2016)127}{10.1007/JHEP05(2016)127},
  \href{http://arxiv.org/abs/1601.06794}{{\tt 1601.06794}}.

\bibitem{Alkalaev:2016rjl}
K.~B. Alkalaev, ``{Many-point classical conformal blocks and geodesic networks
  on the hyperbolic plane},'' {\em JHEP} {\bf 12} (2016) 070,
  \href{https://doi.org/10.1007/JHEP12(2016)070}{10.1007/JHEP12(2016)070},
  \href{http://arxiv.org/abs/1610.06717}{{\tt 1610.06717}}.

\bibitem{Alkalaev:2018nik}
K.~Alkalaev and M.~Pavlov, ``{Perturbative classical conformal blocks as
  Steiner trees on the hyperbolic disk},''
  \href{http://arxiv.org/abs/1810.07741}{{\tt 1810.07741}}.

\bibitem{Alkalaev:2019zhs}
K.~B. Alkalaev and M.~Pavlov, ``{Four-point conformal blocks with three heavy
  background operators},'' \href{http://arxiv.org/abs/1905.03195}{{\tt
  1905.03195}}.

\bibitem{DEramo:1971hnd}
M.~D'Eramo, G.~Parisi, and L.~Peliti, ``{Theoretical Predictions for Critical
  Exponents at the Lambda Point of Bose Liquids},'' {\em Lett. Nuovo Cim.} {\bf
  2} (1971), no.~17 878--880,
  \href{https://doi.org/10.1007/BF02774121}{10.1007/BF02774121}.

\bibitem{Heydeman:2016ldy}
M.~Heydeman, M.~Marcolli, I.~Saberi, and B.~Stoica, ``{Tensor networks,
  $p$-adic fields, and algebraic curves: arithmetic and the AdS$_3$/CFT$_2$
  correspondence},'' {\em Adv. Theor. Math. Phys.} {\bf 22} (2018) 93--176,
  \href{https://doi.org/10.4310/ATMP.2018.v22.n1.a4}{10.4310/ATMP.2018.v22.n1.a4},
  \href{http://arxiv.org/abs/1605.07639}{{\tt 1605.07639}}.

\bibitem{Melzer:1988he}
E.~Melzer, ``{Nonarchimedean Conformal Field Theories},'' {\em Int. J. Mod.
  Phys.} {\bf A4} (1989) 4877,
  \href{https://doi.org/10.1142/S0217751X89002065}{10.1142/S0217751X89002065}.

\bibitem{Gubser:2017vgc}
S.~S. Gubser, C.~Jepsen, S.~Parikh, and B.~Trundy, ``{O(N) and O(N) and
  O(N)},'' {\em JHEP} {\bf 11} (2017) 107,
  \href{https://doi.org/10.1007/JHEP11(2017)107}{10.1007/JHEP11(2017)107},
  \href{http://arxiv.org/abs/1703.04202}{{\tt 1703.04202}}.

\bibitem{Jepsen:2018dqp}
C.~B. Jepsen and S.~Parikh, ``{$p$-adic Mellin Amplitudes},'' {\em JHEP} {\bf
  04} (2019) 101,
  \href{https://doi.org/10.1007/JHEP04(2019)101}{10.1007/JHEP04(2019)101},
  \href{http://arxiv.org/abs/1808.08333}{{\tt 1808.08333}}.

\bibitem{Jepsen:2018ajn}
C.~B. Jepsen and S.~Parikh, ``{Recursion Relations in $p$-adic Mellin Space},''
  {\em J. Phys.} {\bf A52} (2019), no.~28 285401,
  \href{https://doi.org/10.1088/1751-8121/ab227b}{10.1088/1751-8121/ab227b},
  \href{http://arxiv.org/abs/1812.09801}{{\tt 1812.09801}}.

\bibitem{Marcolli:2018ohd}
M.~Marcolli, ``{Holographic Codes on Bruhat--Tits buildings and Drinfeld
  Symmetric Spaces},'' \href{http://arxiv.org/abs/1801.09623}{{\tt
  1801.09623}}.

\bibitem{Heydeman:2018qty}
M.~Heydeman, M.~Marcolli, S.~Parikh, and I.~Saberi, ``{Nonarchimedean
  Holographic Entropy from Networks of Perfect Tensors},''
  \href{http://arxiv.org/abs/1812.04057}{{\tt 1812.04057}}.

\bibitem{Gubser:2016htz}
S.~S. Gubser, M.~Heydeman, C.~Jepsen, M.~Marcolli, S.~Parikh, I.~Saberi,
  B.~Stoica, and B.~Trundy, ``{Edge length dynamics on graphs with applications
  to $p$-adic AdS/CFT},'' {\em JHEP} {\bf 06} (2017) 157,
  \href{https://doi.org/10.1007/JHEP06(2017)157}{10.1007/JHEP06(2017)157},
  \href{http://arxiv.org/abs/1612.09580}{{\tt 1612.09580}}.

\bibitem{Bhattacharyya:2017aly}
A.~Bhattacharyya, L.-Y. Hung, Y.~Lei, and W.~Li, ``{Tensor network and
  ($p$-adic) AdS/CFT},'' {\em JHEP} {\bf 01} (2018) 139,
  \href{https://doi.org/10.1007/JHEP01(2018)139}{10.1007/JHEP01(2018)139},
  \href{http://arxiv.org/abs/1703.05445}{{\tt 1703.05445}}.

\bibitem{Dutta:2017bja}
P.~Dutta, D.~Ghoshal, and A.~Lala, ``{Notes on exchange interactions in
  holographic p -adic CFT},'' {\em Phys. Lett.} {\bf B773} (2017) 283--289,
  \href{https://doi.org/10.1016/j.physletb.2017.08.042}{10.1016/j.physletb.2017.08.042},
  \href{http://arxiv.org/abs/1705.05678}{{\tt 1705.05678}}.

\bibitem{Gubser:2017qed}
S.~S. Gubser, M.~Heydeman, C.~Jepsen, S.~Parikh, I.~Saberi, B.~Stoica, and
  B.~Trundy, ``{Melonic theories over diverse number systems},'' {\em Phys.
  Rev.} {\bf D98} (2018), no.~12 126007,
  \href{https://doi.org/10.1103/PhysRevD.98.126007}{10.1103/PhysRevD.98.126007},
  \href{http://arxiv.org/abs/1707.01087}{{\tt 1707.01087}}.

\bibitem{Qu:2018ned}
F.~Qu and Y.-h. Gao, ``{Scalar fields on $p$AdS},'' {\em Phys. Lett.} {\bf
  B786} (2018) 165--170,
  \href{https://doi.org/10.1016/j.physletb.2018.09.043}{10.1016/j.physletb.2018.09.043},
  \href{http://arxiv.org/abs/1806.07035}{{\tt 1806.07035}}.

\bibitem{Gubser:2018cha}
S.~S. Gubser, C.~Jepsen, and B.~Trundy, ``{Spin in $p$-adic AdS/CFT},'' {\em J.
  Phys.} {\bf A52} (2019), no.~14 144004,
  \href{https://doi.org/10.1088/1751-8121/ab0757}{10.1088/1751-8121/ab0757},
  \href{http://arxiv.org/abs/1811.02538}{{\tt 1811.02538}}.

\bibitem{Hung:2018mcn}
L.-Y. Hung, W.~Li, and C.~M. Melby-Thompson, ``{Wilson line networks in
  $p$-adic AdS/CFT},'' \href{http://arxiv.org/abs/1812.06059}{{\tt
  1812.06059}}.

\bibitem{Hung:2019zsk}
L.-Y. Hung, W.~Li, and C.~M. Melby-Thompson, ``{$p$-adic CFT is a holographic
  tensor network},'' {\em JHEP} {\bf 04} (2019) 170,
  \href{https://doi.org/10.1007/JHEP04(2019)170}{10.1007/JHEP04(2019)170},
  \href{http://arxiv.org/abs/1902.01411}{{\tt 1902.01411}}.

\bibitem{gouvea1997p}
F.~Q. Gouv{\^e}a, ``p-adic Numbers,'' in {\em p-adic Numbers}, pp.~43--85.
\newblock Springer, 1997.

\bibitem{Czech:2016xec}
B.~Czech, L.~Lamprou, S.~McCandlish, B.~Mosk, and J.~Sully, ``{A Stereoscopic
  Look into the Bulk},'' {\em JHEP} {\bf 07} (2016) 129,
  \href{https://doi.org/10.1007/JHEP07(2016)129}{10.1007/JHEP07(2016)129},
  \href{http://arxiv.org/abs/1604.03110}{{\tt 1604.03110}}.

\bibitem{deBoer:2016pqk}
J.~de~Boer, F.~M. Haehl, M.~P. Heller, and R.~C. Myers, ``{Entanglement,
  holography and causal diamonds},'' {\em JHEP} {\bf 08} (2016) 162,
  \href{https://doi.org/10.1007/JHEP08(2016)162}{10.1007/JHEP08(2016)162},
  \href{http://arxiv.org/abs/1606.03307}{{\tt 1606.03307}}.

\bibitem{daCunha:2016crm}
B.~Carneiro~da Cunha and M.~Guica, ``{Exploring the BTZ bulk with boundary
  conformal blocks},'' \href{http://arxiv.org/abs/1604.07383}{{\tt
  1604.07383}}.

\bibitem{Guica:2016pid}
M.~Guica, ``{Bulk fields from the boundary OPE},''
  \href{http://arxiv.org/abs/1610.08952}{{\tt 1610.08952}}.

\bibitem{wolfram2}
 Relations between contiguous functions for hypergeometric ${}_3F_2$ functions,
  \url{http://functions.wolfram.com/07.27.17.0007.01}.

\bibitem{Freedman:1998bj}
D.~Z. Freedman, S.~D. Mathur, A.~Matusis, and L.~Rastelli, ``{Comments on 4
  point functions in the CFT / AdS correspondence},'' {\em Phys. Lett.} {\bf
  B452} (1999) 61--68,
  \href{https://doi.org/10.1016/S0370-2693(99)00229-4}{10.1016/S0370-2693(99)00229-4},
  \href{http://arxiv.org/abs/hep-th/9808006}{{\tt hep-th/9808006}}.

\bibitem{DHoker:1999mic}
E.~D'Hoker, S.~D. Mathur, A.~Matusis, and L.~Rastelli, ``{The Operator product
  expansion of N=4 SYM and the 4 point functions of supergravity},'' {\em Nucl.
  Phys.} {\bf B589} (2000) 38--74,
  \href{https://doi.org/10.1016/S0550-3213(00)00523-X}{10.1016/S0550-3213(00)00523-X},
  \href{http://arxiv.org/abs/hep-th/9911222}{{\tt hep-th/9911222}}.

\bibitem{Giombi:2009wh}
S.~Giombi and X.~Yin, ``{Higher Spin Gauge Theory and Holography: The
  Three-Point Functions},'' {\em JHEP} {\bf 09} (2010) 115,
  \href{https://doi.org/10.1007/JHEP09(2010)115}{10.1007/JHEP09(2010)115},
  \href{http://arxiv.org/abs/0912.3462}{{\tt 0912.3462}}.

\bibitem{Bekaert:2014cea}
X.~Bekaert, J.~Erdmenger, D.~Ponomarev, and C.~Sleight, ``{Towards holographic
  higher-spin interactions: Four-point functions and higher-spin exchange},''
  {\em JHEP} {\bf 03} (2015) 170,
  \href{https://doi.org/10.1007/JHEP03(2015)170}{10.1007/JHEP03(2015)170},
  \href{http://arxiv.org/abs/1412.0016}{{\tt 1412.0016}}.

\bibitem{Bekaert:2015tva}
X.~Bekaert, J.~Erdmenger, D.~Ponomarev, and C.~Sleight, ``{Quartic AdS
  Interactions in Higher-Spin Gravity from Conformal Field Theory},'' {\em
  JHEP} {\bf 11} (2015) 149,
  \href{https://doi.org/10.1007/JHEP11(2015)149}{10.1007/JHEP11(2015)149},
  \href{http://arxiv.org/abs/1508.04292}{{\tt 1508.04292}}.

\bibitem{Fitzpatrick:2011dm}
A.~L. Fitzpatrick and J.~Kaplan, ``{Unitarity and the Holographic S-Matrix},''
  {\em JHEP} {\bf 10} (2012) 032,
  \href{https://doi.org/10.1007/JHEP10(2012)032}{10.1007/JHEP10(2012)032},
  \href{http://arxiv.org/abs/1112.4845}{{\tt 1112.4845}}.

\bibitem{Gadde:2017sjg}
A.~Gadde, ``{In search of conformal theories},''
  \href{http://arxiv.org/abs/1702.07362}{{\tt 1702.07362}}.

\bibitem{wolfram4F3}
 Specific value of the hypergeometric ${}_4F_3$ function,
  \url{http://functions.wolfram.com/07.28.03.0050.01}.

\bibitem{andrews1999special}
G.~E. Andrews, R.~Askey, and R.~Roy, ``Special Functions, volume 71 of
  Encyclopedia of Mathematics and its Applications,'' 1999.

\bibitem{Milgram}
M.~Milgram, ``{447 Instances of Hypergeometric 3F2(1)},''
  \href{http://arxiv.org/abs/1105.3126}{{\tt 1105.3126}}.

\end{thebibliography}\endgroup
\end{document}